\documentclass[aps,pre,twocolumn,a4paper,nofootinbib,longbibliography]{revtex4-1}

\usepackage[bbgreekl]{mathbbol} 
\usepackage{amsfonts,amsthm,amsmath,amssymb,mathtools}
\usepackage{color,graphics,graphicx,hyperref}
\usepackage{upgreek}
\usepackage{afterpage,placeins}
\usepackage{chngcntr,enumitem}
\usepackage[normalem]{ulem}

\newcommand*{\diff}{\mathop{}\!\mathrm{d}}
\newcommand{\derpar}[2]{\frac{\partial #1}{\partial #2}}
\newcommand{\dersecpar}[2]{\frac{\partial^2 #1}{\partial #2^2}}
\renewcommand{\vec}{\mathbf}
\renewcommand{\Re}{\mathrm{Re}\,}
\renewcommand{\Im}{\mathrm{Im}\,}
\newcommand{\mrI}{\mathbb{I}}

\newcommand{\mrD}{\mathbb{D}}

\newcommand{\mcF}{\mathcal{F}}
\newcommand{\vecomega}{\boldsymbol{\omega}}
\newcommand{\vecxi}{\boldsymbol{\Xi}}

\newcommand{\rd}[1] {#1}

\newcommand{\la}{\langle}
\newcommand{\ra}{\rangle}
\newcommand{\norm}[1]{\left\lVert#1\right\rVert}

\counterwithin*{equation}{section}
\renewcommand\theequation{\arabic{section}.\arabic{equation}}

\begin{document}

\title{Active L{\'e}vy Matter: Anomalous Diffusion, Hydrodynamics and Linear Stability}

\author{Andrea Cairoli}\email{andrea.cairoli@crick.ac.uk}
\altaffiliation[current author's affiliation: ]{The Francis Crick Institute, London NW1 1AT, United Kingdom.}
\affiliation{Department of Bioengineering, Imperial College London, London SW7 2AZ, United Kingdom}
\author{Chiu Fan Lee}\email{c.lee@imperial.ac.uk}
\affiliation{Department of Bioengineering, Imperial College London, London SW7 2AZ, United Kingdom}

\begin{abstract}

Anomalous diffusion, 
\rd{manifest as a nonlinear temporal evolution of the position mean square displacement,} 
and/or non-Gaussian features of the position statistics, 
is prevalent in biological transport processes. 
Likewise, collective behavior is often observed to emerge spontaneously from the mutual interactions between constituent motile units in biological systems.
Examples where these phenomena can be observed simultaneously have been identified in recent experiments on bird flocks, fish schools and bacterial swarms.
These results pose an intriguing question, 
which cannot be resolved by existing theories of active matter: 
How is the collective motion of these systems affected by the anomalous diffusion of the constituent units? 
Here, we answer this question for a microscopic model of \textit{active L{\'e}vy matter}
--
a collection of active particles
\rd{that perform superdiffusion akin to a L{\'e}vy flight
and interact by promoting polar alignment of their orientations.}
We present in details the derivation of the hydrodynamic equations of motion of the model, 
obtain from these equations 
\rd{the criteria for a disordered or ordered state,}
\rd{and apply linear stability analysis on these states at the onset of collective motion.} 
Our analysis \rd{reveals} that the disorder-order phase transition in active L{\'e}vy matter is critical, 
in contrast to ordinary active fluids where the phase transition is, instead, first-order.   
\rd{Correspondingly, we estimate the critical exponents of the transition by finite size scaling analysis}
and use these numerical estimates to relate our findings to known universality classes.
\rd{These results highlight the novel physics exhibited by active matter integrating both anomalous diffusive single-particle motility and inter-particle interactions.}

\end{abstract}

\maketitle

Active matter refers to systems comprising constitutive units with 
the ability to harvest energy from the environment 
and employ it to generate motion \cite{Toner2005,Schweitzer2007,Ramaswamy2010,Marchetti2013,Hauser2015,Needleman2017}.   
The mutual interactions of the active constituent units in these systems 
often cause the spontaneous emergence of collective behavior manifest as 
collective motion (flocking) \cite{Vicsek1995,Toner1995,Toner1998,Vicsek2012},
turbulence \cite{Dombrowski2004,Hernandez-Ortiz2005,Sokolov2007,Aranson2007,Saintillan2007,Wolgemuth2008,
Sanchez2012,Wensink2012,Doostmohammadi2018},
and motility-induced phase separation \cite{Tailleur2008,Fily2012,Redner2013,Cates2015}. 
Examples are widespread in nature: 
microorganisms swimming in ambient fluids   
\cite{Dombrowski2004,Sokolov2007,Wensink2012,Lauga2009,Koch2011,Elgeti2015}, 
cell tissues \cite{Poujade2007,Saw2017,Kawaguchi2017,Blanch2018,Xi2018,Trepat2018},  
the cellular cytoskeleton \cite{Mackintosh2010,Prost2015,Needleman2017},    
and social groups of animals 
(from small-scale ones like ants and locusts 
to large-scale ones like birds, fish and even humans) 
\cite{Helbing2001,Buhl2006,Ballerini2008,Vicsek2012,Feinerman2018}.
Furthermore, there has been wide interest in artificially engineering active systems. Numerous examples can be mentioned, such as collections of    
chemically self-propelled Janus particles \cite{Bechinger2016} and   
swarming robots \cite{Brambilla2013}, 
and motility assays of self-assembled cytoskeletal components, 
especially cross-linked actin filaments or microtubules and motor proteins 
\cite{Nedlec1997,Schaller2010,Butt2010}.

Being inherently out of equilibrium due to the continuous energy flux, 
active matter systems cannot be described within the framework of
equilibrium statistical mechanics \cite{Chaikin1995}. 
Instead, several agent-based microscopic models and hydrodynamic continuum theories 
have been formulated so far 
to capture the characteristic collective properties of these systems 
(e.g., see \cite{Marchetti2013}).  
These hydrodynamic theories are usually derived by coarse-graining the many-body dynamics resulting from the underlying interacting active units  \cite{Bertin2006,Peruani2008,Baskaran2008,Bertin2009,Lee2010,Peshkov2014,Thuroff2014,Bertin2017}, 
whose motion is typically assumed to be simple self-propulsion plus Gaussian fluctuations in their orientations and spatial positions	
\cite{Schweitzer2007,Romanczuk2012,Hauser2015,Marchetti2013}.
Alternatively, hydrodynamic theories have also been derived in full generality from first principles based on symmetry arguments \cite{Toner1995,Toner1998,Toner2012,Simha2002,Hatwalne2004,Toner2005,
Kruse2004,Kruse2005,Julicher2018}. 
\rd{To incorporate the aforementioned microscopic fluctuations, these theories typically contain Gaussian noise terms, 
which are a manifestation of the celebrated \textit{central limit theorem} \cite{Gardiner2009}, 
stating that the averaging over many fluctuations following a distribution with finite variance inevitably leads to a Gaussian distribution law. }

In particular, flocking systems have been intensely studied in this respect \cite{Toner2005,Vicsek2012,Marchetti2013}. 
The pioneering Vicsek model \cite{Vicsek1995} first identified that 
(a) overdamped self-propelled particles    
with   
(b) noisy reorientation of the velocities 
and 
(c) short-range interactions favoring polar alignment of the velocities
can self-organize into polar ordered states (flocks), 
where particles move on average in the same direction.  
This phase transition between disordered and ordered states is 
widely believed to be first-order: 
This result was first supported by numerical simulations of microscopic dynamics using
either continuous or lattice based models implementing the basic ingredients (a)--(c),  
\cite{Gregoire2004,Chate2008,Solon2013,Solon2015}
(however their conclusions on the nature of the phase transition have been long debated \cite{Szabo2006,Czirok1997,Czirok1999,Czirok2000,Aldana2007,Gonci2008,Ginelli2016}),   
and only later supported analytically 
by linear stability analysis of hydrodynamic theories \cite{Bertin2006,Baskaran2008,Peruani2008,Peshkov2014,Thuroff2014,Bertin2009,
Lee2010}.  
The Vicsek model has been adapted over the years  
to include additional features such as       
particle cohesion \cite{Gregoire2001a,Gregoire2001};
metric-free \cite{Ginelli2010,Peshkov2012}, 
nematic {\cite{Ramaswamy2003,Chate2006}}
and possibly mixed nematic-polar alignment interactions \cite{Ngo2012,Huber2018}; 
density dependent velocities {\cite{Cates2010,Farrell2012}};
inertial and non-Markovian orientational dynamics 
\cite{Sumino2012,Attanasi2014,Cavagna2015,Yang2015,Nagai2015};   
particle chirality \cite{Kummel2013,Kaiser2013,Wensink2014,Nguyen2014,Denk2016,Liebchen2016,Liebchen2017}; 
and velocity reversals \cite{Mahault2018}. 


Nevertheless, an important setup has remained so far largely unexplored: 
Consider a population of active particles that,   
when their mutual interactions are switched off,
can perform anomalous diffusive dynamics other than self-propulsion 
(see, e.g., \cite{Bouchaud1990,Metzler2000,Metzler2004,Klages2008,Zaburdaev2015} 
and references below).
The following question arises naturally:
How does this different single-particle diffusive dynamics of the active units  
affect the emergence, stability and universality of collective behavior, 
when particle interactions are switched back on? 

Anomalous diffusion,  
which is generically characterized by a non linear position mean-square displacement (MSD) 
exhibiting possibly multiple scaling regimes \cite{Cairoli2015} 
where $\text{MSD}(t)\sim t^{\beta}$ with 
$0<\beta<1$ for subdiffusion and 
$1<\beta<2$ for superdiffusion 
($\beta=1$ is normal diffusion),     
has been indeed widely observed experimentally for 
both transport processes in physical systems 
\cite{Metzler2000,Metzler2004,Klages2008}
and biological motion    
\cite{Dieterich2008,Harris2012,Bressloff2013,Hofling2013,Zaburdaev2015}. 
While subdiffusion is typical for single particles whose motion is impeded, 
e.g. because they move in a crowded environment \cite{Sokolov2012,Meroz2015}, 
superdiffusion has been shown to optimize the search and foraging strategies 
of living organisms in specific environmental conditions
\cite{Viswanathan1999,Lomholt2008levy,Benichou2011,Viswanathan2011},  
such that it can be potentially advantageous 
not only for single organisms per se but also 
when they belong to large interacting groups.    
Recent experiments have indeed measured superdiffusive dynamics manifest as power-law displacement distributions for individual organisms in bird flocks, fish schools and swarming bacteria \cite{Cavagna2013,Murakami2015,Ariel2015}, 
thus supporting the biological relevance of the scenario just depicted.
\rd{Furthermore, such superdiffusive single-particle dynamics has been also shown to emerge 
as a result of non linear and non-Markovian effects \cite{Fedotov2017}. 
Remarkably, these dynamics can be described, to first approximation, by models 
incorporating fluctuations with infinite variances, 
whose average has been shown to converge again to a universal distribution, called the $\alpha$-stable L\'evy distribution, 
as guaranteed by the {\it generalized} central limit theorem \cite{Gnedenko1954}.}

Therefore, \rd{in this paper}, we investigate the consequences of the anomalous superdiffusive dynamics on the collective properties of \rd{active matter}. 
\rd{As a distinct notation, we will denote systems integrating anomalous superdiffusive behaviour and many-body particle interactions as \textit{active L{\'e}vy matter} (ALM) \cite{CairoliL}.}   
\rd{Collective properties can be generically understood by formulating a hydrodynamic theory.}  
However, 
while in ordinary active matter, symmetry considerations 
alone can sometimes be sufficient for the derivation of the hydrodynamic equations of motion (EOM), 
it is in contrast unclear how symmetry constrains the form of the hydrodynamic EOM for ALM.    
In particular, since fractional derivatives are expected 
to appear in these equations, 
the underlying model equations becomes non local and effectively long range 
and the typical way of truncating 
them by ignoring higher order derivatives may not be applicable.

To elucidate this issue, we will thus present the detailed derivation of the hydrodynamic EOM for ALM \cite{CairoliL}, 
by coarse-graining a microscopic model of polar alignment interacting active particles 
that perform superdiffusion \rd{akin to} L{\'e}vy flights \cite{Mandelbrot1982,Hughes1981,Shlesinger1995}. 
The L{\'e}vy flight is a Markovian random-walk model where superdiffusive behaviour is captured by fat-tailed jump-length distributions.     
\rd{Our setup differs from other models of ALM that have been previously proposed \cite{Grossmann2016, Estrada2018}, 
as ours is the only one that can permit the emergence of collective motion. 
}

The Article is organized as follows. 
	In Section~\ref{sec-single-particle}, we characterize the single-particle diffusive dynamics of our microscopic model of ALM in terms of the position mean-square displacement and the position statistics along any arbitrary direction. 
	In Section~\ref{si-sec-BBGKY}, we switch on the polar alignment interactions and derive the effective one-body Fokker-Planck equation that determines the joint statistics of the position and direction of motion of the active L{\'e}vy particles  
	by employing the Bogoliubov-Born-Green-Kirkwood-Yvon (BBGKY) hierarchical method \cite{Huang1987} and the stochastic calculus of L{\'e}vy processes \cite{Applebaum2009}. 
	In Section~\ref{si-sec-hydro}, we show how to perform a Fourier angular expansion on this equation, 
	and discuss the relevant terms that survive in the hydrodynamic limit, i.e., 
	the limit $k\to0$ with $k\equiv|\vec{k}|$ and $\vec{k}$ the independent variable in Fourier space.
	This procedure yields the hydrodynamic description of our microscopic model of ALM, which is our first main result.   
	Section~\ref{si-sec-lsa} presents the linear stability analysis on the characteristic disordered and ordered phases 
	\rd{that are} predicted by the hydrodynamic theory. 
	Specifically, in Section~\ref{si-sec-disPh} we show that the disordered phase is stable at linear level against small perturbations in all directions.     
	\rd{Likewise, in Section~\ref{si-sec-ordPh} we discuss the stability of the ordered phase. 
	First, we consider small longitudinal perturbations (i.e., in the direction of spontaneous symmetry breaking).} 
	Differently from ordinary active fluids, 
	where a characteristic banding instability emerges at the onset of collective motion 
	that renders the phase-transition first-order {\cite{Gregoire2004,Chate2008,Solon2013,Solon2015}}, 
	our hydrodynamic theory predicts a stable ordered phase \rd{in the hydrodynamic limit}
	\rd{(see Eq.~\eqref{asymp2}).}  
	\rd{Secondly, we demonstrate that the phase is stable at the onset also against transversal perturbations 
	(i.e., in the direction orthogonal to the collective motion).}
\rd{These results, therefore,} suggest that the order-disorder phase transition can be \rd{potentially} critical in ALM.
	This prediction is the second main result of our study. 
	\rd{A further confirmation of the criticality of the transition is presented in Sec.~\ref{si-sec-fss}, 
	where we perform a finite size scaling analysis through extensive numerical simulations of the underlying microscopic dynamics.} 
	Correspondingly, we characterize numerically both static and dynamic critical exponents of the disorder-to-collective motion transition and 
	use these numerical estimates to relate our findings to known universality classes.
	This numerical characterization of the critical properties of the transition is our third main result. 
	Finally, in Section~\ref{sec-conclusions} we draw conclusions and outline future perspectives.  


\begin{figure*}[!htb]
\centering
\includegraphics[width=180mm,keepaspectratio]{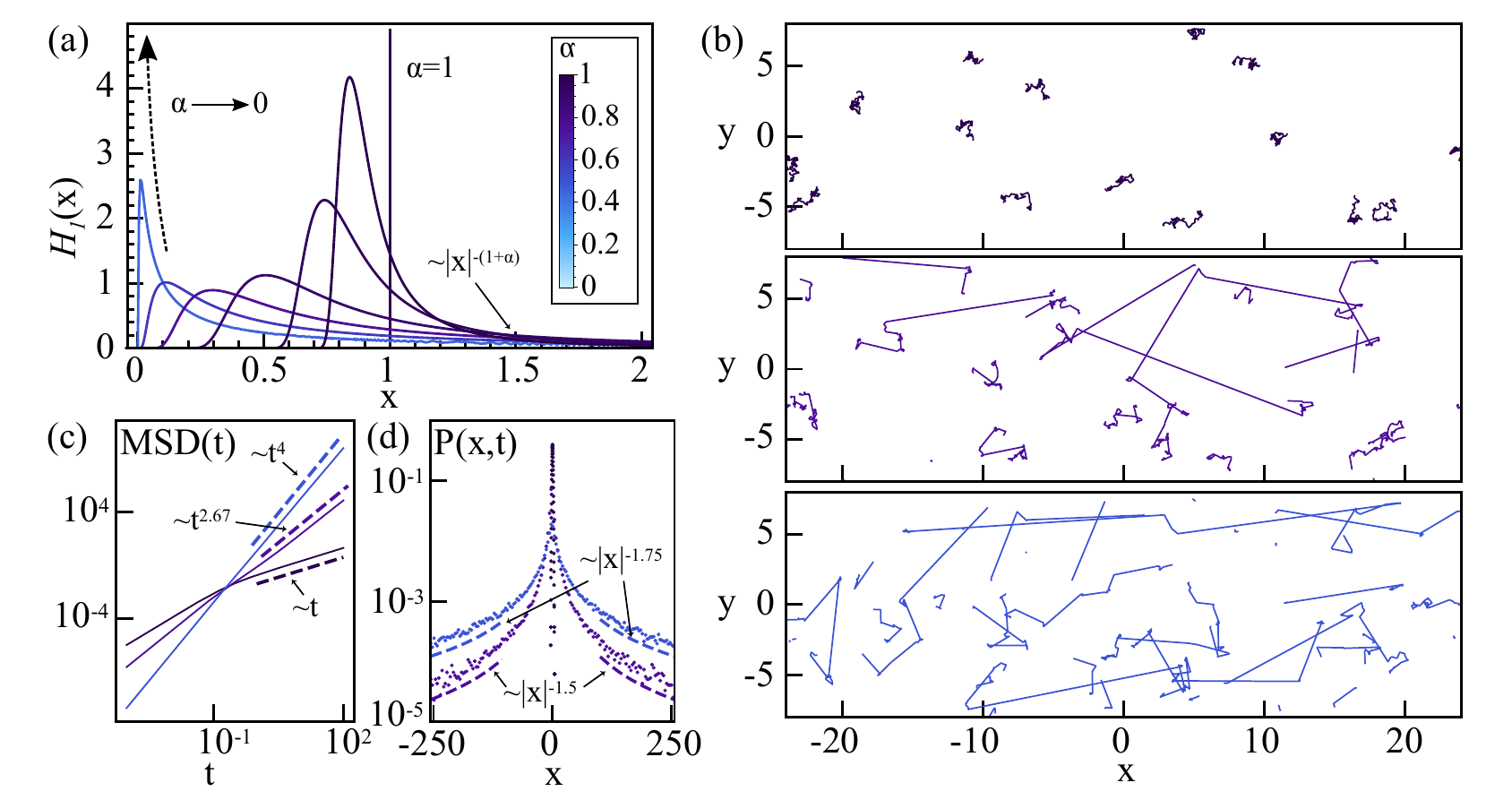}
\caption{
(a) Scaling function $H_1$ for different values of the stability index $0<\alpha<1$. 
The special case $\alpha=1$ where $H_1(x)=\delta(x-1)$ is plotted for reference.   
(b) Exemplary trajectories of particles moving according to the dynamical model~\eqref{ni-model} in an asymmetric box $6L\times2L$ (here $L=8$) with periodic boundary conditions.  
Characteristic parameters are $\alpha=\{1,0.75,0.5\}$ (top to bottom panel) and $\sigma=10$.
Large particle jumps induced by the L{\'e}vy stable displacement statistics are evident.
(c) Mean-square displacement and 
(d) position statistics projected in the $x$-direction 
of the active L{\'e}vy particles. 
As we are interested in the characteristic features of the free diffusion of the active particles, we run the simulations for these panels without periodic boundary conditions.
All the simulation parameters are defined as in panel (b). 
}\label{figure1}
\end{figure*} 

\section{Single particle dynamics}
\label{sec-single-particle}

We consider a single particle whose position $\vec{r}(t)$ on a planar surface is described by the Langevin equations:
\begin{align}
\dot{\vec{r}}(t)&=\eta(t) \vec{n}(\theta(t)), & 
\dot{\theta}(t)&=\xi(t)  
\label{ni-model}
\end{align}
with the unit vector $\vec{n}(\theta)\equiv(\cos{\theta},\sin{\theta})$ prescribing its direction of motion. 
We define the noises $\xi$ and $\eta$, 
and respectively $\langle \cdot \rangle$ and $\prec \cdot \succ$ 
\rd{the ensemble averages over their stochastic realizations}, 
as follows:    
(i) $\xi$ is a white Gaussian noise of variance $\sigma$, i.e., $\langle \xi(t)\rangle=0$ and $\langle \xi(t)\xi(t^{\prime})\rangle=2\sigma\delta(t-t^{\prime})$; 
(ii) $\eta$ is the formal derivative of 
the one-sided positive L{\'e}vy process 
$L(t)\equiv \int_0^t \eta(t^{\prime})\diff{t^{\prime}}$. 
This process is thus a subordinator 
(i.e., a strictly non-decreasing L\'{e}vy process) 
and is specified by the characteristic function   
$\prec e^{i k L(t)} \succ=e^{\,t \Psi(k)}$,  
with the so called L{\'e}vy symbol \cite{Applebaum2009} 
\begin{align}
\Psi(k)&=i v k+\int_{0}^{\infty} (e^{\,i k z}-1)\nu(\diff{z}) , 
\label{LevySymbol} 
\end{align}
where $v \geq 0$ is \rd{the constant} drift  
and $\nu(\diff{z})$ is the L{\'e}vy measure  satisfying 
$\nu(-\infty,0)=0$ and $\int_0^{\infty}\min{(z,1)}\nu(\diff z)<\infty$. 
\rd{An important example of such processes is represented by} stable subordinators with stability parameter $0<\alpha<1$. These are defined by setting    
\begin{align}
v&=0, &
\nu(\diff{z})&=\frac{\alpha}{\Gamma(1-\alpha)}\frac{\diff{z}}{z^{1+\alpha}},
\label{stable-measure}
\end{align}
which yields $\Psi(k)=k^{\alpha}$ \cite{Applebaum2009}. 
\rd{When an explicit upper cutoff $H$ is introduced in the L{\'e}vy measure, 
we obtain tempered L{\'e}vy stable subordinators; in this case}
\begin{equation}
\nu(\diff{z})=\frac{\alpha}{\Gamma(1-\alpha)} e^{\,-H z} \frac{\diff{z}}{z^{1+\alpha}}
\label{tempered-stable-measure}
\end{equation}
and 
consequently 
$\Psi(k)=(k+H)^{\alpha}-H^{\alpha}$.   
The characteristic functional of $\eta$ is thus defined as 
\cite{Tankov2003}
\begin{equation}
G[h(t^{\prime})]\equiv\langle e^{-\int_0^{\infty} h(t^{\prime})\eta(t^{\prime})\diff{t^{\prime}}} \rangle =e^{-\int_0^{\infty} \Psi(h(t^{\prime})) \diff{t^{\prime}}}   
\label{chF}
\end{equation} 
with $h$ an arbitrary test function. 
The noises $\xi$, $\eta$ are assumed independent, and 
It{\^o} prescription is chosen for the multiplicative term of Eq.~\eqref{ni-model}.

The process $L$ measures the total distance traveled by the active particles. 
Its statistics, $Q(l,t)\equiv\langle \delta(l-L(t))\rangle$, is easily calculated by Fourier inversion of its characteristic function. 
For stable subordinators, in particular, we find the following characterization:  
For $0<\alpha<1$, $Q(l,t)=t^{-1/\alpha}H_1(l t^{-1/\alpha})$ 
with the scaling function 
$H_1(x)\equiv(2\pi)^{-1}\int_{-\infty}^{\infty}e^{-i\omega x-(-i\omega)^{\alpha}}\diff{\omega}$  
(see Fig.~\ref{figure1}a). 
The function $H_1$ is unimodal with maximum at the point $x_m(\alpha)$ and  
heavy-tailed asymptotics for large $x$, i.e., $H_1(x)\sim x^{-(1+\alpha)}$ for $x \to \infty$, 
and satisfies $H_1(x)\to0$ for $x\to0^+$. 
As $\alpha\to0$, $x_m\to0$ and $H_1$ simultaneously sharpens at this point \cite{Mikusinski1959,Bendler1984,Hilfer2002,Penson2010}.
For $\alpha=1$, $Q(l,t)=\delta(l-t)$ \cite{Bouchaud1990,Fogedby1994}. In this case Eqs.~\eqref{model} reduce to self-propelled dynamics with constant particle velocity \cite{Vicsek1995}. 
Exemplary particle trajectories for different values of the characteristic parameter $\alpha$ are shown in Fig.~\ref{figure1}b.

When $L$ is a stable subordinator, 
the dynamics described by the Langevin Eqs.~\eqref{ni-model}
exhibits the following super-diffusive features 
(see Appendix~\ref{app-sdynamics}): 
(a) fractional moments $\prec\langle |\vec{r}^{\prime}(t)|^{\delta} \rangle\succ$ 
($0<\delta<\alpha$) with $\vec{r}^{\prime}(t)\equiv \vec{r}(t)-\vec{r}(0)$
scale as $\sim t^{\delta/\alpha}$;
(b) moments of order $\geq 1$ do not exist. 
Nevertheless, upon rescaling suitably the fractional moments, we obtain that
the mean-square displacement  
$\text{MSD}(t)\equiv\prec\langle [\vec{r}^{\prime}(t)]^2 \rangle\succ$ 
scales as $\sim t^{2/\alpha}$ for long times (see Fig.~\ref{figure1}c);
(c) the projected position statistics, e.g. in the direction $\vec{e}_1\equiv(1,0)$, 
which is defined as $P(x,t)\equiv \prec\langle \delta(x-\vec{r}^{\prime}(t)\cdot\vec{e}_1) \rangle\succ$,
has power-law asymptotic tail behaviour with the same characteristic exponent of $H_1$, 
i.e., $P(x,t)\sim |x|^{-(1+\alpha)}$ (see Fig.~\ref{figure1}d).


\begin{figure}[!tb]
\centering
\includegraphics[width=80mm,keepaspectratio]{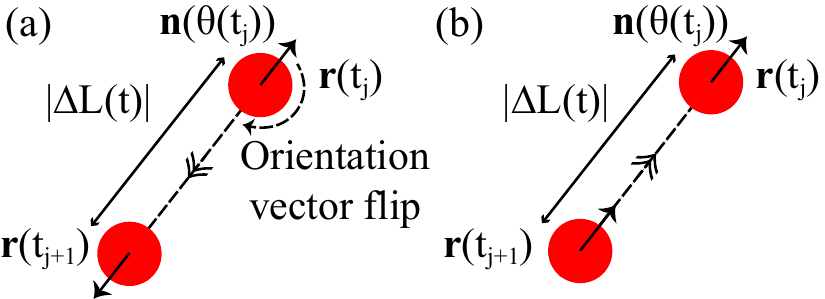}
\caption{\rd{Schematic of different interpretations for negative distributed displacements of active L\'evy particles.}}\label{fig:symmetric}
\end{figure}

\rd{
We remark that  
the stochastic increments 
$\Delta L(t)\equiv\int_t^{t+\Delta t} \eta(t^{\prime})\diff{t^{\prime}}=|\vec{r}(t+\Delta t)-\vec{r}(t)|$ 
with $\Delta t$ an arbitrarily small time step, 
which measure the distance traveled by the active particle 
over the time interval $[t,t+\Delta t]$ in the direction $\vec{n}(\theta(t))$, 
are non-negative 
with the definition chosen for the step size process $\eta$;  
thus, the particle orientation vector is determined only by the dynamics of the angular variable $\theta$ as prescribed by Eqs.~\eqref{ni-model}. 
Nevertheless, more general L\'evy processes generating both positive and negative increments, such as, e.g., symmetric L\'evy stable processes with 
$\Psi(k)=-|k|^{\beta}$ ($0<\beta<2$),      
can also be used to define the step-size process  
by adopting a suitable interpretation of the negative increments.  
In fact, if $\Delta L(t)$ is negative, 
two alternative scenarios can be identified 
(see schematic in Fig.~\ref{fig:symmetric}).   
On the one hand, we can flip the orientation vector of the active L\'evy particle, 
which is then displaced by $|\Delta L(t)|$ (Fig.~\ref{fig:symmetric}a); 
this choice inevitably increases the noise in the system, thus disallowing collective motion 
(verified numerically; not shown).  
On the other hand, we can keep the orientation vector unchanged and 
displace the particle by $|\Delta L(t)|$ in the direction $-\vec{n}(\theta(t))$ 
(Fig.~\ref{fig:symmetric}b). 
This latter case is reminiscent of the model studied by Mahault and collaborators \cite{Mahault2018}, 
where ordinary self-propelled active particles can move in the direction opposite to their intrinsic polarity with some constant probability $p$.   
}


\section{Many-body system with polar alignment interactions and the one-body Fokker-Planck equation}
\label{si-sec-BBGKY}

We now consider $N$ active L{\'e}vy particles and switch on interactions between them that promote polar alignment of their \rd{orientation vectors}. 
These are modeled by modifying the angular dynamics in \eqref{ni-model} as 
\begin{align}
\dot{\vec{r}}_i(t)&=\eta_i(t) \vec{n}(\theta_i(t)), & 
\dot{\theta}_i(t)&= \rd{F_{i}(t)} + \xi_i(t) , 
\label{model}
\end{align}
\rd{where the force $F_i$ is defined as \cite{Vicsek1995,Peruani2008}
\begin{equation}
F_i(t)\equiv \frac{\gamma}{\pi d^2} \sum_{j=1}^N \mathrm{H}(d-|\vec{r}_{ij}|)\sin{(\theta_j-\theta_i)}
\label{force-model}
\end{equation}
with $\vec{r}_{ij}\equiv \vec{r}_i-\vec{r}_j$ the distance between particle $i$-th and $j$-th, 
$d>0$ the interaction range, 
and $\mathrm{H}$ the Heaviside function
($\mathrm{H}(x)=1$ for $x \geq 0$; $\mathrm{H}(x)=0$ otherwise).
}

The one-body Fokker-Planck equation for the stochastic dynamics described by the Langevin equations~\eqref{model} is derived by using the BBGKY hierarchical formalism \cite{Huang1987}.  
Schematically this method consists in the following steps:  
(1.) one computes the $N$-body probability density function (PDF) 
$P_N(\vec{X},\boldsymbol{\Phi},t)\equiv \prec\langle \prod_{j=1}^N \delta(\vec{x}_j-\vec{r}_j(t)) \delta(\phi_j - \theta_j(t)) \rangle\succ$, 
where we introduce the shorthand vector notation 
$\vec{X}\equiv(\vec{x}_1,\ldots,\vec{x}_N)$, 
$\boldsymbol{\Phi}\equiv(\phi_1,\ldots,\phi_N)$.   
(2.) One then derives an exact equation for the one-body PDF $P(\vec{x}_1,\phi_1,t)\equiv N\int \diff{^2} \vec{x}_2 \diff{} \phi_2 \cdots  \diff{^2} \vec{x}_N \diff{} \phi_N P_N(\vec{X},\boldsymbol{\Phi},t)$ by integrating out $N-1$ other degrees of freedom. 
The resulting equation naturally depends on the two-body PDF 
$P_{2}(\vec{x}_1,\phi_1,\vec{x}_2,\phi_2,t)\equiv N(N-1) \int \diff{^2} \vec{x}_3 \diff{} \phi_3 \cdots  \diff{^2} \vec{x}_N \diff{} \phi_N P_N(\vec{X},\boldsymbol{\Phi},t) $.
(3.) Finally, one adopts a suitable approximation of the two-body distribution in terms of the one-body PDF in order to close the equation. 
We adopt here the \textit{molecular chaos} approximation from the Boltzmann kinetic theory \cite{Brilliantov2010}; however we apply the approximation directly in the BBGKY formalism to close the equation. 
Specifically, in mathematical terms, the two-body distribution \rd{function} is factorized into a product of two independent one-body densities 
\begin{equation}
P_{2}(\vec{x}_1,\phi_1,\vec{x}_2,\phi_2,t)=P(\vec{x}_1,\phi_1,t)P(\vec{x}_2,\phi_2,t).
\label{vlasov}
\end{equation}
This approximation is particularly suited for dilute systems with long-range interactions between particles \cite{Vlasov1968}, 
but is nevertheless widely applied to 
self-propelled particle models for active matter, where in contrast the interactions are typically short-range 
\cite{Bertin2006,Peruani2008,Baskaran2008,Bertin2009,Lee2010,Peshkov2014,Bertin2017}. 
\rd{In light of this, the approximation is, in fact, even more justified for ALM than for ordinary (self-propelled) active matter, 
because the L\'evy displacement statistics renders the short-range alignment interactions effectively long-range. 
Two arguments support the validity of this statement: 
First, the large relocations of the active particles 
that are induced by their fat-tailed displacement statistics 
greatly facilitate the removal of the spatial correlations between them. 
This is evidenced by the rapid homogenization of the pair correlation function of the system observed in numerical simulations (see Sec.~\ref{si-sec-fss}, Fig.~\ref{fig:equilibration}(a)).   
Second, these relocations ultimately invalidate the assumption that only 
interactions local in time and space between the constituent particles determine the macroscopic behaviour of the system, which is at the basis of the standard kinetic theory of dilute gases \cite{Brilliantov2010}. 
As a further consequence of the L\'evy displacement statistics, in fact, even in dilute conditions there is a finite probability for more than one particle to jump in one time-step in the vicinity of another particle.  
}  

Step (1.) can be solved by employing the stochastic calculus of L{\'e}vy processes \cite{Applebaum2009}. 
We \rd{thus} rewrite the Langevin equations~\eqref{model} in vector notation, i.e.,  
\begin{align}
&\left(
\begin{matrix}
\{ \diff{\vec{r}_i(t)} \} \\
\{ \diff{\theta_i(t)} \}
\end{matrix}
\right)
=
\left(
\begin{matrix}
\vec{0}_N \\
\{ F_i(t) \} 
\end{matrix}
\right)\diff{t} \\ 
& \quad 
+ \left(
\begin{matrix}
\mathrm{diag}(\{\vec{n}(\theta_i(t))\}) & \vec{0}_N \\
\vec{0}_N & \sqrt{2\sigma}\boldsymbol{\mathrm{I}}_N) 
\end{matrix}
\right)
\left(
\begin{matrix}
\{\diff{L_i(t)}\} \\
\{\diff{W_i(t)}\}
\end{matrix}
\right) \notag
\end{align}
with 
$\diff L_i(t)\equiv \eta_i \diff{t}$ the traveled distance of the active particle during the infinitesimal time interval $[t,t+\diff{t}]$,   
and $\diff{W}_i(t)\equiv \xi_i \diff{t}$ the increment of \rd{the angular noise} 
over the same time interval.  
The Fokker-Planck equation of the $2N$ dimensional process 
$(\{\vec{r}_i(t)\},\{\theta_i(t)\})$ is  
\begin{equation}
\derpar{}{t} P_N(\vec{X},\boldsymbol{\Phi},t)=(A^{\dagger} P_N)(\vec{X},\boldsymbol{\Phi},t)  
\label{NbodyFP}
\end{equation}
with $A^{\dagger}$ the adjoint of its characteristic generator $A$. 
Because the noises $\{ L_i \}, \{W_i\}$ are independent, $A$ can be written as the sum of its continuous $A_c$ and c{\`a}dl{\`a}g $A_d$ parts
($A_d$ is right continuous with left limits).
In other words, 
$(A h)(\vec{X},\boldsymbol{\Phi})=(A_c h)(\vec{X},\boldsymbol{\Phi})+(A_d h)(\vec{X},\boldsymbol{\Phi})$
for an arbitrary smooth function $h(\vec{X},\boldsymbol{\Phi})$ within its domain.     
\rd{These operators} are respectively defined as   
\begin{align}
A_c h
&\equiv \sum_{j=1}^N \left[ v_j \vec{n}(\phi_j)\cdot\nabla_{j} 
+F_j(t)\derpar{}{\phi_j}+\sigma \dersecpar{}{\phi_j}\right] h , \\ 
A_d h
&\equiv \sum_{j=1}^N \int_0^{\infty} [ -1 + \mathcal{T}^+_{z \vec{n}(\phi_j)}] h \, \nu(\diff{z})   
\end{align}
with the translation operator  
\begin{equation}
\mathcal{T}^{\pm}_{z \vec{n}(\phi_j)} h 
\equiv 
h(\{\vec{x}_i\}_{i=1}^{j-1},\vec{x}_j\pm z \vec{n}(\phi_j),\{\vec{x}_i\}_{i=j+1}^N,\boldsymbol{\Phi}).
\end{equation}
Their adjoint operators are easily computed. 
Considering two arbitrary smooth functions $h_1$, $h_2$ and adopting the notation 
$\int_{\mathbb{R}^2} \int_{-\pi}^{\pi}\diff{\boldsymbol{\Phi}}\diff{\vec{X}}=\prod_{j=1}^N \int_{\mathbb{R}^2} \int_{-\pi}^{\pi}\diff{\phi_j}\diff{\vec{x}_j}$ 
we can calculate  
\begin{align}
(A_c h_1,h_2)&=\int_{\mathbb{R}^2} \int_{-\pi}^{\pi} [(A_c h_1)h_2] \diff{\boldsymbol{\Phi}}\diff{\vec{X}} \notag\\
&=\sum_{j=1}^N \int_{\mathbb{R}^2} \int_{-\pi}^{\pi} h_2 
\left[ \mathcal{L}_j + F_j(t)\derpar{}{\phi_j} \right]h_1 \diff{\boldsymbol{\Phi}}\diff{\vec{X}} \notag\\
&=\sum_{j=1}^N \int_{\mathbb{R}^2} \int_{-\pi}^{\pi} h_1
\left[ \mathcal{L}_j^{\dagger} -\derpar{}{\phi_j}F_j(t) \right]h_2 \diff{\boldsymbol{\Phi}}\diff{\vec{X}}
\notag\\&
=(h_1,A^{\dagger}_c h_2) , 
\end{align}
where the second and third lines are related by integration by parts, 
and we introduce the auxiliary operators  
\begin{align}
\mathcal{L}_j(\vec{x}_j,\phi_j)&\equiv v_j \vec{n}(\phi_j)\cdot\nabla_{j} + \sigma \dersecpar{}{\phi_j}\ , \\
\mathcal{L}^{\dagger}_j(\vec{x}_j,\phi_j)&\equiv -v_j \vec{n}(\phi_j)\cdot\nabla_{j} + \sigma \dersecpar{}{\phi_j} \ . 
\end{align}
Similarly, for the operator $A_d$ we obtain  
\begin{align}
&(A_d h_1,h_2)=\int_{\mathbb{R}^2} \int_{-\pi}^{\pi} [(A_d h_1)h_2] \diff{\boldsymbol{\Phi}}\diff{\vec{X}} \notag\\
&\qquad =\sum_{j=1}^N \int_{\mathbb{R}^2} \int_{-\pi}^{\pi} h_2 
\int_0^{\infty} [ -1 + \mathcal{T}^+_{z \vec{n}(\phi_j)}] h_1 \nu(\diff{z})
\diff{\boldsymbol{\Phi}}\diff{\vec{X}}
\notag\\ 
&\qquad =\sum_{j=1}^N \int_{\mathbb{R}^2} \int_{-\pi}^{\pi} h_1 
\int_0^{\infty} [ -1 + \mathcal{T}^-_{z \vec{n}(\phi_j)}] h_2 \nu(\diff{z}) 
\diff{\boldsymbol{\Phi}}\diff{\vec{X}^{\prime}} \notag\\ 
&\qquad =(h_1,A^{\dagger}_d h_2) , 
\end{align}
where we use the change of variables 
$\vec{x}_j^{\prime}=\vec{x}_j + z \vec{n}(\phi_j)$, 
$\vec{x}_i^{\prime}=\vec{x}_i$ for $i \neq j$ 
to rearrange the integrand terms. 
Thus, 
$A^{\dagger}P_N\equiv(A_c^{\dagger}+A_d^{\dagger})P_N$ with
\begin{align}
A_c^{\dagger}P_N &\equiv 
\sum_{j=1}^N \left[ \mathcal{L}^{\dagger}_j -\derpar{}{\phi_j} F_j(t) \right] P_N , \label{AdagC} \\
A^{\dagger}_d P_N &\equiv
\sum_{j=1}^N \int_0^{\infty} [ -1 + \mathcal{T}^-_{z \vec{n}(\phi_j)}]P_N \, \nu(\diff{z}).  
\label{AdagD}
\end{align}
Finally, substituting $F_j$ \rd{as specified in Eq.~\eqref{force-model}} and Eqs.~(\ref{AdagC}, \ref{AdagD}) into Eq.~\eqref{NbodyFP}, 
we obtain 
\begin{multline}
\left( \derpar{}{t} 
- \sum_{j=1}^N \mathcal{L}^{\dagger}_j \right) P_N = 
- \sum_{j=1}^N \int_0^{\infty} [ 1 - \mathcal{T}^-_{z \vec{n}(\phi_j)}]P_N\, \nu(\diff{z}) \\
-\frac{\gamma}{\pi d^2} \sum_{j,m} \mathrm{H}{(d-|\vec{x}_{mj}|)} \derpar{}{\phi_j}[\sin{(\phi_m-\phi_j)}P_N]  
\label{Nbody}
\end{multline}
\rd{with the shorthand notation $\vec{x}_{mj}\equiv \vec{x}_m-\vec{x}_j$.}

Step (2.) is straightforward. Only the integration of the fractional term requires special care. In details, we calculate the quantity      
$N\prod_{m=2}^N \int_{\mathbb{R}^2} \int_{-\pi}^{\pi}
\int_0^{\infty} [ 1 - \mathcal{T}^-_{z \vec{n}(\phi_j)}] P_N \, \nu(\diff{z})\diff{\phi_m}\diff{\vec{x}_m}$.  
If $j=1$ this formula reduces to  
$\int_0^{\infty} [ 1 - \mathcal{T}^-_{z \vec{n}(\phi_1)}] P \, \nu(\diff{z})$.
Instead, if $j\neq 1$ we obtain   
\begin{align}
&\int_{\mathbb{R}^2} \int_{-\pi}^{\pi} \int_0^{\infty} 
[ 1 - \mathcal{T}^-_{z \vec{n}(\phi_j)}]
P(\vec{x}_j,\phi_j,t) \nu(\diff{z}) 
\diff{\phi_j}\diff{\vec{x}_j} \notag\\
=&
\int_{\mathbb{R}^2} 
\int_0^{\infty} 
\prec \langle
[ 1 - \mathcal{T}^-_{z \vec{n}(\theta_j(t))}]
\delta(\vec{x}_j-\vec{r}_j(t)) 
\rangle \succ
\nu(\diff{z}) 
\diff{\vec{x}_j}
 \notag\\ 
 =& \notag
\left.
\int_0^{\infty} 
\prec \langle 
e^{\,i \vec{k} \cdot \vec{r}_j(t)} [1-e^{i z \vec{k}\cdot\vec{n}(\theta_j(t))}] 
\rangle \succ
\nu(\diff{z})
\right|_{\vec{k}=0}
\\
=& \ 0  \ . 
\end{align}
Therefore, we obtain the equation     
\begin{align}
&\left(\derpar{}{t} - \mathcal{L}^{\dagger}_1 \right) P + \int_0^{\infty} [ 1 - \mathcal{T}^-_{z \vec{n}(\phi_1)}]P\,\nu{(\diff{z})} = -\frac{\gamma}{\pi d^2} \times  
\notag \\
& \times \derpar{}{\phi_1}\int_{-\pi}^{\pi} \int_{\mathbb{R}^2} \mathrm{H}(d-|\vec{x}_{12}|)\sin{(\phi_2-\phi_1)}P_2 \diff{\vec{x}_2}\diff{\phi_2} . 
\label{si-1bodyeq}  
\end{align}
Implementing the approximation \eqref{vlasov} as prescribed in step (3.) finally yields 
\begin{equation}
\left(\derpar{}{t} - \widetilde{\mathcal{L}}_1 \right) P 
+ \int_0^{\infty} [ 1 - \mathcal{T}^-_{z \vec{n}(\phi_1)}]P\,\nu{(\diff{z})}=0
\label{vlasoveq}
\end{equation}
where we define the operator $\widetilde{\mathcal{L}}_1$ as 
\begin{equation}
\widetilde{\mathcal{L}}_1\equiv \mathcal{L}^{\dagger}_1 + \derpar{}{\phi_1} M[P]
\label{Loperator}
\end{equation} 
and the functional of $P$  
\begin{align}
M[P]&\equiv -\frac{\gamma}{\pi d^2} \int_{-\pi}^{\pi} \int_{\mathbb{R}^2} \mathrm{H}(d-|\vec{x}_{12}|)\times \notag\\ 
&\quad \times \sin{(\phi_2-\phi_1)} P(\vec{x}_2,\phi_2,t) \diff{\vec{x}_2}\diff{\phi_2} \ . 
\label{MoperatorF}   
\end{align}
Taking the large-scale limit where $d\simeq0$, this reduces to 
(see Appendix~\ref{sec-lscale} \rd{for a formal derivation})
\begin{equation}
M[P]\simeq -\gamma \int_{-\pi}^{\pi} \sin{(\phi_2-\phi_1)} P(\vec{x}_1,\phi_2,t) \diff{\phi_2} .
\label{Moperator}   
\end{equation}
Eq.~\eqref{vlasoveq} with the specifics~(\ref{Loperator}, \ref{Moperator}) is the reduced one-body Fokker-Planck equation of the microscopic model~\eqref{model}. 
The remaining non local operator is specified by the prescribed statistics of the noise $L_1$. 

For stable subordinators in particular 
Eq.~\eqref{vlasoveq} can be rewritten as (see Eq.~\eqref{stable-measure})
\begin{align}
\left(\derpar{}{t} + \mathcal{D}^{\alpha}_{\vec{n}(\phi)} - \sigma\dersecpar{}{\phi} - \derpar{}{\phi}M[P] \right) P&=0 , & 
\label{LSvlasoveq}
\end{align}
where $\mathcal{D}^{\alpha}_{\vec{n}(\phi)}$ is the fractional directional derivative \cite{Samko1993}
\begin{equation}
\mathcal{D}_{\vec{n}(\phi)}^{\alpha} P \equiv \frac{\alpha}{\Gamma(1-\alpha)}\int_0^{\infty} [1-\mathcal{T}^-_{z \vec{n}(\phi)}]P\, \frac{\diff{z}}{z^{1+\alpha}}.   
\label{si-fracder}
\end{equation}

Fractional advection-diffusion equations similar to Eq.~\eqref{vlasoveq} were first discussed by Meerschaert and collaborators \cite{Meerschaert1999}. 
Their model however did not account for interactions between the moving particles, 
which could affect the time evolution of the velocity directions.  
Thus, the statistics of the velocity directions of the particles 
can be prescribed \textit{a priori} in their case by specifying a probability measure $M(\mathrm{d}\vec{n})$ over the unit circle. 
The fractional operator thus simplifies to 
\begin{equation}
\int_{\norm{\vec{n}}=1} \int_0^{\infty} [ 1 - \mathcal{T}^-_{z \vec{n}}]P\nu{(\diff{z})} M(\mathrm{d}\vec{n})
\end{equation}
with $P$ the particle position statistics 
(indeed the velocity direction is no longer needed to build a statistical description of the anomalous diffusion process). 
This is different from the scenario discussed here, 
where the velocity directions of the active particles are primitive statistical variables for the diffusion process, 
and their statistics must be inferred from the many-body dynamics.

\section{Derivation of the hydrodynamic equations}
\label{si-sec-hydro}

The hydrodynamic description for the microscopic model \eqref{model} is obtained by performing a Fourier 
expansion of the distribution $P$ \rd{with respect to the angular argument} 
\cite{Bertin2006,Peruani2008,Bertin2009,Lee2010,Peshkov2014,Bertin2017}. 
We thus write  
\begin{align}
P(\vec{x},\phi,t)&\equiv 
\frac{1}{2\pi}\sum_{m\in \mathbb{Z}} f_m(\vec{x},t) e^{\,-i m \phi} 
\label{eq:Fexp}
\end{align}
with the $m$-th order angular mode 
\begin{align}
f_m(\vec{x},t)&\equiv
\int_{-\pi}^{\pi} e^{\,i m \phi} P(\vec{x},\phi,t)\diff{\phi}.  
\end{align}
The slow macroscopic fields are the density 
\begin{equation}
\rho(\vec{x},t)\equiv \int_{-\pi}^{\pi} P(\vec{x},\phi,t) \diff{\phi} \ ,
\end{equation}
the mean direction 
\begin{equation}
\vec{p}(\vec{x},t)\equiv \int_{-\pi}^{\pi} \vec{n}(\phi) P(\vec{x},\phi,t) \diff{\phi} \ ,
\end{equation} 
\rd{and the apolar nematic tensor 
\begin{equation}
\vec{Q}(\vec{x},t)\equiv \int_{-\pi}^{\pi} \left[ \vec{n}(\phi)\vec{n}(\phi) -\frac{1}{2}\vec{1}\right] P(\vec{x},\phi,t) \diff{\phi}  
\end{equation}
with $\vec{1}$ the $2 \times 2$ identity matrix,
which are determined by the lower order Fourier angular modes 
$f_0$, $f_1$ and $f_2$, respectively, 
as prescribed by the following relations:}   
\begin{align}
\rd{\rho(\vec{x},t)}\,&\rd{=f_0(\vec{x},t),} \label{eq:rhotof0} \\
\rd{\vec{p}(\vec{x},t)}\,&
\rd{=\left(
\begin{matrix}
\Re f_1(\vec{x},t) \\
\Im f_1(\vec{x},t) 
\end{matrix}
\right),} \label{eq:ptof1} \\
\rd{\vec{Q}(\vec{x},t)}\,&
\rd{=\frac{1}{2}
\left(
\begin{matrix}
\Re f_2(\vec{x},t) & \Im f_2(\vec{x},t) \\
\Im f_2(\vec{x},t) & -\Re f_2(\vec{x},t) 
\end{matrix}
\right).} \label{eq:Qtof2}
\end{align}

To determine the Fourier coefficients $f_{m}$ we first take the spatial Fourier transform of Eq.~\eqref{vlasoveq} 
\footnote{The Fourier transform of a function 
$f_1(\vec{x})$ is written as   
$\skew{3.5}\hat f_1(\vec{k})\equiv \mathcal{F}\{f_1(\vec{x})\}(\vec{k})$.
Correspondingly, $\mathcal{F}^{-1}$ is the inverse transform. 
The symbol $\star$ denotes the convolution of two such functions, i.e.,   
$f_1(\vec{x})\star f_2(\vec{x})\equiv\int_{-\infty}^{\infty} f_1(\vec{x}-\vec{x^{\prime}}) f_2(\vec{x^{\prime}}) \diff{\vec{x^{\prime}}}$.}. 
\rd{The latter transform is particularly advantageous to expand the non local integral operator.
In details, we obtain} 
\begin{equation}
\derpar{}{t} \hat P - \mathcal{F}\{\widetilde{\mathcal{L}}_1 P\} 
+ \int_0^{\infty} [1-e^{\,i \zeta \vec{k}\cdot\vec{n}(\phi)}]\hat P\, \nu(\diff{\zeta})
=0 \ . 
\end{equation}
Secondly, we multiply both its sides by $e^{\,i m \phi}$ and integrate them in the angular variable $\phi$.  
This yields 
\begin{multline}
\derpar{}{t} \hat f_m 
-\int_{-\pi}^{\pi} e^{\,i m \phi} \mathcal{F}\{\widetilde{\mathcal{L}}_1P\}\diff{\phi} \\
+\int_{-\pi}^{\pi} e^{\,i m \phi} 
\int_0^{\infty} [1-e^{\,i \zeta \vec{k}\cdot\vec{n}(\phi)}]\hat P\, \nu(\diff{\zeta})
\diff{\phi}=0\ . 
\end{multline} 
\rd{To relate analytically the integrals appearing in this equation to the Fourier modes $\hat f_m$, 
we explicitly write out the angle specifying the direction of the Fourier variable, 
i.e., we substitute $\vec{k}\equiv k \, \vec{n}(\phi)$ with $k\equiv |\vec{k}|$ \cite{taylor2016}.}
On the one hand, the first integral in its rhs is given by 
\begin{align}
&\int_{-\pi}^{\pi} e^{\,i m \phi} \mathcal{F}\{\widetilde{\mathcal{L}}_1P\}\diff{\phi}=\frac{v_1}{2} i k (e^{-\,i\psi}\hat f_{\rd{m+1}} + e^{\,i\psi}\hat f_{\rd{m-1}}) \notag\\
&\qquad - \sigma m^2 \hat f_{m} + \frac{m\gamma}{2} (\hat f_{m-1}\star \hat f_{1} - \hat f_{m+1}\star \hat f_{-1})  .
\label{fintegral}
\end{align}
On the other hand, using Eq.~\eqref{eq:Fexp} we can expand the non local advective term as    
\begin{align}
&\int_{-\pi}^{\pi} e^{\,i m \phi} 
\int_0^{\infty} [1-e^{\,i \zeta \vec{k}\cdot\vec{n}(\phi)}]\hat P\, \nu(\diff{\zeta})
\diff{\phi}= \\
&\frac{1}{2\pi}\sum_{m^{\prime}\in \mathbb{Z}} \hat f_{m^{\prime}+m}
\int_{-\pi}^{\pi} e^{\,-i m^{\prime} \phi} 
\int_0^{\infty} [1-e^{\,i \zeta k \cos{(\psi-\phi)}}]\nu(\diff{\zeta})
\diff{\phi} , \notag
\end{align} 
\rd{where we used the relation}
$\vec{n}(\psi)\cdot\vec{n}(\phi)=\cos{(\psi-\phi)}$. 
Finally, changing angular coordinate as $\phi^{\prime}=\psi-\phi$, 
we obtain  
\begin{multline}
\int_{-\pi}^{\pi} e^{\,i m \phi} 
\int_0^{\infty} [1-e^{\,i \zeta \vec{k}\cdot\vec{n}(\phi)}]\hat P\, \nu(\diff{\zeta})
\diff{\phi}= \\
\sum_{m^{\prime}\in \mathbb{Z}} i^{m^{\prime}} \hat f_{m^{\prime}+m}e^{\,-i m^{\prime}\psi} C_{m^{\prime}}(k,\alpha), 
\label{expansion}
\end{multline} 
where we define the $(k,\alpha)$-dependent coefficients 
\begin{equation}
C_{m^{\prime}}\equiv \frac{(-i)^{m^{\prime}}}{2\pi}
\int_{-\pi}^{\pi} e^{\,i \phi^{\prime} m^{\prime}}
\int_0^{\infty} [1-e^{\,i \zeta k \cos{\phi^{\prime}}}]\nu(\diff{\zeta}) 
\diff{\phi^{\prime}}. 
\label{CoeffC}
\end{equation}

\begin{figure}[!bt]
\centering
\includegraphics[width=85mm,keepaspectratio]{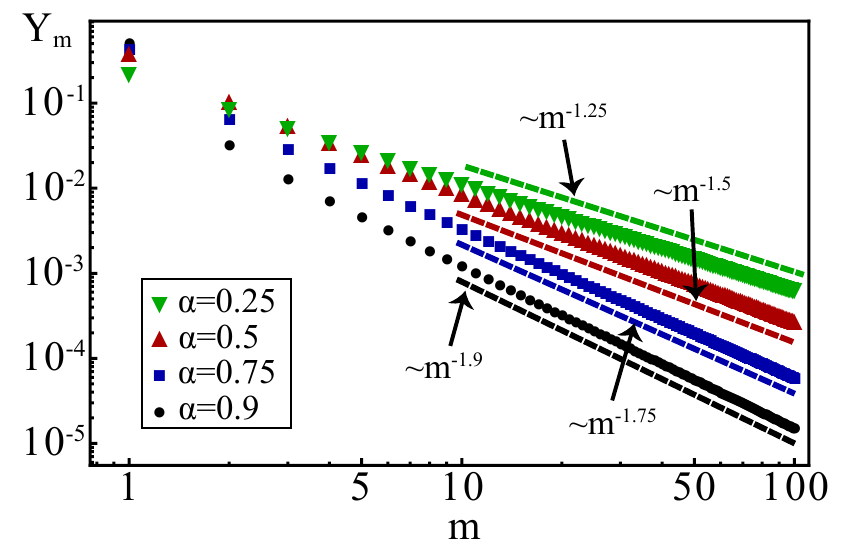}
\caption{
Plot of the coefficients $\Upsilon_{m}$ \rd{($m>0$)} for L{\'e}vy stable distributed step sizes (see Eq.~\eqref{upsilon_n}) for different values of the characteristic parameter $\alpha$. 
Numerical results suggest that $\Upsilon_{m}$ scales as $m^{-1-\alpha}$ for $m \gg 1$ (dashed lines).
}\label{fig3}
\end{figure}

In the hydrodynamic limit, the scaling behaviour of $C_{m^{\prime}}$ is determined by the corresponding scaling of the L{\'e}vy measure $\nu$ for large displacements. 
For a heavy-tailed measure $\nu(\diff{z})\sim z^{-1-\alpha} \diff{z}$, 
such as for the L{\'e}vy stable subordinator~\eqref{stable-measure},      
we can rescale $z$ in Eq.~\eqref{CoeffC} such that   
\begin{equation}
C_{m^{\prime}}\sim k^{\alpha} \Upsilon_{m^{\prime}}(\alpha) 
\label{CToUps}
\end{equation}
with $\alpha$-dependent numerical coefficients 
\begin{align}
\Upsilon_{m^{\prime}}&\propto \frac{(-i)^{m^{\prime}}}{2\pi}
\int_{-\pi}^{\pi} e^{\,i \phi^{\prime} m^{\prime}}
\int_0^{\infty} [1-e^{\,i z^{\prime} \cos{\phi^{\prime}}}]\frac{\diff{z^{\prime}}}{z^{\prime\,1+\alpha}} \diff{\phi^{\prime}} \notag\\
&\propto \frac{(-i)^{m^{\prime}}}{2\pi}
\int_{-\pi}^{\pi} e^{\,i \phi^{\prime} m^{\prime}}
(-i\cos{\phi^{\prime}})^{\alpha} \diff{\phi^{\prime}},   
\label{UpsilonC}
\end{align}
where any multiplicative constant in front of the integral is determined by the exact definition of the measure $\nu$.  
\rd{The scaling behaviour in the hydrodynamic limit~\eqref{CToUps} reveals that the collective properties of the system are only determined by the corresponding scaling properties of the tails of the microscopic step size distribution.
In fact, while for $0<\alpha<1$ this term is dominant, in the hydrodynamic limit, against the conventional advective term 
(i.e., the term $\propto v_1$ in Eq.~\eqref{fintegral}), it becomes of the same order or sub-leading for $\alpha \geq 1$.  
In this latter case, therefore, no matter what specific step size distribution is chosen, 
the model is equivalent to an ordinary active fluid with constant self-propulsion velocity 
(again in the hydrodynamic limit).}
\rd{For L{\'e}vy stable distributed step sizes analytical formulas can be obtained straightforwardly 
(see Appendix~\ref{sec:exact-Upsilon} Eqs.~(\ref{upsilon_0}, \ref{upsilon_n})).} 
\rd{In this specific case,} numerical results suggest that these coefficients decay as $m^{\prime\,-1-\alpha}$ for $m^{\prime} \to \infty$ (see Fig.~\ref{fig3}), thus ensuring the convergence of the series expansion~\eqref{expansion}.
These coefficients \rd{also} satisfy the property   
\begin{equation}
\Upsilon_{-m^{\prime}}=(-1)^{m^{\prime}} \Upsilon_{m^{\prime}},  
\label{upsprop}
\end{equation} 
or alternatively   
$i^{-m^{\prime}}\Upsilon_{-m^{\prime}}=i^{m^{\prime}}\Upsilon_{m^{\prime}}$.

In contrast, when $\nu$ exhibits an upper cutoff $H$ on its tails, 
such as for the tempered L{\'e}vy stable subordinator 
\eqref{tempered-stable-measure},
different scenarios are predicted according to 
how the cutoff depends on the system size.  
Upon rescaling the integration variable, we obtain   
\begin{equation}
C_{m^{\prime}}\sim k^{\alpha}
\int_{-\pi}^{\pi} e^{\,i \phi^{\prime} m^{\prime}}
\int_0^{\infty} [1-e^{\,i z^{\prime} \cos{\phi^{\prime}}}]e^{\,-\frac{z^{\prime}}{H k}} \frac{\diff{z^{\prime}}}{z^{\prime\,1+\alpha}}
\diff{\phi^{\prime}}. 
\end{equation}
The scaling behaviour of this integral in the hydrodynamic limit thus depends on the exponential tempering factor, 
which in terms of the system size $L$ can be written as 
$e^{-z^{\prime}L/H }$. 
Three different scenarios are identified: 
(I) If $H\gg L$, in the hydrodynamic limit we have 
$e^{-z^{\prime}L/H}\sim 1$. 
Thus, the model is effectively equivalent to the case of heavy-tailed $\nu$. 
(II) If $H\ll L$, in the hydrodynamic limit we obtain 
$e^{-z^{\prime}L/H}\sim 0$, 
\rd{meaning that} the terms originating from the non local advection do not contribute to the hydrodynamic EOM. 
In this case, the model is equivalent to ordinary active matter with self-propulsion particle velocity $v_1$ \cite{Vicsek1995,Toner1995,Toner1998}. 
(III) If $H\sim L$, in the hydrodynamic limit we find 
$e^{-z^{\prime}L/H}\sim e^{-z^{\prime}}$.  
In this case, the model is expected to exhibit distinctive behaviour in the hydrodynamic limit than the other two regimes, but this \rd{special} case will be discussed in future publications. 

We now derive the hydrodynamic equations for a heavy-tailed measure $\nu$, and specifically we choose the L{\'e}vy stable subordinator~\eqref{stable-measure}. 
Using the relations~(\rd{\ref{fintegral}) and (\ref{expansion}}), 
\rd{the scaling formulas for the coefficients $C_m$~(\ref{CToUps}) and (\ref{UpsilonC}), 
and their property~(\ref{upsprop}),} 
we obtain 
the equations for the Fourier transform of the angular modes of $P$
\begin{multline}
\derpar{}{t} \hat f_m= - (\Upsilon_0 k^{\alpha}+\sigma m^2) \hat f_m + 
\frac{\gamma m}{2} (\hat f_{m-1}\star \hat f_{1} - \hat f_{m+1}\star \hat f_{-1})
\\ 
-k^{\alpha}\sum_{m^{\prime}=1}^{\infty} i^{m^{\prime}} \Upsilon_{m^{\prime}} (e^{-i m^{\prime} \psi} \hat f_{m+m^{\prime}}+e^{i m^{\prime} \psi} \hat f_{m-m^{\prime}}).
\label{fncup}
\end{multline}
\rd{These coupled non-linear equations clearly cannot be solved analytically for the entire hierarchy of angular modes; a suitable approximation scheme is thus required. 
In analogy with ordinary active matter 
\cite{Peruani2008,Bertin2006,Bertin2009,Lee2010,Peshkov2014,Bertin2017}, 
Eqs.~\eqref{fncup} prescribe that higher order angular modes are suppressed by the "mass" term $\propto m^2$, 
thus highlighting that only lower order modes are relevant to describe the macroscopic dynamical behaviour of the system.   
The fractional advection similarly enhances this damping by a term $\propto \Upsilon_0 k^{\alpha}$, 
which is however equal for all modes. 
In addition, it strengthens the coupling between the equations,   
which now extends to the entire hierarchy of angular modes, 
while for ordinary active matter it is limited only to the modes $\hat f_1$, $\hat f_{m-1}$ and $\hat f_{m+1}$.  
We remark that this coupling between the equations leaves the damping effect unaltered.          
All these observations suggest that we can adopt here the same approximation strategy usually employed in ordinary active matter 
\cite{Peruani2008,Bertin2006,Bertin2009,Lee2010,Peshkov2014,Bertin2017}. 
Therefore, we assume $\hat f_{m}\approx 0$ for $m>2$. 
Further recalling that $\hat f_{-m}=\hat f_{m}^*$ 
(the superscript ${}^*$ denotes complex conjugation), 
we obtain the set of coupled equations}
\begin{align}
\derpar{}{t} \hat f_0&=-\Upsilon_0 k^{\alpha} \hat f_0 - i\Upsilon_{1} k^{\alpha}(e^{-i\psi} \hat f_1 + e^{i\psi} \hat f_{1}^*) 
\notag\\ & \quad 
+\Upsilon_{2} k^{\alpha}(e^{-i 2\psi} \hat f_2 + e^{i2\psi} \hat f_{2}^*), \label{f0eq} \\
\derpar{}{t} \hat f_1&=-(\Upsilon_0 k^{\alpha}+\sigma)\hat f_1+\frac{\gamma}{2}(\hat f_{0}\star \hat f_{1} - \hat f_{2}\star \hat f_{1}^*) 
\notag\\ & \quad
-i\Upsilon_{1} k^{\alpha}(e^{-i\psi}\hat f_2 + e^{i\psi}\hat f_{0})+\Upsilon_{2} k^{\alpha}e^{i 2\psi} \hat f_{1}^* 
\notag\\ & \quad
+i\Upsilon_{3} k^{\alpha}e^{i3\psi} \hat f_{2}^*, \label{f1eq} \\
\derpar{}{t} \hat f_2&=-(\Upsilon_0 k^{\alpha}+4\sigma)\hat f_2 + \gamma(\hat f_{1}\star \hat f_{1})-i\Upsilon_{1} k^{\alpha}e^{i \psi} \hat f_1 
\notag\\ & \quad
+ \Upsilon_{2} k^{\alpha}e^{i 2\psi} \hat f_{0} + i\Upsilon_{3} k^{\alpha}e^{i3\psi} \hat f_{1}^* - \Upsilon_{4} k^{\alpha}e^{i4\psi} \hat f_{2}^*. \label{f2eq}
\end{align} 
\rd{By further imposing $\partial_t \hat f_2\approx0$, 
Eq.~\eqref{f2eq} can be solved analytically for $\hat f_2$ 
(see Appendix~\ref{derivationf2}). 
In detail,    
\begin{align}
\hat f_2&= \gamma \hat D_0(\hat f_{1}\star \hat f_{1}) 
- \hat D_1 k^{\alpha}ie^{i\psi}\hat f_{1}
+ \hat D_2 k^{\alpha}e^{i2\psi}\hat f_{0}
\notag\\ & \quad
+ \hat D_3 k^{\alpha}ie^{i3\psi}\hat f^*_{1}
- \gamma \hat D_4 e^{i4\psi}(\hat f_{1}\star \hat f_{1})^* ,
\label{eq:f2sol}
\end{align}  
with $(k,\alpha)$-dependent coefficients 
$\hat D_j$ ($j=0,\ldots,4$)  
(see Appendix~\ref{derivationf2}, Eqs.~(\ref{eq:D0}), (\ref{eq:D4}) and (\ref{eq:D1}--\ref{eq:D3})). 
Their dependence on the Fourier variable $k$ is studied in Fig.~\ref{fig4} (see Appendix~\ref{derivationf2} for asymptotic formulas). 
Finally, by substituting Eq.~\eqref{eq:f2sol} into Eqs.~(\ref{f0eq}) and (\ref{f1eq}) 
we derive closed analytical equations for the lower order modes $\hat f_0$ and $\hat f_1$ valid at all wavelengths (i.e., $\forall k \in \mathbb{R}$).

\begin{figure}[!htb]
\centering
\includegraphics[width=85mm,keepaspectratio]{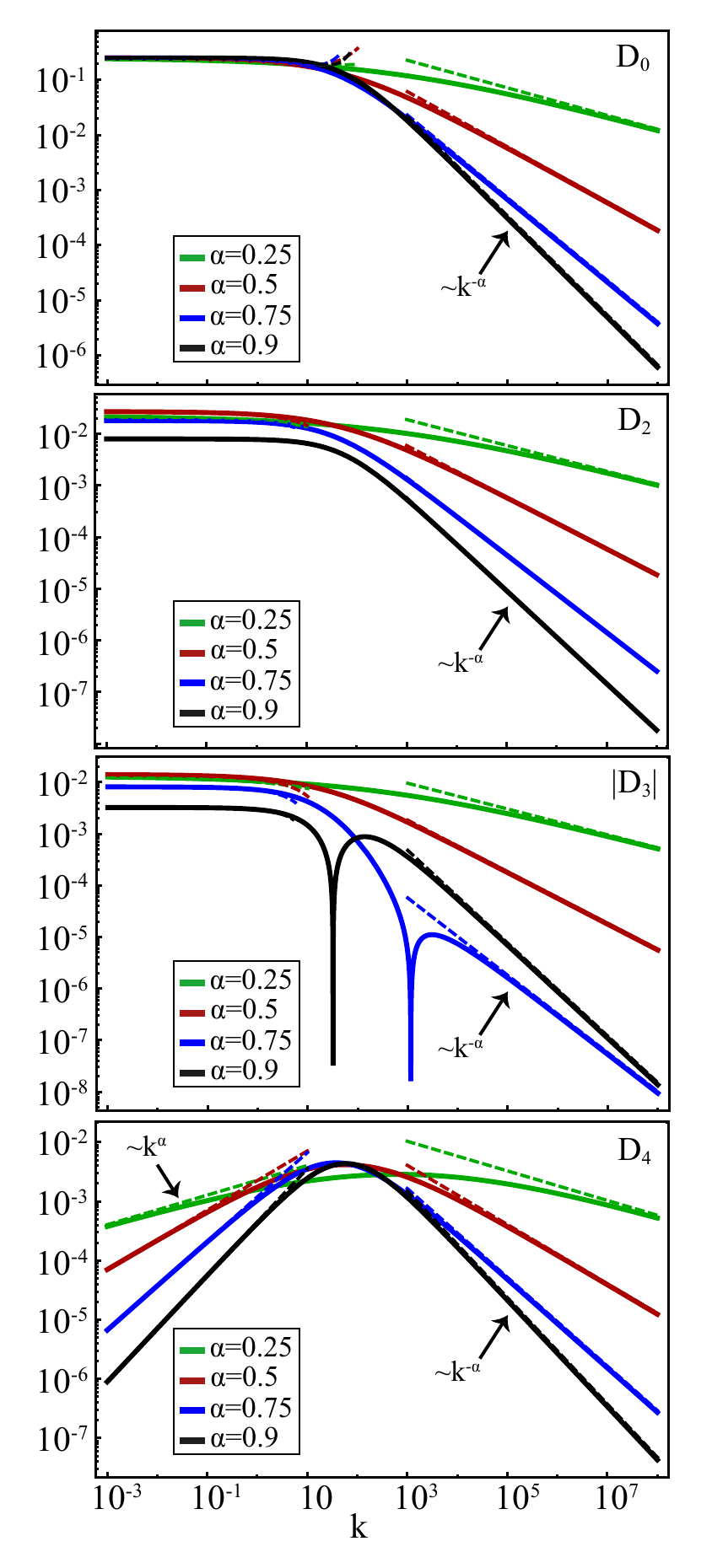}
\caption{\rd{Coefficients $\hat D_j$ ($j=0,\ldots,4$) vs. the Fourier variable $k$ for different values of $\alpha$. $\hat D_1$ is not shown as it is qualitatively similar to $\hat D_0$. Their asymptotic behaviour in the limit of long and short wavelengths is highlighted (see Appendix~\ref{derivationf2}, Eqs.~(\ref{eq:A0}--\ref{eq:A4}) and (\ref{eq:B0}--\ref{eq:B4}) respectively).}}
\label{fig4}
\end{figure}

These equations potentially contain an infinite number of higher order terms in the wavelength number $k$, 
which are encapsulated in the coefficients $\hat D_j$. 
Our interest, however, is in the hydrodynamic limit of these equations, 
where only few of these terms are in fact relevant. 
To reveal these terms, we will thus expand all coefficients $\hat D_j$ as power series of $k^{\alpha}$
and keep only the first order terms, i.e., those that are $\mathcal{O}(k^{2\alpha})$.    
Within this expansion scheme, we can then write 
$\hat f_2\sim \hat f_2^{\,(0)}+k^{\alpha}\hat f_2^{\,(1)}$   
with 
(see Appendix~\ref{derivationf2}, Eqs.~(\ref{eq:A0}--\ref{eq:A4}))
\begin{align}
\hat f_2^{\,(0)}&=\frac{\gamma}{4\sigma} (\hat f_{1}\star \hat f_{1}) \\
\hat f_2^{\,(1)}&=
-\frac{\gamma\Upsilon_0}{(4\sigma)^2} (\hat f_{1}\star \hat f_{1})
-\frac{\Upsilon_1}{4\sigma}ie^{i\psi}\hat f_{1}
+\frac{\Upsilon_2}{4\sigma}e^{i2\psi}\hat f_{0}
\notag\\ & \quad
+\frac{\Upsilon_3}{4\sigma}ie^{i3\psi}\hat f^*_{1}
-\frac{\gamma\Upsilon_4}{(4\sigma)^2} e^{i4\psi}(\hat f_{1}\star \hat f_{1})^* . 
\end{align}
This corresponds to an expansion of the nematic tensor as 
$\vec{Q}\sim \vec{Q}^{(0)}+\vec{Q}^{(1)}$  
with, respectively, 
\begin{align}
\rd{\hat{\vec{Q}}^{(0)}}\,&
\rd{=\frac{1}{2}
\left(
\begin{matrix}
\Re \hat f_2^{(0)} & \Im \hat f_2^{(0)} \\
\Im \hat f_2^{(0)} & -\Re \hat f_2^{(0)}
\end{matrix}
\right),} \label{eq:Q0tof2} \\
\rd{\hat{\vec{Q}}^{(1)}}\,&
\rd{=\frac{1}{2}
\left(
\begin{matrix}
\Re \hat f_2^{(1)} & \Im \hat f_2^{(1)} \\
\Im \hat f_2^{(1)} & -\Re \hat f_2^{(1)} 
\end{matrix}
\right).} \label{eq:Q1tof2}
\end{align}
The hydrodynamic equations for $\hat f_0$ and $\hat f_1$ are thus 
\begin{align}
&\derpar{}{t} \hat f_0=-\Upsilon_0 k^{\alpha} \hat f_0 
- \Upsilon_{1} k^{\alpha}i(e^{-i\psi} \hat f_1 + e^{i\psi} \hat f_{1}^*)
\notag\\ & \quad 
+ \frac{\gamma \Upsilon_{2}}{4\sigma}k^{\alpha}[e^{-i 2\psi}(\hat f_{1}\star \hat f_{1})+ e^{i 2\psi}(\hat f_{1}\star \hat f_{1})^{*}] , 
\label{f0eq2}\\
&\derpar{}{t} \hat f_1=-\left(\sigma+\Upsilon_0 k^{\alpha}
\right)\hat f_1 
+ \frac{\gamma}{2}(\hat f_{0}\star \hat f_{1}) 
-\frac{\gamma^2}{8\sigma}(\hat f_{1}\star \hat f_{1}^*)\star \hat f_{1}
\notag\\ & \quad
- \Upsilon_{1} k^{\alpha}ie^{i\psi} \hat f_{0}
+\Upsilon_{2} k^{\alpha}e^{i 2\psi}\hat f_{1}^*
-\frac{\gamma\Upsilon_1}{4\sigma}k^{\alpha}ie^{-i\psi}(\hat f_{1}\star \hat f_{1})
\notag\\ & \quad
+\frac{\gamma\Upsilon_3}{4\sigma}k^{\alpha}ie^{i3\psi}(\hat f_{1}\star \hat f_{1})^*
-\frac{\gamma}{2}\hat f_{1}^*\star(k^{\alpha}\hat f_{2}^{\,(1)}).
\label{f1eq2} 
\end{align}
}

Finally, we perform the inverse Fourier transform of Eqs.~(\ref{f0eq2}, \ref{f1eq2}). 
To this aim, we define the fractional Riesz integral operator $\mrI^{\beta}$ 
in terms of its Fourier transform $\mathcal{F}\{(\mrI^{\beta} h)(\vec{x})\}\equiv k^{-\beta} \hat h(\vec{k})$ (for $\Re{\beta}>0$).
Its expression as a non local spatial integral in two dimensions is \cite{Tarasov2011}
\begin{equation}
(\mrI^{\beta} h)(\vec{x})\equiv \frac{1}{\mu_2^{\prime}(\beta)}\int_{\mathbb{R}^2} h(\vec{x}-\boldsymbol{\zeta})
\frac{\diff{\boldsymbol{\zeta}}}{|\boldsymbol{\zeta}|^{2-\beta}}
\label{DiInt}
\end{equation}
with 
$\mu_2^{\prime}(\beta)\equiv 2^{\beta} \pi \Gamma(\beta/2)/\Gamma(2-\beta/2)$
for $\beta \neq 2(1+\rd{l})$ ($\rd{l}>0$, $\rd{l} \in \mathbb{Z}$). 
Correspondingly, we introduce the fractional Riesz derivative 
$\mrD^{\beta}$, 
as the operator with Fourier representation
$\mathcal{F}\{(\mrD^{\beta} h)(\vec{x})\}\equiv k^{\beta} \hat h(\vec{k})$.
Its integral representation is \rd{(see \cite{Samko1993}, Sec.~25.4)}  
\begin{equation}
(\mrD^{\beta}h)(\vec{x})=\frac{1}{c_{l}}\int_{\mathbb{R}^2}(\Delta^l_{\boldsymbol{\zeta}}h)(\vec{x})\frac{\diff{\boldsymbol{\zeta}}}{|\boldsymbol{\zeta}|^{2+\beta}}
\label{DbInt}
\end{equation}
with $l>\beta$, 
\rd{$\beta$-dependent normalizing coefficients $c_{l}$
and finite differences of the scalar function $h$
either centered, 
\begin{equation}
\Delta^l_{\boldsymbol{\zeta}}h(\vec{x})\equiv\sum_{m=0}^l (-1)^m \binom{l}{m} \mathcal{T}^-_{\left(\frac{l}{2}-m\right) \boldsymbol{\zeta}}h(\vec{x}),
\end{equation} 
in which case
$c_{l}\equiv 2^{l-\beta} i^l \int_{\mathbb{R}^2}(\sin{\zeta_x})^l |\boldsymbol{\zeta}|^{-2-\beta}\diff{\boldsymbol{\zeta}}$;
or non centered,  
\begin{equation}
\Delta^l_{\boldsymbol{\zeta}}h(\vec{x})\equiv\sum_{m=0}^l (-1)^m \binom{l}{m} \mathcal{T}^-_{m \boldsymbol{\zeta}}h(\vec{x}),
\end{equation}
in which case 
$c_{l}\equiv \int_{\mathbb{R}^2}(1-e^{i \zeta_x})^l |\boldsymbol{\zeta}|^{-2-\beta}\diff{\boldsymbol{\zeta}}$.
} 
\rd{These operators are the inverses of each other.}
\rd{Furthermore,} we use the relations presented in Table~\ref{identities} 
(see Section~\ref{auxiliary} for detailed proofs),  
with the vorticity field 
$\vecomega(\vec{r},t)\equiv\nabla\times\vec{p}$
and 
the auxiliary non-linear \rd{$(\vec{r},t)$-dependent} vector fields 
\rd{$\vecxi_0$ \rd{and} $\vecxi_1$ (below we introduce the notation $p \equiv |\vec{p}|$)}
\begin{align}
\vecxi_0&\equiv 2[(\vec{p}\cdot \nabla)\vec{p}+(\nabla\cdot\vec{p})\vec{p}]-\nabla p^2, \label{si-xi0} \\
\vecxi_1&\equiv (\vec{p}\cdot\nabla)\nabla^2\vec{p}+(\vec{p}\times\nabla)\times\nabla^2\vec{p} \notag\\ 
& \quad 
+3 [(\nabla\cdot\vec{p})\nabla^2\vec{p} + (\vecomega\times\nabla^2\vec{p}) - (\vecomega\cdot\nabla)\vecomega] . \label{si-xi1} 
\end{align} 
\rd{The hydrodynamic equations for the density and mean direction fields are thus}
\begin{widetext}
\begin{align}
\left( \derpar{}{t} + \Upsilon_0 \mrD^{\alpha} \right) \rho&=
2\Upsilon_1 \mrI^{1-\alpha} (\nabla \cdot \vec{p}) 
- \lambda_2 \mrI^{2-\alpha} (\nabla\cdot\vecxi_0);
\label{rhoeq} \\
\left( \derpar{}{t} + \Upsilon_0 \mrD^{\alpha} \right)\vec{p}&=
\left[\kappa_0(\rho) - \xi p^2\right]\vec{p}
+\Upsilon_1\mrI^{1-\alpha}\nabla \rho
+\lambda_1 \mrI^{1-\alpha} \vecxi_0
-\Upsilon_2\mrI^{2-\alpha}\nabla^2\vec{p}
+\lambda_3\mrI^{3-\alpha} \vecxi_1 
-\vec{Q}^{(1)} \vec{p}
\label{meandireq}
\end{align}
\end{widetext}
\rd{with the components of the tensor $\vec{Q}^{(1)}$ specified by 
$\text{Q}^{(1)}_{xx}=\mathcal{Q}^{(1)}_1=-\text{Q}^{(1)}_{yy}$ and   
$\text{Q}^{(1)}_{xy}=\mathcal{Q}^{(1)}_2=\text{Q}^{(1)}_{yx}$ with
(note that we absorbed a multiplicative factor $\gamma$ in its definition)
\begin{align}
\mathcal{Q}_{j}^{(1)} 
&=
-B_0\mrD^{\alpha}
\text{M}_{jl}
p_{l}
+\frac{\lambda_1}{2} \mrI^{1-\alpha}
\text{T}_{jl}p_l
-\frac{\lambda_2}{4}\mrI^{2-\alpha}
\text{T}_{jl}\partial_l \rho
\notag \\
& \quad
+\frac{\lambda_3}{4}\mrI^{3-\alpha}
\text{T}_{jl}\nabla^2 p_l
-B_4 \mrI^{4-\alpha}
\text{T}_{jl}\Xi_{1l} , 
\end{align}
where $j, l = 1, 2$ and we identify $p_1$ and $p_2$ with the components of $\vec{p}$ along the $x$ and $y$ axis, respectively. In the above, we introduce the $(p_x,p_y)$-dependent tensor function       
$\text{M}_{jl}\equiv (p_{x}\delta_{jl} + p_{y} \epsilon_{jl})$, 
the differential matrix operator 
$\text{T}_{jl}\equiv \partial_{x}\delta_{jl} + \partial_{y}\epsilon_{jl}$, 
the Kronecher symbol $\delta_{jl}$ and the generator of counter-clock wise rotations $\epsilon_{jl}$,  
defined by $\epsilon_{11}=0=\epsilon_{22}$ and $\epsilon_{21}=1=-\epsilon_{12}$ 
(see Appendix~\ref{auxiliary}, Eq.~\eqref{eq:pauli}). 
In addition, we defined the function of the density field 
\begin{equation}
\kappa_0(\rho)\equiv -\sigma+ \frac{\gamma \rho}{2}, 
\label{si-kappa0}
\end{equation}
the $\alpha$-dependent coefficients
\begin{align}
\lambda_1&\equiv \frac{\gamma \Upsilon_{1}}{4\sigma} , \ &
\lambda_2&\equiv \frac{\gamma \Upsilon_{2}}{2\sigma} , \ &
 \lambda_3&\equiv \frac{\gamma \Upsilon_{3}}{2\sigma} ,
\label{si-alphaP}
\end{align}
and $\alpha$-independent ones
\begin{align}
B_0&\equiv \frac{\gamma^2 \Upsilon_0}{32\sigma^2} ,\ &
B_4&\equiv \frac{\gamma^2 \Upsilon_4}{16\sigma^2} , \ &  
\xi&\equiv \frac{\gamma^2}{8\sigma} \ . 
\label{si-alphaP}
\end{align}
}

As a check on the validity of our hydrodynamic EOM,	
we write out explicitly these equations 
for $\alpha=1$.  
In this case, one can show that (see \rd{Appendix~\ref{sec:exact-Upsilon}, Eq.~\eqref{rep1}})
\begin{equation}
\Upsilon_{m^{\prime}}=
(-i)^{m^{\prime}+1}\frac{m^{\prime} \sin{(m^{\prime}\pi)}}{\pi (1-m^{\prime\,2})}=
\left\{
\begin{matrix}
0 & & m^{\prime}\neq 1 \\
-1/2 & & m^{\prime}=1
\end{matrix}
\right. . 
\label{UpsilonVicsek}
\end{equation}
Therefore, introducing further auxiliary parameters     
\begin{eqnarray}
\zeta_0\equiv \frac{3\gamma}{16\sigma} \ &,& \
\rd{\lambda_0\equiv \frac{5\gamma}{16\sigma} \ , }
\end{eqnarray} 
\rd{we obtain 
$\mathcal{Q}_{1}^{(1)}=\lambda_0(\partial_y p_y - \partial_x p_x)/5$
and 
$\mathcal{Q}_{2}^{(1)} =-\lambda_0(\partial_y p_x + \partial_x p_y)/5$.
Using these formulas, we can show that 
$\vec{Q}^{(1)}\vec{p}=\lambda_0[(\nabla\cdot\vec{p})\vec{p}-(\vec{p}\cdot\nabla)\vec{p}-\nabla p^2/2]/5$. Therefore,}
Eqs.~(\ref{rhoeq}) and (\ref{meandireq}) reduce to a version of the hydrodynamic equations of ordinary active fluids derived by Toner and Tu \cite{Toner1995} 
\begin{align}
\derpar{}{t} \rho &= - \nabla \cdot \vec{p}\ , \label{TTeq1} \\ 
\derpar{}{t}\vec{p}&=
\left[\kappa_0(\rho)-\xi p^2\right]\vec{p}
-\frac{1}{2}\nabla\left(\rho - \lambda_0 p^2\right) \notag\\
&\quad -\lambda_0(\nabla\cdot\vec{p})\vec{p}
-\zeta_0(\vec{p}\cdot \nabla)\vec{p} \ . 
\label{TTeq2}
\end{align}
\rd{These are equations for an active inviscid fluid. 
In fact, the passive viscous term in ordinary active matter is of higher order 
(specifically, it is of order $k^2$). Therefore, it cannot be captured within our approximation scheme.}

\rd{Our hydrodynamic EOM are manifestly more complex than the Toner-Tu equations.
Nevertheless, Eq.~\eqref{rhoeq} can be interpreted as the counterpart of the conventional continuity equation expressing mass conservation. 
In fact, the rate of mass change over an infinitesimal time interval is given by    
\begin{equation}
\derpar{}{t} \int_{\mathbb{R}^2} \rho(\vec{r},t) \diff{\vec{r}}=
\left. \derpar{}{t} \hat\rho(\vec{k},t)\right|_{\vec{k}=0}=0 \ , 
\end{equation}
because  
$\partial_t \hat\rho(\vec{k},t)\propto k^{\alpha}$ (see Eq.~\eqref{f0eq2}). 
Therefore, our hydrodynamic model conserves mass as required.  
Eq.~\eqref{meandireq} instead describes the time evolution of the director field $\vec{p}$ by accounting for contributions due to the rotational diffusion of active particles, their anomalous displacement statistics and polar alignment interactions.} 

\rd{
In more details, we can understand Eqs.~\eqref{rhoeq} and \eqref{meandireq} term by term. 
Starting from the former one, 
the term $\mathbb{D}^{\alpha}\rho$ in its lhs describes the isotropic superdiffusion of the density field, 
which is a direct consequence of the L{\'e}vy stable distributed particle displacements. 
Its effect can be understood by calculating the solution of the equation with $\vec{p}=0$ for the initial condition 
$\rho(\vec{r},t=0)=\delta(\vec{r}-\vec{r}_0)$, which is
\begin{equation}
\rho(\vec{r},t)=\frac{1}{(\Upsilon_0 t)^{2/\alpha}}H_2\!\left(\frac{|\vec{r}-\vec{r}_0|}{(\Upsilon_0 t)^{1/\alpha}}\right)
\end{equation}
with $H_2(x)\equiv (2\pi)^{-1} \int_{0}^{\infty} k J_0(k x)e^{-k^{\alpha}} \diff{k}$
and $J_0$ a Bessel function of the first kind \cite{Fouxon2017}. 
This is a L\'evy distribution with variance increasing over time; 
therefore, it displays the anomalous diffusive spreading of the underlying microscopic particles. 
Conversely, for ordinary active matter, the advective term is null in the absence of global polarization order 
and, consequently, the density remains localized at the origin forever.  
In addition, by recalling \eqref{DbInt}, we note that the kernel function specifying the non local character of this term can be defined as 
$\mu_0(\diff{\boldsymbol{\zeta}})\propto |\boldsymbol{\zeta}|^{-2-\alpha}\diff{\boldsymbol{\zeta}}$, 
which is the distribution of rotationally invariant stable variables in two dimensions with stability parameter $\alpha$.  
Interestingly, this is an isotropic probability distribution with the same tail scaling properties of the microscopic step size distribution.   

The first term in the rhs of Eq.~\eqref{rhoeq} is a fractional divergence, 
thus quantifying the flux of active particles. 
Indeed, taking its integral over some prescribed volume $V\subset{\mathbb{R}}^2$, we can write  
\begin{multline}
\int_V \mrI^{1-\alpha}(\nabla \cdot \vec{p}) \diff{\vec{r}}=
\int_{S} \mrI^{1-\alpha}\vec{p}\cdot\diff{\vec{S}} 
\\
=
\int_{\mathbb{R}^2} \left( \int_{S} \vec{p}(\vec{r} + \boldsymbol{\zeta} ,t)\cdot\diff{\vec{S}}\right) |\boldsymbol{\zeta}|^{-1-\alpha} \frac{\diff{\boldsymbol{\zeta}}}{\mu^{\prime}_1}
\end{multline}
with $S$ the volume boundary 
and the positive coefficient
$\mu^{\prime}_1(\alpha)\equiv \Gamma((1-\alpha)/2)/[2^{\alpha}\Gamma((3+\alpha)/2)]$  
\cite{Tarasov2011}.
This expression can be interpreted straightforwardly:
the inner integral measures the flux of active particles jumping  into $V$ from an arbitrary  position $\vec{r} + \boldsymbol{\zeta}$; 
the outer integral is the average over all such jumps whose statistics is specified by the probability measure 
$\mu_1(\diff{\boldsymbol{\zeta}})\equiv (2\Upsilon_1/\mu^{\prime}_1) |\boldsymbol{\zeta}|^{-1-\alpha}\diff{\boldsymbol{\zeta}}$,
which we can show to be the distribution of one-dimensional $\alpha$-stable random variables.   
Thus, the non local character of this term is fully specified by the microscopic step size statistics.  
The second term 
can be interpreted, analogously, as a non-linear contribution to the active flux.  
In fact, we can write  
\begin{multline}
\int_V \mrI^{2-\alpha}(\nabla \cdot \vecxi_0) \diff{\vec{r}}=
\int_{S} \mrI^{2-\alpha}\vecxi_0\cdot\diff{\vec{S}}= \\
\int_{\mathbb{R}^2} \left( \int_{S} \vecxi_0(\vec{r} + \boldsymbol{\zeta} ,t)\cdot\diff{\vec{S}} \right) |\boldsymbol{\zeta}|^{-\alpha} \frac{\diff{\boldsymbol{\zeta}}}{\mu^{\prime}_2}
\end{multline}
with    
$\mu^{\prime}_2(\alpha)\equiv 2^{1-\alpha}\Gamma(1-\alpha/2)/\Gamma(1+\alpha/2)$ \cite{Tarasov2011}.
We can recognize the flux of the vector field $\vecxi_0$, 
which depends non-linearly on $\vec{p}$ [see Eq.~\eqref{si-xi0}], 
averaged over all jumps with probability measure 
$\mu_2(\diff{\boldsymbol{\zeta}})\equiv (\gamma\Upsilon_2/2\sigma\mu^{\prime}_2) \zeta^{-\alpha}\diff{\boldsymbol{\zeta}}$.
Differently from the other terms, $\mu_2$ cannot be related as clearly to the microscopic step size distribution.


Likewise, we can also understand the term-by-term structure of Eq.~\eqref{meandireq}.  
The first term in its rhs typically characterizes alignment interacting systems, but is here
augmented by the anomalous diffusive term $\Upsilon_0 \mrD^{\alpha} \vec{p}$, similar to \eqref{rhoeq}.    
The hydrodynamic pressure gradient  
$\mrI^{1-\alpha}\nabla\left[\rho - (\gamma/4\sigma) p^2 \right]$, 
the convective term $\mrI^{1-\alpha}(\vec{p}\cdot \nabla)\vec{p}$, 
and the divergence-induced flow
$\mrI^{1-\alpha}(\nabla\cdot\vec{p})\vec{p}$, 
which are encapsulated in the term $\mrI^{1-\alpha}\vecxi_0$,
all have their counterparts in the Toner-Tu EOM, 
albeit modified with fractional derivative operators and possibly different multiplicative coefficients. 
Interestingly, their corresponding non local integral expressions all possess kernel functions 
that directly relate to the microscopic step size distribution, 
akin to the flux term in~\eqref{rhoeq}.

Conversely, the viscous-like term $\mrI^{2-\alpha}\nabla^2\vec{p}$,
the non linear term $\mrI^{3-\alpha}\vecxi_1$ and all those included through
the nematic term $\vec{Q}^{(1)}\vec{p}$, have no counterparts in ordinary active fluids.
While for some of these terms the corresponding non-local integral expression is integrated over the same probability measures discussed previously,
for other ones (such as those involving the operators $\mathbb{I}^{3-\alpha}$ and $\mathbb{I}^{4-\alpha}$) we need to introduce other non-trivial measures (not discussed further here).   
All these latter terms correspond to higher-order terms in the Fourier angular expansion.   

Overall, this analysis highlights that, on the one hand,
many terms in the hydrodynamic EOM of ALM can be inferred directly by simply knowing the Toner-Tu equations for ordinary active matter and the scaling properties of the microscopic step size distribution.  
Therefore, all these terms can potentially be derived by using symmetry considerations only. 
On the other hand, however, it shows that these equations also contain several higher order terms 
that exhibit not only non-linear fields contributions but also highly non trivial jump statistics 
(in the corresponding integral expression).  
These terms can thus be correctly identified only by applying a formal coarse-graining procedure. 
}

\begin{table*}[!htb]
\newcommand*{\SetEqNum}{\refstepcounter{equation}\thetag\theequation}
\begin{tabular}{@{\hskip 0.1in}l@{\hskip 0.1in}l@{\hskip 0.1in}l@{\hskip 0.1in}l@{\hskip 0.1in}l@{\hskip 0.1in}} 
Original term & Manipulated term & Inv. Fourier transf. & $\mathbb{C}\to\mathbb{R}^2$ & Eq.  \\ 
\hline\hline
& & & & \\ [\dimexpr-\normalbaselineskip+0pt]
$i k^{\alpha}(e^{-i\psi} \hat f_1 + e^{i\psi} \hat f_{1}^*)$ & $-k^{\alpha-1}(\hat{\nabla}_{\vec{k}}^* \hat f_{1}+\hat{\nabla}_{\vec{k}} \hat f_{1}^*)$ & $-2\mrI^{1-\alpha}\Re{(\hat{\nabla}^*f_{1})}$ & $-2\mrI^{1-\alpha}(\nabla\cdot\vec{p})$ & \SetEqNum\label{aux1} \\
$k^{\alpha}[e^{-i 2\psi}(\hat f_{1}\star \hat f_{1})+ c.c.]$ & $-2k^{\alpha-2}\Re{[\hat{\nabla}_{\vec{k}}^{*2}(\hat f_{1}\star \hat f_{1})]}$ & $-2 \mrI^{2-\alpha} \Re{(\hat{\nabla}^{*2} f_1^2)}$ & $-2\mrI^{2-\alpha}(\nabla\cdot\vecxi_0)$ & \SetEqNum\label{aux2} \\ 
$ik^{\alpha}e^{i\psi} \hat f_{0}$ & $-k^{\alpha-1}\hat{\nabla}_{\vec{k}} \hat f_{0}$ & $-\mrI^{1-\alpha}\nabla f_{0}$ & $-\mrI^{1-\alpha}\nabla \rho$ & \SetEqNum\label{aux3} \\
$i k^{\alpha}e^{-i\psi} (\hat f_{1}\star \hat f_{1})$ & $-k^{\alpha-1}\hat{\nabla}_{\vec{k}}^*(\hat f_{1}\star \hat f_{1})$ & $-\mrI^{1-\alpha}\hat\nabla^* f_1^2$ & 
$-\mrI^{1-\alpha}\vecxi_0$ & \SetEqNum\label{aux4} \\
$k^{\alpha}e^{i2\psi} \hat f_{1}^*$ & $-k^{\alpha-2}\hat{\nabla}_{\vec{k}}^2 \hat f_{1}^*$ & $-\mrI^{2-\alpha}\hat\nabla^2 f_1^*$ & $-\mrI^{2-\alpha}\nabla^2 \vec{p}$ & \SetEqNum\label{aux6} \\ 
\rd{$i k^{\alpha}e^{i\psi} \hat f_{1}$} & \rd{$-k^{\alpha-1}(\hat{\nabla}_{\vec{k}}\hat f_{1})$} & \rd{$-\mrI^{1-\alpha} (\hat \nabla f_1)$} & \rd{$-\mrI^{1-\alpha}\textbf{T}\vec{p}$} & \SetEqNum\label{aux5} \\
\rd{$k^{\alpha}e^{i2\psi} \hat f_{0}$} & \rd{$-k^{\alpha-2}(\hat{\nabla}_{\vec{k}}^2 \hat f_{0})$} & \rd{$-\mrI^{2-\alpha}(\hat\nabla^2 f_0)$} & \rd{$-\mrI^{2-\alpha} \textbf{T}\nabla\rho$} & \SetEqNum\label{aux7} \\
\rd{$i k^{\alpha}e^{i3\psi} \hat{f}^*_{1}$} & \rd{$k^{\alpha-3}(\hat{\nabla}_{\vec{k}}^3 \hat{f}^*_{1})$} & \rd{$\mrI^{3-\alpha}(\hat\nabla^3 f_1^*)$} & \rd{$\mrI^{3-\alpha} \textbf{T} \nabla^2 \vec{p}$} & \SetEqNum\label{aux10} \\
$i k^{\alpha}e^{i3\psi} (\hat{f}_{1}\star\hat{f}_{1})^*$ & $k^{\alpha-3}\hat{\nabla}_{\vec{k}}^3 (\hat{f}_{1}\star\hat{f}_{1})^*$ & $\mrI^{3-\alpha}\hat\nabla^3 f_1^{*2}$ & $2\mrI^{3-\alpha} \vecxi_1 $ & \SetEqNum\label{aux8} \\
\rd{$k^{\alpha}e^{i4\psi} (\hat{f}_{1}\star\hat{f}_{1})^*$} & \rd{$k^{\alpha-4}[\hat{\nabla}_{\vec{k}}^4 (\hat{f}_{1}\star\hat{f}_{1})^*]$} & \rd{$\mrI^{4-\alpha}(\hat\nabla^4 f_1^{*2})$} & \rd{$2\mrI^{4-\alpha}\textbf{T}\vecxi_1 $} & \SetEqNum\label{aux9} \\
& & & & \\ [\dimexpr-\normalbaselineskip+0pt]
\hline
\end{tabular}
\caption{Summary of the identities used to calculate the inverse Fourier transform and the vector representation of Eqs.~(\ref{f0eq2}, \ref{f1eq2}). We define complex derivatives 
$\hat\nabla\equiv \partial_x + i\partial_y$ and 
$\hat\nabla^*\equiv \partial_x-i\partial_y$, 
such that their Fourier transforms are 
$\hat\nabla_{\vec{k}}\equiv -i k e^{i \psi}$, 
$\hat\nabla^*_{\vec{k}}\equiv -i k e^{-i \psi}$.
\rd{In addition, we have introduced the differential matrix operator 
$\text{T}_{ij}\equiv \partial_{x}\delta_{ij} + \partial_{y}\epsilon_{ij}$ 
with $\delta_{ij}$ the Kronocker symbol and $\epsilon_{ij}$ the generator of counter-clock wise rotations 
(see main text).}  
Proofs of these relations are presented in Appendix~\ref{auxiliary}. }\label{identities}
\end{table*}

\section{Mean field solutions and Linear stability analysis \label{si-sec-lsa}}

The mean-field solutions of Eqs.~(\ref{rhoeq}, \ref{meandireq}), 
which we call $\rho^*$ and $\vec{p}^*$,  
are obtained by solving the EOM for spatially homogeneous hydrodynamic fields. 
Clearly, not only ordinary but also fractional derivatives are null when applied on spatially homogeneous functions.   
Therefore, ALM is completely equivalent to ordinary active matter at the mean field level. 
In particular, their hydrodynamic EOM have the same mean field solutions: 
On the one hand, there exists a disordered gas phase with constant density and $\vec{p}^*=0$ for noise strengths $\sigma\geq\sigma_t$  
with the density dependent threshold 
$\sigma_t(\rho^*)\equiv \gamma\rho^*/2$; 
on the other hand, for $\sigma<\sigma_t$ a ordered liquid phase spontaneously emerges 
where collective motion is observed along an arbitrary direction $\vec{e}_1$ with 
$\vec{p}^*=\sqrt{\kappa_0(\rho^*)/\xi}\vec{e}_1$.

We then consider a generic perturbation of these solutions 
$\rho(\vec{r},t)=\rho^*+\delta\rho(\vec{r},t)$ and 
$\vec{p}(\vec{r},t)=\vec{p}^*+\delta\vec{p}(\vec{r},t)$, 
with $\delta \rho$, $\delta \vec{p}$ expressed as plane waves   
\begin{align}
\delta\rho(\vec{r},t)&=\delta \rho_0 e^{\,s t +i \vec{q}\cdot\vec{r}}, &
\delta\vec{p}(\vec{r},t)&=\delta\vec{p}_0 e^{\,s t +i \vec{q}\cdot\vec{r}}.
\label{perturbs}  
\end{align}
Using the linearity of the fractional operators,  
and the spatial homogeneity of the fields $\rho^*$ \rd{and} $\vec{p}^*$, we obtain the following equations for the perturbations:     
\begin{align}
\derpar{}{t} \delta\rho&= -\Upsilon_0 \mrD^{\alpha} \delta\rho  + 2\Upsilon_1 \mrI^{1-\alpha} (\nabla \cdot \delta\vec{p}) 
\notag\\ & \quad 
- 2 \lambda_2 \mrI^{2-\alpha} (\nabla\cdot\delta\vecxi_0),
\label{drho} \\
\derpar{}{t}\delta\vec{p}&=
\left[\frac{\gamma}{2}\delta\rho - 2\xi (\vec{p}^*\cdot\delta\vec{p})\right]\vec{p}^*
\rd{- \Upsilon_0 \mathbb{D}^{\alpha} \delta\vec{p}}
\notag  \\ & \quad
+\rd{\Upsilon_1\mrI^{1-\alpha}\nabla\delta\rho}
-\rd{\Upsilon_2 \mrI^{2-\alpha}\nabla^2 \delta\vec{p}}
\notag  \\ & \quad
+\rd{2\lambda_1\mrI^{1-\alpha} \delta\vecxi_0}
+\rd{\lambda_3\mrI^{3-\alpha} \delta\vecxi_1}
-\rd{\delta\vec{Q}^{(1)}\vec{p}^*},
\label{dp} 
\end{align}
\rd{where the perturbations of the vector fields $\vecxi$ are} 
\begin{align}
\delta\vecxi_0&=
(\nabla\cdot\delta\vec{p})\vec{p}^*+(\nabla\times\delta\vec{p})\times\vec{p}^*, \label{dzeta0} \\
\delta\vecxi_1&=(\vec{p}^*\cdot\nabla)\nabla^2\delta\vec{p}+(\vec{p}^*\times\nabla)\times\nabla^2\delta\vec{p} \label{dzeta1}   
\end{align}
\rd{and that of the nematic tensor $\vec{Q}^{(1)}$ is specified by
\begin{align}
\delta \text{Q}_{j}^{(1)}
&=
-B_0\mrD^{\alpha}
\delta \text{M}_{jl} \,
p_l^*  
-B_0\mrD^{\alpha}
\text{M}_{jl}^*\delta p_l
\notag \\
& \quad 
+\frac{\lambda_1}{2} \mrI^{1-\alpha}
\text{T}_{jl}\delta p_l
- \frac{\lambda_2}{4}\mrI^{2-\alpha}
\text{T}_{jl}\partial_l \delta\rho
\notag \\
& \quad
+\frac{\lambda_3}{4}\mrI^{3-\alpha}
\text{T}_{jl}\nabla^2\delta p_l
-B_4 \mrI^{4-\alpha}
\text{T}_{jl}\delta\Xi_{1 l} 
\label{dQ1}
\end{align}  
with the short-hand notations 
$\delta \text{M}_{jl} \equiv (\delta p_{x}\delta_{jl} + \delta p_{y} \epsilon_{jl})$ and 
$\text{M}_{jl}^* \equiv (p^*_{x}\delta_{jl} + p^*_{y} \epsilon_{jl})$.
In this equation, we adopt the same index convention defined in Sec.~\ref{si-sec-hydro}.
To further simplify the notation, in the next paragraphs, we will suppress all explicit dependencies on the homogeneous density $\rho^*$. 
}

\subsection{Disordered fluid state \label{si-sec-disPh}}

In the disordered phase we set $\vec{p}^*=\vec{0}$ and choose $\rho^*$ such as $\kappa_0(\rho^*)<0$. Therefore, Eqs.~(\ref{drho}, \ref{dp}) reduce to 
\begin{align}
\derpar{}{t} \delta\rho&=-\Upsilon_0 \mrD^{\alpha} \delta\rho + 2\Upsilon_1 \mrI^{1-\alpha} (\nabla \cdot \delta\vec{p}) , \\
\derpar{}{t} \delta\vec{p}&=\rd{[\kappa_0 - \Upsilon_0 \mathbb{D}^{\alpha}]\delta\vec{p}}+\Upsilon_1\mrI^{1-\alpha}\nabla\delta\rho 
-\Upsilon_2\mrI^{2-\alpha}\nabla^2 \delta\vec{p}.
\end{align}
We apply the ansatz~\eqref{perturbs} and compute the corresponding fractional terms by taking their Fourier transforms. 
Recalling that 
$\mathcal{F}\{e^{i \vec{q}\cdot\vec{r}}\}= 2\pi \delta(\vec{k}+\vec{q})$, 
we can write, e.g., for the terms in the first equation, 
\begin{align}
\mrD^{\alpha} \delta\rho &= 
\mathcal{F}^{-1}\{k^{\alpha} [2\pi e^{s t} \delta(\vec{k}+\vec{q})]\delta \rho_0\} \notag\\ 
&=\mathcal{F}^{-1}\{q^{\alpha} [2\pi e^{s t}\delta(\vec{k}+\vec{q})] \delta \rho_0\}= 
q^{\alpha} \delta \rho, 
\label{lsa1.1}
\end{align}
\rd{where we denote $q\equiv |\vec{q}|$,} and 
\begin{align}
\mrI^{1-\alpha} (\nabla \cdot \delta\vec{p})&= 
\mathcal{F}^{-1}\{ k^{\alpha-1} (-i \vec{k}\cdot\delta\vec{p}_0)[2\pi e^{s t} \delta(\vec{k}+\vec{q})]\} \notag\\
&=\mathcal{F}^{-1}\{ q^{\alpha-1} (i\vec{q}\cdot\delta\vec{p}_0)[2\pi e^{s t}\delta(\vec{k}+\vec{q})]\} \notag\\ 
&=q^{\alpha-1}(i\vec{q}\cdot\delta\vec{p}). 
\label{lsa1.2}
\end{align}
Thus, \rd{after eliminating the multiplicative factor $e^{s t + i \vec{q}\cdot\vec{r}}$,} we obtain the equation    
\begin{align}
(s + \Upsilon_0 q^{\alpha})\delta\rho_0-2\Upsilon_1 q^{\alpha-1} (i \vec{q}\cdot\delta\vec{p}_0)&=0.
\label{gasLeq1}   
\end{align}
Likewise, for the terms in the second equation we obtain    
\begin{align}
\mrI^{1-\alpha}\nabla\delta\rho&= 
\mathcal{F}^{-1}\{ k^{\alpha-1}(-i\vec{k})[2\pi e^{s t} \delta(\vec{k}+\vec{q})]\delta\rho_0\} \notag\\ 
&=\mathcal{F}^{-1}\{ q^{\alpha-1}(i\vec{q})[2\pi e^{s t} \delta(\vec{k}+\vec{q})]\delta\rho_0 \} \notag\\ 
&=q^{\alpha-1}(i\vec{q})\delta\rho,
\end{align}
and 
\begin{align} 
\mrI^{2-\alpha} \nabla^2 \delta\vec{p}&= 
\mathcal{F}^{-1}\{ k^{\alpha-2} (-i\vec{k})^2[2\pi e^{s t} \delta(\vec{k}+\vec{q})]\delta\vec{p}_0 \} \notag\\ 
&=\mathcal{F}^{-1}\{ -q^{\alpha} [2\pi e^{s t}\delta(\vec{k}+\vec{q})]\delta\vec{p}_0 \} \notag\\ 
&=-q^{\alpha}\delta\vec{p}. 
\end{align}
\rd{Therefore, the second equation yields}  
\begin{align}
\rd{\left[s -\kappa_0 + (\Upsilon_0-\Upsilon_2) q^{\alpha} \right]\delta\vec{p}_0}&=\Upsilon_1 q^{\alpha-1} (i\vec{q})\delta\rho_0.  
\label{gasLeq2} 
\end{align}
Multiplying Eq.~\eqref{gasLeq2} by $(i\vec{q})$ and solving it for $i\vec{q}\cdot\delta \vec{p}_0$ we obtain 
\begin{equation}
i\vec{q}\cdot\delta \vec{p}_0=\frac{-\Upsilon_1 q^{1+\alpha}\delta\rho_0}{\rd{\left[s -\kappa_0 + (\Upsilon_0-\Upsilon_2) q^{\alpha} \right]}}, 
\end{equation}
which can be combined with Eq.~(\ref{gasLeq1}) to yield the dispersion relation 
\begin{multline}  
\rd{s^2+s\left[ -\kappa_0 + (2\Upsilon_0-\Upsilon_2)q^{\alpha} \right] +2\Upsilon_1^2 q^{2\alpha}} \\
\rd{+ \Upsilon_0 q^{\alpha}\left[ -\kappa_0 + (\Upsilon_0-\Upsilon_2)q^{\alpha} \right]=0.}
\label{gasDrel}   
\end{multline}
Because by definition $\Upsilon_0>\Upsilon_2$ (for all $0<\alpha<1$)
\rd{and $\kappa_0<0$ by assumption}, the coefficient of the linear term is always positive. 
The solutions of Eq.~\eqref{gasDrel} are then given by    
\begin{equation} 
s_{\pm}=\rd{\frac{1}{2}\left[ \kappa_0 - (2\Upsilon_0-\Upsilon_2)q^{\alpha} \pm \sqrt{\Delta}\right]}, 
\label{GasSol}
\end{equation}
with the discriminant given by 
\begin{align}
\Delta&=\rd{\left[\kappa_0 - (2\Upsilon_0-\Upsilon_2 )q^{\alpha} \right]^2 + 4 \Upsilon_0 \kappa_0 q^{\alpha}} \notag \\ 
& \quad \rd{- 4 \left[\Upsilon_0(\Upsilon_0-\Upsilon_2)+ 2\Upsilon_1^2\right]q^{2\alpha}. }   \label{delta1} 
\end{align}
The last three terms in its rhs are negative for all $\alpha$, i.e, the inequality      
\rd{$\sqrt{\Delta}<-\kappa_0 + (2\Upsilon_0-\Upsilon_2)q^{\alpha}$} 
holds.  
This relation implies that $s_+<0$. 
Therefore, the disordered \rd{fluid} phase is stable with respect to perturbations in all directions, as in conventional polar active fluids.

\subsection{Ordered fluid state \label{si-sec-ordPh}}

In the ordered phase, 
\rd{we assume} the stationary homogeneous flow 
\rd{$\vec{p}^*\equiv p^* \vec{e}_{\parallel}$ 
with the unit vector $\vec{e}_{\parallel}\equiv(\cos \upvartheta_{\parallel}, \sin \upvartheta_{\parallel})$ specifying the arbitrary direction of symmetry breaking. 
We denote the corresponding orthogonal direction with the unit vector $\vec{e}_{\perp}\equiv(-\sin \upvartheta_{\parallel}, \cos \upvartheta_{\parallel})$.
These two vectors form a basis on which we can project the wave amplitude $\delta\vec{p}_0$, 
such that we can write $\delta\vec{p}_0 = \delta p_{0\parallel} \vec{e}_{\parallel} + \delta p_{0\perp} \vec{e}_{\perp}$, }  
\rd{Furthermore, we assume the wave number vector $\vec{q}=q \, \vec{e}_{\upvartheta}$ 
with the angle $\upvartheta$ defined with respect to $\vec{e}_{\parallel}$. 
Therefore, the unit vector is $\vec{e}_{\upvartheta} \equiv \cos{\upvartheta} \, \vec{e}_{\parallel} + \sin{\upvartheta} \, \vec{e}_{\perp}$.} 
Under these assumptions Eqs.~(\ref{drho}, \ref{dp}) can be solved similarly to the disordered phase. 
For the first equation in particular the only term \rd{left to be calculated} is 
$\mrI^{2-\alpha}(\nabla\cdot\,\delta\vecxi_0)$ with $\delta\vecxi_0$ specified by Eq.~\eqref{dzeta0}, 
which is given by (see Appendix~\ref{sec:lsa_aux1})  
\begin{align}
\mrI^{2-\alpha}(\nabla\cdot\delta\vecxi_0)&=\rd{-q^{\alpha}p^*[(\vec{e}_{\upvartheta}\cdot\vec{e}_{\parallel})(\delta\vec{p}\cdot\vec{e}_{\upvartheta})} 
\notag \\ & \quad 
\rd{+(\vec{e}_{\upvartheta}\cdot\vec{e}_{\perp})(\delta \vec{p}\cdot\vec{e}_{\upvartheta+\frac{\pi}{2}})] .}
\label{lsa1.3}
\end{align}  
\rd{All other terms have been already computed in the previous paragraph (see Eqs.~(\ref{lsa1.1}, \ref{lsa1.2})).}
\rd{Combining all these equations and removing the multiplicative factor $e^{s t + i\vec{q}\cdot\vec{r}}$, we derive} 
\rd{the following relation between $\delta \rho_0$ and $\delta\vec{p}_0$}
\begin{multline}
(s + \Upsilon_0 q^{\alpha}) \delta\rho_0 =
\rd{2 q^{\alpha} [i \Upsilon_1  + \lambda_2 p^*(\vec{e}_{\upvartheta}\cdot\vec{e}_{\parallel})](\delta\vec{p}_0 \cdot \vec{e}_{\upvartheta})} \\
\rd{+2q^{\alpha}\lambda_2 p^*(\vec{e}_{\upvartheta}\cdot\vec{e}_{\perp})(\delta \vec{p}\cdot\vec{e}_{\upvartheta+\frac{\pi}{2}}).}  
\label{linearEq1}
\end{multline}
\rd{For the second equation, instead, 
by applying the same technique we calculate the terms  
(see Appendix~\ref{auxiliary-lsa}) 
\begin{align}
\mrI^{1-\alpha}\delta\vecxi_0&=
i q^{\alpha}p^*[(\delta\vec{p} \cdot \vec{e}_{\upvartheta})\vec{e}_{\parallel}
+(\delta\vec{p} \cdot \vec{e}_{\upvartheta+\frac{\pi}{2}})\vec{e}_{\perp}], \label{lsa2.1} \\
\mrI^{3-\alpha}\delta\vecxi_1&=
-i q^{\alpha}p^*\left[(\vec{e}_{\upvartheta}\cdot\vec{e}_{\parallel})\delta\vec{p} 
+(\vec{e}_{\upvartheta}\cdot\vec{e}_{\perp})\textbf{R}\left(\frac{\pi}{2}\right)\delta \vec{p}\right] \label{lsa2.2} 
\end{align}
with $\textbf{R}$ denoting a two-dimensional rotation matrix.} 
%
\rd{Substituting these expressions into Eq.~(\ref{dp}) and again eliminating the exponential multiplicative factor yields} 
\begin{align}
s\delta\vec{p}_0&= 
\rd{\left[ \frac{\gamma}{2}\delta\rho_0 - 2\xi p^* (\delta \vec{p}_{0}\cdot\vec{e}_{\parallel}) \right] p^* \vec{e}_{\parallel}}
\rd{+\Upsilon_1 i q^{\alpha} \delta\rho_0 \vec{e}_{\upvartheta}} 
\notag\\ 
&\quad 
\rd{-\Upsilon_0 q^{\alpha} \delta\vec{p}_0 }
\rd{+\Upsilon_2 q^{\alpha} \delta\vec{p}_0}
\rd{+ 2 \lambda_1 i q^{\alpha} p^* [ (\delta\vec{p}_0\cdot\vec{e}_{\upvartheta})\vec{e}_{\parallel}} 
\notag \\ 
& \quad
\rd{+ (\delta \vec{p}_{0}\cdot\vec{e}_{\upvartheta+\frac{\pi}{2}})\vec{e}_{\perp} ]}
\rd{- \lambda_3 i q^{\alpha} p^* \left[(\vec{e}_{\upvartheta}\cdot\vec{e}_{\parallel})\delta\vec{p}_0 \right. }
\notag \\ 
& \quad
\rd{+ \left.(\vec{e}_{\upvartheta}\cdot\vec{e}_{\perp})\textbf{R}\left(\frac{\pi}{2}\right)\delta\vec{p}_0\right]}
\rd{-p^* \delta \vec{Q}^{(1)}\vec{e}_{\parallel},}
\label{linearEq2}  
\end{align}
\rd{where the components of the tensor $\delta \vec{Q}^{(1)}$ are given by (see Appendix~\ref{auxiliary-lsa}) 
\begin{align}
&
\delta \text{Q}_{j}^{(1)} 
=
-2 B_0 q^{\alpha}p^* \delta p_{0 l} 
+ i \frac{\lambda_1}{2} q^{\alpha}
\text{R}_{jl}(\upvartheta + \upvartheta_{\parallel}) \delta p_{0 l}
\notag \\
& \qquad
+ \frac{\lambda_2}{4} q^{\alpha}
\delta\rho_0 \text{R}_{jl}(\upvartheta + \upvartheta_{\parallel}) e_{\upvartheta l}
- i \frac{\lambda_3}{4}q^{\alpha}
\text{R}_{jl}(\upvartheta + \upvartheta_{\parallel})\delta p_{0 l}
\notag\\
& \qquad
- B_4 q^{\alpha}p^* (\vec{e}_{\upvartheta}\cdot\vec{e}_{\perp}) \text{R}_{jm}(\upvartheta + \upvartheta_{\parallel}) \text{R}_{ml}\left(\frac{\pi}{2}\right) \delta p_{0 l}
\notag\\
& \qquad
- B_4 q^{\alpha}p^* (\vec{e}_{\upvartheta}\cdot\vec{e}_{\parallel}) \text{R}_{jl}(\upvartheta + \upvartheta_{\parallel}) \delta p_{0 l}.
\label{lsa2.3}
\end{align} 
}


\rd{\subsubsection{Longitudinal perturbations}}

\rd{We first investigate the linear stability of the ordered phase in the case of longitudinal perturbations.
We thus set $\upvartheta=0$
(i.e., $\vec{e}_{\upvartheta}=\vec{e}_{\parallel}$ and $\vec{e}_{\upvartheta+\frac{\pi}{2}}=\vec{e}_{\perp}$)
in Eqs.~(\ref{linearEq1}, (\ref{linearEq2}) and (\ref{lsa2.3}).
In addition, because also $\delta \vec{p}_0 \propto \vec{e}_{\parallel}$, we set $\delta p_{0\perp}=0$.    
Under these assumptions, we obtain the reduced equations   
}
\begin{align}
&(s + \Upsilon_0 q^{\alpha})\delta\rho_0 = 
2q^{\alpha}(i \Upsilon_1+\lambda_2 p^*)\delta p_{0\parallel}, 
\\
&\rd{[s  + 2 \kappa_0 + (\Upsilon_0-\Upsilon_2 - \Lambda_{\parallel} p^{*2}) q^{\alpha}]\delta p_{0\parallel}=} 
\\ & \qquad
 \rd{\left[\frac{p^*}{2}\left(\gamma - \frac{\lambda_2}{2} q^{\alpha} \right)+ i \Upsilon_1 q^{\alpha} \right] \delta\rho_0 
+ i q^{\alpha} p^* \zeta_{\parallel} \delta p_{0\parallel}} 
\notag
\end{align}
\rd{with auxiliary parameters 
$\Lambda_{\parallel} \equiv B_4+2B_0\cos{\phi_{\parallel}}$ 
and 
$\zeta_{\parallel}=3(2\lambda_1-\lambda_3)/4$.}
Solving the first equation for \rd{$\delta p_{0\parallel}/\delta\rho_0$} and substituting it in the second one, we obtain the dispersion relation  
\rd{
\begin{align}
& s^2 + s \left[2 \kappa_0 + \left(2\Upsilon_0-\Upsilon_2-\rd{\Lambda_{\parallel} p^{*\,2}}\right) q^{\alpha}  
- i q^{\alpha} p^* \zeta_{\parallel} \right] 
\notag \\ 
& \,
- i q^{2\alpha} p^* \Upsilon_0 \zeta_{\parallel}
+ \Upsilon_0 q^{\alpha} \left[ 2 \kappa_0 + \left(\Upsilon_0-\Upsilon_2-\rd{\Lambda_{\parallel} p^{*\,2}}\right)  q^{\alpha} \right] 
\notag \\
& \,\, 
-2q^{\alpha}(i\Upsilon_1+\lambda_2 p^*)\left[\frac{p^*}{2}\left(\gamma - \frac{\lambda_2}{2} q^{\alpha} \right)+ i \Upsilon_1 q^{\alpha}\right]
=0 . 
\end{align}
}
The real part of its solutions can be calculated. In detail,
\begin{align}
\Re{(s_{\pm})}&=
\rd{-\frac{1}{2}\left[2 \kappa_0 + \left(2\Upsilon_0-\Upsilon_2-\rd{\Lambda_{\parallel} p^{*\,2}} \right) q^{\alpha}  
\right]} 
\notag \\  
& \quad  
\pm \frac{1}{2}\sqrt{\frac{1}{2}\left(J_1+\sqrt{J_1^2+J_2^2}\right)} 
\label{eq:LWdispersion}
\end{align}  
with auxiliary functions 
\begin{align}
J_1&= \rd{[ 2 \kappa_0 + (2\Upsilon_0-\Upsilon_2 - \Lambda_{\parallel} p^{*\,2}) q^{\alpha} ]^2 }
\rd{- 8(\Upsilon_0-2\Upsilon_2)\kappa_0 q^{\alpha}}
\label{eq:J1}
\notag \\ & \quad 
\rd {+ 4 \Upsilon_0( \Upsilon_2 - \Upsilon_0 + \rd{\Lambda_{\parallel} p^{*\,2}}) q^{2\alpha}}
\notag \\ & \quad
\rd{- (8 \Upsilon_1^2 + \zeta_{\parallel}^2 p^{*2} + 2 \lambda_2^2 p^{*2}) q^{2\alpha},} 
\\
J_2&=\rd{4 p^* q^{\alpha} \left[\gamma \Upsilon_1 - \zeta_{\parallel} \kappa_0 \right] }
\notag \\ & \quad 
\rd{+ 2 p^* q^{2\alpha} \left[ 3\Upsilon_1\lambda_2 + \zeta_{\parallel} (\Upsilon_2 + \Lambda_{\parallel} p^{*\,2}) \right].}
\label{eq:J2}
\end{align}
%
In the hydrodynamic limit $q \to 0$\rd{, we can expand these equations up to terms $\mathcal{O}(q^{2\alpha})$ 
consistently with the order of the expansion employed to derive the full set of hydrodynamic equations. 
At this order of approximation, therefore, the linear stability of the model can only be assessed for long wavelength perturbations.}
\rd{In details,} we find the asymptotic behaviour  
\begin{align}
\Re{(s_-)}&\simeq 
\rd{-2\kappa_0 - (\Upsilon_0 + \Upsilon_2 - \Lambda_{\parallel} p^{*\,2}) q^{\alpha}} \label{asymp1} \\
\Re{(s_+)}&\simeq
\rd{(-\Upsilon_0+2\Upsilon_2)q^{\alpha} }
\label{asymp2}. 
\end{align}
\rd{For comparison, we recover the stability picture of ordinary active fluids.} 
Setting $\alpha=1$ \rd{and expanding Eqs.~(\ref{eq:LWdispersion}--\ref{eq:J2}) up to terms $\mathcal{O}(q^{4})$ 
(in this case the expansion is valid because the only higher order term in the hydrodynamic equation, $\nabla^2 \vec{p}$, can be shown relevant only for short wavelength perturbations), 
we find
$\Re{(s_-)} \simeq \rd{-2\kappa_0(\rho^*)}$
and 
$\Re{(s_+)}\simeq \rd{\frac{\sigma}{16 \kappa_0^2}\left[ 4 - 7 \frac{\kappa_0}{\sigma} \right] q^{2}}$.} 
The mode $s_-$ has negative real part such that it represents the fast relaxation of small perturbations from the stationary solution $\vec{p}^*$. For the mode $s_+$, instead, 
by writing 
$4-7\kappa_0/\sigma=11-7 \sigma_t/\sigma$,
we can see that
its real part is negative for $\sigma<7\sigma_t/11$.  
In this region of the parameter space, the ordered phase is thus stable against longitudinal perturbations at the linear level. 
Conversely,
when $7\sigma_t/11\leq\sigma<\sigma_t$
the mode becomes unstable 
and the perturbation can destabilize the system. 
The eigenmode $s_+$ 
thus corresponds to the banding instability characteristic of ordinary active fluids, which renders the transition first-order \cite{Gregoire2004,Chate2008,Solon2013,Solon2015}.
 
In the L{\'e}vy case $\alpha\neq1$, the mode $s_-$ only differs by a term of higher order $\sim q^{\alpha}$, 
which is thus negligible in the hydrodynamic limit.
In contrast, the mode $s_+$ \rd{exhibits} a term of lower order $\sim q^{\alpha}$, \rd{which is thus dominant in the hydrodynamic limit}.  
Because $\Upsilon_0>2\Upsilon_2$ for all $0<\alpha<1$ (not shown), 
\rd{this term} has negative sign and thus exerts a stabilizing effect on the perturbed phase.
\rd{Therefore, 
in ALM the banding instability is suppressed, 
such that the critical behaviour of the order-disorder phase transition predicted by the mean field theory can manifest.}


This stabilization of the order phase in ALM is caused by the terms in the hydrodynamic EOM representing the enhanced diffusion of the fields $\rho$ and $\vec{p}$. 
In fact, if we switched them off by setting $\Upsilon_0=0$, 
\rd{the sign of Eq.~\eqref{asymp2} would still be positive} 
in the hydrodynamic limit. 
The banding instability would thus still disrupt the order phase at the onset of collective motion, and the phase transition would remain first-order as in ordinary active fluids.

\rd{We remark that the previous results on the stability of the order phase at the onset of collective motion are strictly valid only in the hydrodynamic limit, i.e., for long-wavelength perturbations. 
This calculation as is, thus, can not rule out the possibility that other instabilities could emerge in the order phase for intermediate wavelength perturbations.  
To perform a reliable linear stability analysis in this regime, in fact, 
the hydrodynamic equations~\eqref{rhoeq} and \eqref{meandireq} must be first corrected to account for, potentially, the entire hierarchy of higher-order expansion terms. 
This analysis, however, goes beyond the scope of the present manuscript and will not be pursued. 
Further definite validation of the previous predictions will be provided, instead, through numerical simulations of the microscopic dynamics (see Sec.~\ref{si-sec-fss}).}

\rd{
\subsubsection{Transversal perturbations}

We now study the stability of the ordered phase against perturbations orthogonal to the direction of the collective motion. 
Therefore, we set $\upvartheta=\pi/2$
(such that $\vec{e}_{\upvartheta}=\vec{e}_{\perp}$ and $\vec{e}_{\upvartheta+\frac{\pi}{2}}=-\vec{e}_{\parallel}$) 
in Eqs.~(\ref{linearEq1}) and (\ref{linearEq2}). 
Furthermore, because also $\delta \vec{p}_0 \propto \vec{e}_{\perp}$, we assume $\delta p_{0\parallel}=0$.  
With these assumptions, the contribution from the alignment tensor $\delta \vec{Q}^{(1)}$ is determined by the relations 
$\vec{e}_{\perp}\cdot[\delta \vec{Q}^{(1)}\vec{e}_{\parallel}]=\cos{(2\upvartheta_{\parallel})} \delta Q^{(1)}_{xy} - \sin{(2\upvartheta_{\parallel})} \delta Q^{(1)}_{xx}$, 
$\vec{R}(\upvartheta_{\parallel}+\pi/2)\vec{e}_{\perp}=-(\cos{2\upvartheta_{\parallel}},\sin{2\upvartheta_{\parallel}})$ and 
$\vec{R}(\upvartheta_{\parallel}+\pi/2)\vec{e}_{\parallel}=(\sin{2\upvartheta_{\parallel}},-\cos{2\upvartheta_{\parallel}})$.
Finally, we obtain the equations 
\begin{align}
(s + \Upsilon_0 q^{\alpha})\delta\rho_0&= 
2 \Upsilon_1 i q^{\alpha} \delta p_{0\perp}, 
\label{lsaT1} \\
[s + (\Upsilon_0-\Upsilon_2 + \Lambda_{\perp} p^{*2}) q^{\alpha}]\delta p_{0\perp}&=\Upsilon_1 i q^{\alpha} \delta\rho_0 
\label{lsaT2}
\end{align}
with the auxiliary parameter 
$\Lambda_{\perp} \equiv 2B_0\cos{\upvartheta_{\parallel}} - B_4$.  
Solving Eq.~\eqref{lsaT1} for $\delta p_{0\perp}/\delta \rho_0$ and substituting it into Eq.~\eqref{lsaT2} yields the dispersion relation 
\begin{multline}
s^2 + s (2\Upsilon_0-\Upsilon_2 - \Lambda_{\perp} p^{*\,2} ) q^{\alpha} \\
+ [\Upsilon_0 ( \Upsilon_0 - \Upsilon_2 - \Lambda_{\perp} p^{*\,2}) + 2 \Upsilon_1^2 ]q^{2\alpha} =0 . 
\end{multline}
This equation can be solved similarly to the previous case.  
The real parts of its solutions are thus given by 
\begin{align}
\Re{(s_{\pm})}&=
-\frac{q^{\alpha}}{2}[(2\Upsilon_0-\Upsilon_2-\Lambda_{\perp} p^{*\,2})  \pm \sqrt{\Delta} ]
\label{eq:TWdispersion}
\end{align}  
with $\Delta$ prescribed as
\begin{align}
\Delta&= [(\Upsilon_2 + \Lambda_{\perp}p^{*\,2})^2
- 8 \Upsilon_1^2] .
\end{align}
At the onset of collective motion, where $\kappa_0\approx 0$, we find that  
$\Delta\approx \Upsilon_2^2 - 8\Upsilon_1^2$, which is negative  
for all $0<\alpha<1$ (not shown).  
Consequently, we obtain 
\begin{align}
\Re{[s_{\pm}]}&\approx -\left( \frac{2 \Upsilon_0 - \Upsilon_2}{2}  \right) q^{\alpha} , 
\end{align}   
which is negative as for longitudinal perturbations.  
Therefore, in ALM, at the onset of collective motion, the ordered phase is stable against long-wavelength transversal perturbations from the stationary fields. 
This result agrees with the picture valid for ordinary active matter, 
where one obtains non authorized modes (i.e., imaginary; set $\alpha=1$ above) 
indicating that no instabilities whatsoever can arise from perturbations in this direction \cite{Bertin2006}.   
}


\begin{figure*}[!t]
\includegraphics[width=180mm,keepaspectratio]{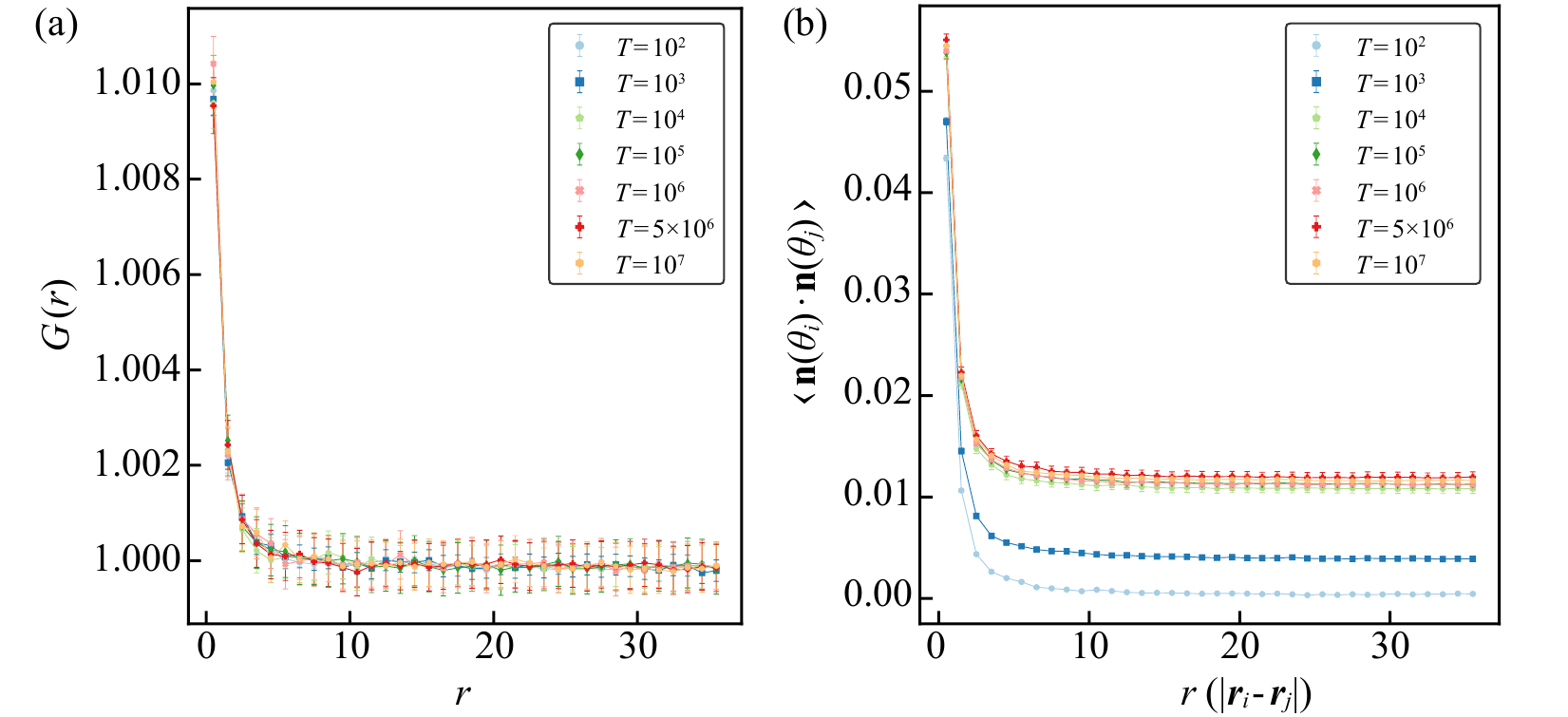}
\caption{Determination of the equilibration timescale of the microscopic model~\eqref{model} of active L{\'e}vy matter at criticality. 
Other model parameters are $\rho=2$, $\gamma=0.25$ and $\alpha=1/2$. 
The model is initialized in the homogeneously disorder phase and updated for different total time sweeps $T$. 
For each data point, we run 300 independent simulations.   
Error bars are 1 s.e.m; lines are also plotted as a guide to the eye. 
(a) Pair correlation function $G$ at criticality vs. the inter-particle distance $r$. 
(b) Correlation function of the orientation vector $\vec{n}(\theta_i)\equiv (\cos{\theta_i},\sin{\theta_i})$ at criticality vs. the distance between active particles.  
}\label{fig:equilibration}
\vspace{0.5cm}
\end{figure*}
\begin{figure*}[!htb]
\includegraphics[width=180mm,keepaspectratio]{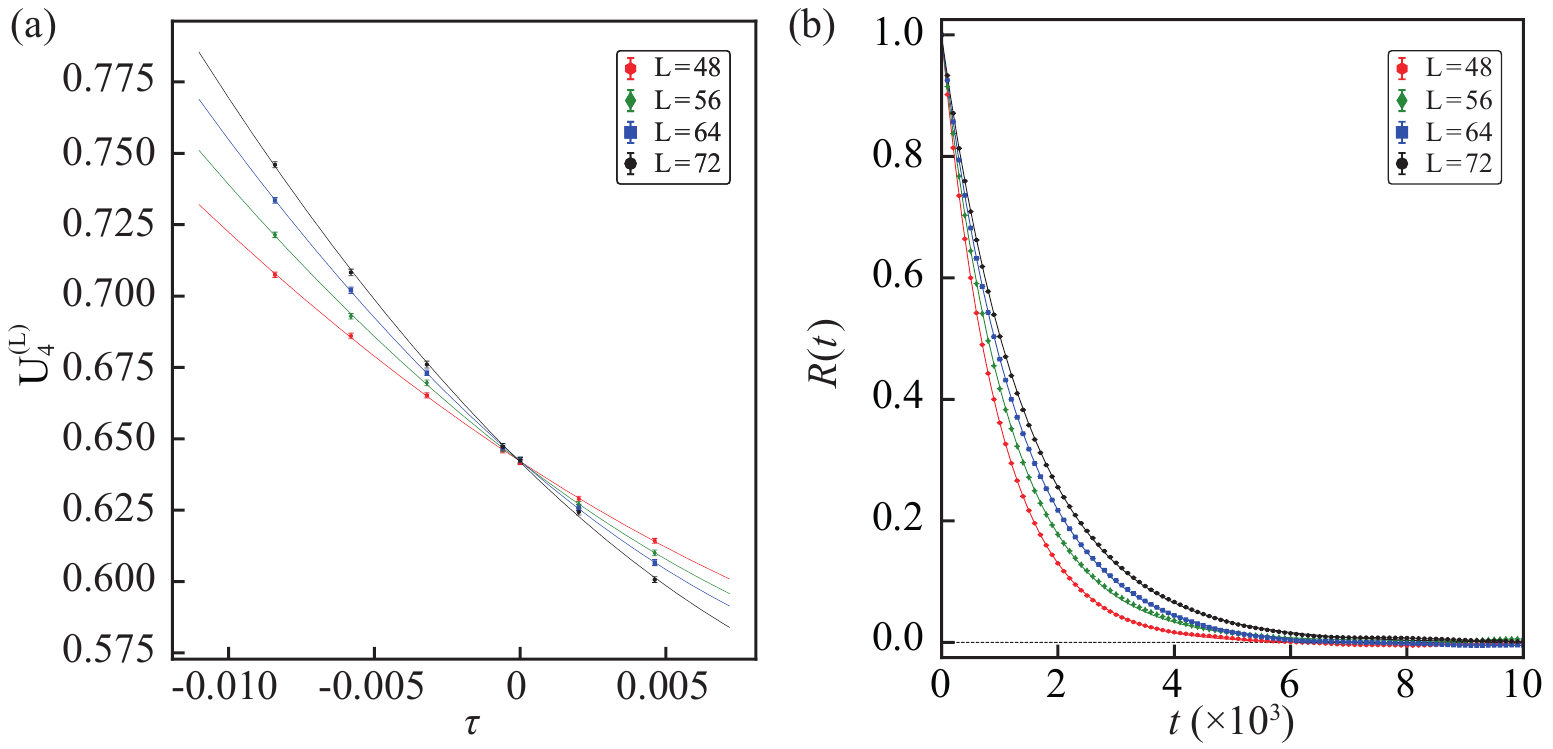}
\caption{Finite size scaling analysis of the microscopic model~\eqref{model} of active L{\'e}vy matter. 
Model parameters are $\rho=2$, $\gamma=0.25$ and $\alpha=1/2$.
For each data point, we run 300 independent simulations of the microscopic model.
(a) Estimation of the critical point through quadratic fits of Binder cumulant curves for different system sizes $L$ as a function of the distance from criticality $\tau\equiv -1+\sigma/\sigma_{\star}$. 
(b) Estimation of the correlation time of the polar order parameter at criticality through exponential fits of the relaxation function $R$ for different system sizes $L$. 
The fits are obtained by using the first $75$ data points until the curve for the largest $L$ starts oscillating.
In both panels, we plot error bars denoting 1 s.e.m. and 95\% confidence bands on the fit parameters. 
Fits are performed by applying a weighted least square method.  
}\label{fig:fits}
\end{figure*}
\begin{figure*}[!t]
\includegraphics[width=180mm,keepaspectratio]{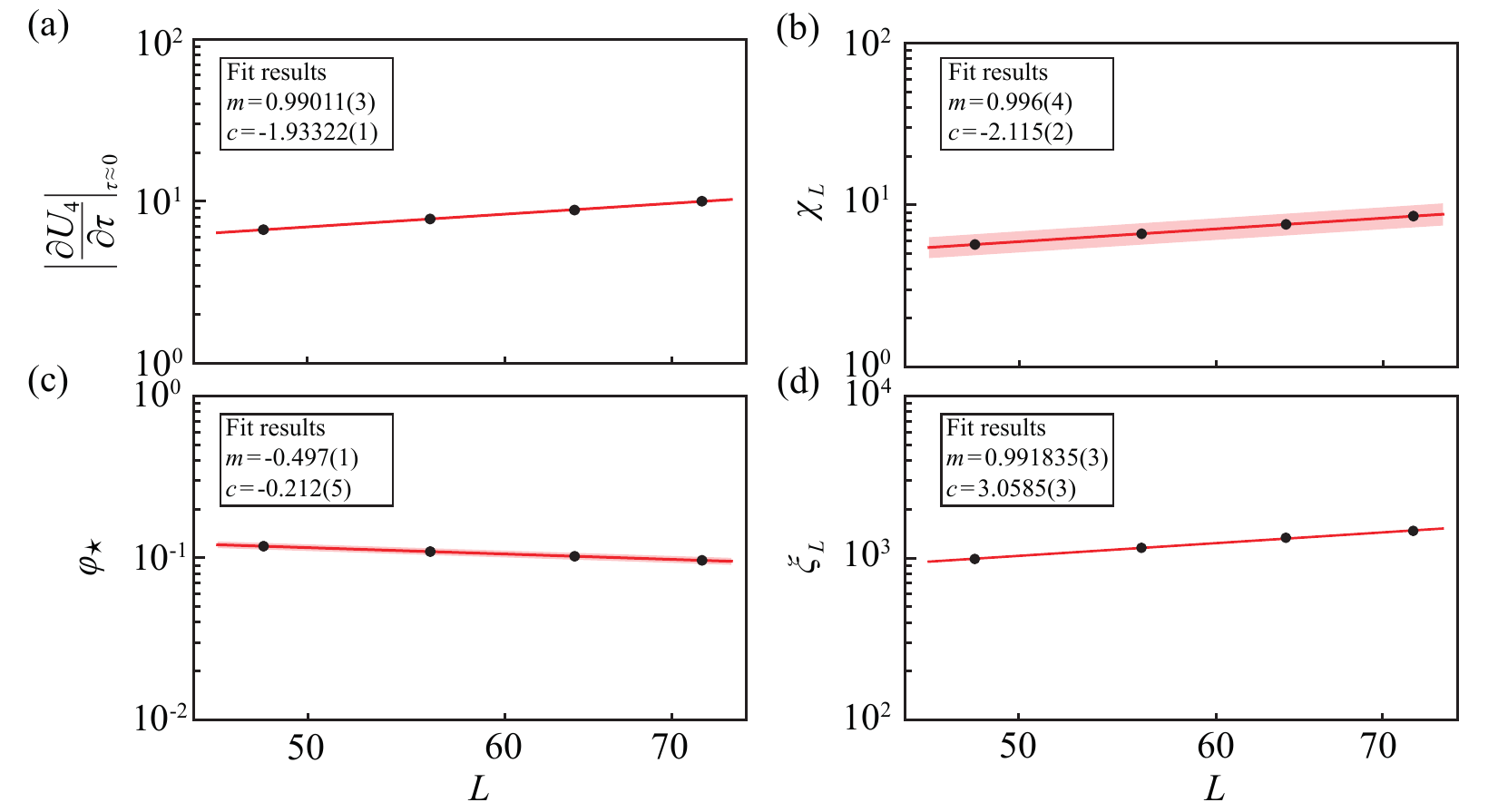}
\caption{Numerical estimation of static and dynamic critical exponents for the order-disorder phase transition in active L{\'e}vy matter. 
Model parameters are $\rho=2$, $\gamma=0.25$ and $\alpha=0.5$.
The simulation data (in logarithmic scales) are fitted with the linear function $m\,L+c$ by using a weighted least squares method. 
We plot error bars denoting 1 s.e.m. and 95\% confidence bands on fit parameters. 
(a) Gradient of the Binder cumulant at criticality $|\partial U_4^{(L)}/\partial \tau|_{\tau \simeq 0}$ vs. the system size $L$. 
(b) Susceptibility at criticality $\chi_L$ vs. $L$. 
(c) Polar order parameter at criticality $\varphi_{\star}$ vs. $L$.
(d) Correlation time of the polar order parameter $\xi_L$ at criticality vs. $L$.
}\label{fig:exps_a05}
\end{figure*}

\rd{
\section{Finite size scaling analysis and critical properties of the transition
\label{si-sec-fss}}

The potential criticality of the order-disorder phase transition in ALM, 
which is consequence of the suppression of the long-wavelength banding instability at the onset of collective motion, 
can be further ascertained by performing a finite size scaling analysis of the microscopic model~\eqref{model} \cite{Landau2014}.    
For this purpose, we perform numerical simulations of this dynamics in a square box of edge length $L$ with periodic boundary conditions (see Appendix~\ref{si-sec-numerics} for details on the numerical implementation). 
We choose model parameters: $\rho=2$, $\gamma=0.25$ and $\alpha=1/2$, which are kept fixed throughout the analysis. 
Conversely, in order to identify the critical noise strength, which we denote as $\sigma_{\star}$, we scan through a wide range of values of $\sigma$ and repeat all simulations for different system sizes $L=\{48, 56, 64, 72\}$. 
Each simulation is initialized in the homogeneously disorder phase and updated for $T$ time sweeps, 
with $T$ sufficiently large that the system can equilibrate (see below).   
At each time step $t_j=j\Delta t$, we calculate the polar order parameter
\begin{equation}
\varphi(t_j)\equiv \frac{1}{N} \left| \sum_{i=1}^N \vec{n}(\theta_i(t_j)) \right| \ , 
\end{equation} 
which measures the time-evolution of the average orientation of the active L\'evy particles. 
Furthermore, we define the time-averaged polar order parameter  
\begin{equation}
\varphi \equiv \frac{1}{T-j_0+1} \sum_{j=j_0}^T \varphi(t_j) \ , 
\label{eq:taveraging}
\end{equation}
again with $j_0$ large enough for the system to equilibrate. 
In our simulations, we choose $T=2.5 \times 10^6$ and $j_0=2.45 \times 10^6$ 
(i.e., the time average is made over the last $5 \times 10^4$ time steps). 
We can confirm, in fact, that these time sweeps are sufficient for the system to equilibrate. 
This is shown in Fig.~\ref{fig:equilibration}, where we plot the pair correlation function $G$ (a) and 
the correlation function for the unit vector $\vec{n}$ (b) as a function of the inter-particle distance $r\equiv |\vec{r}_i - \vec{r}_j|$ ($i,j=1,\ldots,N$; $i\neq j$) for several different values of $T$ at the critical point (estimated numerically; see below) and for the largest system size considered $L=72$. 
We note that, if the system is equilibrated in these conditions, it is also equilibrated for any different $\sigma$ and/or smaller $L$.       
While the pair correlation function is equilibrated for all $T$ considered
(and is thus not informative in this sense), 
the correlation function of the orientation vector suggests that the equilibration time is about $T=10^5$, 
where the different curves start overlapping. 
The value that we use in our simulations is, in fact, one order of magnitude larger than this estimate, 
which thus allows us to safely assume the system to be equilibrated when we perform the time averaging~\eqref{eq:taveraging}.

The conventional technique to assess numerically the criticality of a phase transition, and 
simultaneously locate the critical point, consists in the estimation of the Binder cumulant, 
in our case of the time-averaged polar order parameter, for different system sizes $L$. 
This observable is defined as 
\begin{equation}
U_4^{(L)}\equiv \frac{\langle \varphi^2 \rangle_{L}^2}{\langle \varphi^4 \rangle_{L}} \ , 
\end{equation}
where $\langle \cdot \rangle_L$ denotes the ensemble average taken at finite system size.
Specifically, if the phase transition is of second order (i.e., continuous), the Binder cumulant curves for different $L$ must cross at the critical point \cite{Binder1981}. 
Evidently, by measuring the intersection point of these curves, if it exists, 
we can also obtain an estimate of the critical point. 
We plot the Binder cumulant $U_4^{(L)}$ for our microscopic model of ALM in Fig.~\ref{fig:fits}a. 
Remarkably, the curves intersect, thus highlighting that the order-disorder phase transition in ALM is indeed critical, 
in agreement with the prediction of the linear stability analysis on the hydrodynamic EOM at the onset 
(see Sec.~\ref{si-sec-lsa}).    
Furthermore, by identifying the crossing point, we obtain the numerical estimate of the asymptotic critical noise strength 
\begin{equation}
\sigma_{\star}=0.239816(2) \ ,
\end{equation}
where the error denotes 1 s.e.m..

Having established the criticality of the transition, and the critical point (at fixed density), 
we can now characterize numerically the critical exponents.  
To estimate the static exponent $\nu$, we use the scaling relation 
for the gradient of Binder cumulant at the onset of the transition 
\begin{equation}
\left| \frac{\partial U_4^{(L)}}{\partial \tau} \right|_{\tau \simeq 0}\propto L^{1/\nu} \ ,
\label{scalingNu}
\end{equation} 
with the distance from the critical noise strength $\tau\equiv -1 + \sigma/\sigma_{\star}$.  
To estimate the gradient, we fit $U_4^{(L)}$ around the critical point with a quadratic polynomial. 
The coefficient of the linear term thus estimates $\partial U_4^{(L)}/\partial \tau$. 
Fits for all considered $L$ are shown in Fig.~\ref{fig:fits}a. 
The fit of the relation~\eqref{scalingNu} to these estimates of the Binder cumulant gradient is presented in Fig.~\ref{fig:exps_a05}a.  
The other static critical exponents, $\gamma$ and $\beta$, can be also estimated as ratios by the exponent just calculated, $\nu$.  
In fact, we can determine the ratio $\gamma/\nu$ from the scaling relation for the susceptibility of the time-averaged polar order parameter at criticality, $\varphi_{\star}$, which is defined (in two spatial dimensions) as 
\begin{equation}
\chi_L \equiv L^2 (\langle \varphi_{\star}^2 \rangle_L - \langle \varphi_{\star} \rangle_L^2) \ .
\end{equation} 
This observable scales with the system size $L$ as 
\begin{equation}
\chi_L \propto L^{\gamma/\nu} \ . 
\label{scalingGamma}
\end{equation}
%
\newline\newline
The fit of the relation~\eqref{scalingGamma} to numerical estimates of $\chi_L$ from simulation data is presented in Fig.~\ref{fig:exps_a05}b.
To determine the ratio $\beta/\nu$, instead, we use the scaling relation of the time-averaged polar order parameter at criticality  
\begin{equation}
\varphi_{\star} \propto L^{-\beta/\nu} . 
\label{scalingBeta}
\end{equation}
The fit of the relation~\eqref{scalingBeta} to numerical estimates of $\varphi_{\star}$ from simulation data is presented in Fig.~\ref{fig:exps_a05}c.
Finally, we can estimate the dynamic critical exponent $z$, by applying the technique illustrated in \cite{Landau2014}. 
From simulation data at criticality, we measure the relaxation function $R$, 
defined for a general time lag $t>0$ as  
\begin{equation}
R(t)\equiv \frac{\langle \varphi_{\star}(t)\varphi_{\star}(0) \rangle_L - \langle \varphi_{\star}(0) \rangle_L^2}{\langle \varphi_{\star}^2(0) \rangle_L - \langle \varphi_{\star}(0) \rangle_L^2 } \ . 
\end{equation}
By definition, $R(0)=1$ and $R(t)\to 0$ for $t\gg 1$. 
In addition, this quantity is expected to decay exponentially, i.e., 
$R(t) \propto e^{-t/\xi}$. 
By fitting our numerical data with this exponential function, we can then estimate the correlation length of the time-averaged polar order parameter at criticality, $\xi_L$, which obeys the scaling relation  
\begin{equation}
\xi_L \propto L^z \ .
\label{scalingZ}
\end{equation}
Therefore, we can use this expression to estimate directly the exponent $z$.
Relaxation functions and corresponding exponential fits for all $L$ considered are presented in Fig.~\ref{fig:fits}b.  
Finally, we present the fit of the relation~\eqref{scalingZ} to numerical estimates in Fig.~\ref{fig:exps_a05}d.
This finite size scaling analysis is also repeated for a different microscopic step size distribution, $\alpha=3/4$, (Fig.~\ref{fig:ffs_a075}). 
In this case, we obtain the estimate for the critical noise strength 
\begin{equation}
\sigma_{\star}=0.24343(6) \ .
\end{equation}
All the numerical estimates of static and dynamic critical exponents obtained are summarized in Table~\ref{tab:critical}. 

\begin{table}[!b]
\setlength{\tabcolsep}{0.75em}
\begin{tabular}{cccccc} 
$\alpha$ & $\nu$ & $\beta/\nu$ & $\gamma/\nu$ & $z$ \\ 
\hline\hline
& & & & \\ [\dimexpr-\normalbaselineskip+2pt]
1/2 
& 	$1.00999(3)$
& 	$0.497(1)$ 
& 	$0.996(4)$ 
& 	$0.991835(4)$ 
\\ 
3/4 
& 	$1.0163(1)$ 
& 	$0.496(1)$
& 	$1.011(4)$  
& 	$0.991333(2)$ 
\\ 
& & & & \\ [\dimexpr-\normalbaselineskip+2pt]
\hline
\end{tabular}
\caption{Numerical estimates of the critical exponents for two exemplary values of the parameter $\alpha$. 
Errors are expressed as 1 s.e.m.. 
}\label{tab:critical}
\end{table}

\begin{figure*}[!h]
\includegraphics[width=180mm,keepaspectratio]{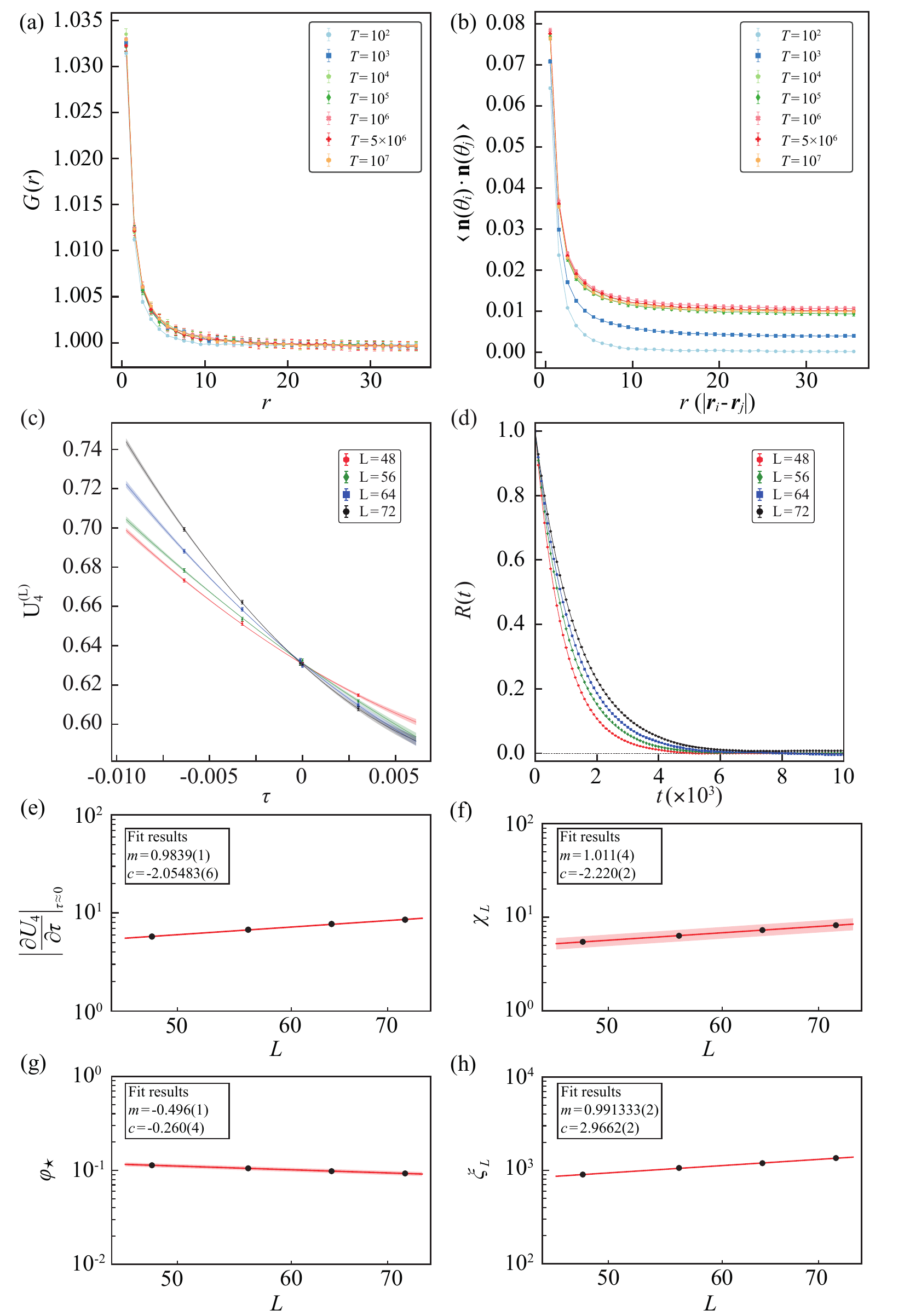}
\caption{(Caption on page 20.)}
\end{figure*}\addtocounter{figure}{-1}\clearpage
\begin{figure*}[!ht]
\caption{Summary of the finite size scaling analysis of the microscopic model~\eqref{model} of active L{\'e}vy matter for $\alpha=3/4$. Other model parameters are $\rho=2$ and $\gamma=0.25$. We refer to the main text and to the captions of Figs.~\ref{fig:equilibration}, \ref{fig:fits} and \ref{fig:exps_a05} for details on the numerical protocol and statistical analysis employed. 
(a) Pair correlation function $G$ at criticality vs. the inter-particle distance $r$.
(b) Correlation function of the orientation vector $\vec{n}(\theta_i)\equiv (\cos{\theta_i},\sin{\theta_i})$ at criticality vs. the distance between active particles.
(c) Estimation of the critical point. 
(d) Estimation of the correlation time of the polar order parameter at criticality. 
(e) Gradient of the Binder cumulant at criticality $|\partial U_4^{(L)}/\partial \tau|_{\tau \simeq 0}$ vs. the system size $L$. 
(f) Susceptibility at criticality $\chi_L$ vs. $L$. 
(g) Polar order parameter at criticality $\varphi_{\star}$ vs. $L$.
(h) Correlation time of the polar order parameter $\xi_L$ at criticality vs. $L$.
}
\label{fig:ffs_a075}
\end{figure*}

Intriguingly, the estimates obtained for the static critical exponents 
(at least for the first case, $\alpha=1/2$) 
suggest that the critical properties of ALM may be related to those of  
equilibrium two-dimensional models with long-range interaction energy $\propto 1/|\vec{r}_{ij}|^{2+2\alpha}$ and
$n$-component isotropic order parameter \cite{Fisher1972,Suzuki1973}.  
For these systems, two regimes with distinct critical properties, 
separated by the threshold $\alpha_c\equiv d/4$ ($d$ are the spatial dimensions; thus, in our specific case, $\alpha_c=1/2$),
have been demonstrated: 
For $\alpha < \alpha_c$, the so called \textit{classical} regime, the critical exponents are   
\begin{align}
\gamma&=1 \ , \ & \beta&=\frac{1}{2} \ , \ & \nu&=\frac{1}{2\alpha} \ .
\label{classical}
\end{align}
These formulas are valid also at the threshold $\alpha_c$, except for logarithmic corrections for both the correlation length and susceptibility of the order parameter.  
The case $\alpha=1/2$, therefore, belongs to this regime, and 
Eqs.~\eqref{classical} predict exponents 
$\gamma=1$, $\beta=1/2$ and $\nu=1$. 
Our numerical estimates are clearly consistent with these predictions (see Table~\ref{tab:critical}). 

For $\alpha_c < \alpha < 1$, the \textit{nonclassical} regime, 
to which our second numerical case study, $\alpha=3/4$, belongs, 
the critical exponents have been characterized as expansions in the distance from the threshold $\Delta \alpha\equiv \alpha-\alpha_c$. 
In details, setting $n=2$ for our polar order parameter, the following results hold to order $\Delta \alpha^2$ \cite{Fisher1972} 
\begin{align}
\gamma&\simeq 1 + \frac{8}{5} \Delta \alpha + \frac{668}{125} \Delta \alpha^2 ,
& \beta&=\frac{\gamma}{2}\left(\frac{1-\alpha}{\alpha}\right) , & 
\nu&=\frac{\gamma}{2\alpha} ,
\label{nonclassical}
\end{align}  
which predict 
$\gamma=1.734$, $\beta=0.289$ and $\nu=1.156$.
Our estimates are thus inconsistent with these predictions. 


}

Having established the criticality of the disorder-to-collective motion transition in our model of ALM, 
we now compare the associated critical behaviour to other critical phenomena previously discussed in this context. 
Critical phenomena have been analyzed, either analytically
or computationally, in the following active matter systems: 
(i) self-propelled particles with long-ranged metric
alignment interactions \cite{Ginelli2010,Peshkov2012}
(ii) incompressible active
fluids \cite{Chen2015,Chen2016,Chen2018}, 
(iii) critical motility-induced phase separation 
\cite{Siebert2018,Caballero2018,Partridge2019},
(iv) self-propelled particles with velocity
reversals and alignment interactions \cite{Mahault2018}, and 
(v) dense collections of active particles with contact inhibition of locomotion \cite{Nesbitt2019}.
Our results indicate that the critical behavior in ALM is different from all of the above systems because our critical exponents depend on the parameter $\alpha$ that quantifies the scaling properties of the L\'{e}vy particle dynamics, and such a parameter does not exist in the aforementioned systems. 
In addition, our observations seem to suggest that this dependence on $\alpha$ may relate the static critical properties of ALM to the universality class of long-range attractive interactions in the classical regime. In contrast, they do not support this relation in the non classical regime. 
However, our numerical considerations are not exhaustive in this matter. 
To further elucidate this connection, in fact, an analytical derivation of the critical exponents for general $\alpha$ from the hydrodynamic equations must be developed by employing dynamic renormalisation group methods.  
This is an interesting open problem that 
we aim to elucidate in future investigations.

\section{Conclusions \label{sec-conclusions}}    


In this paper we studied in details a microscopic model of ALM \cite{CairoliL}, 
where the active particles perform superdiffusive motion, 
\rd{akin to a} L{\'e}vy flight \cite{Mandelbrot1982,Hughes1981,Shlesinger1995}, 
with short-range interactions 
that promote the \rd{polar} alignment of their velocity directions \cite{Vicsek1995,Peruani2008}. 
First, we switched off the inter-particle interactions 
and characterized the diffusion process of individual active particles 
by their position mean-square displacement and statistics.  
We then switched on the alignment interactions and
applied a coarse-graining procedure, 
based on the BBGKY hierarchical method \cite{Huang1987} and on 
the expansion of the reduced single-particle probability density  
in terms of Fourier angular modes \cite{Bertin2006,Peruani2008,Bertin2009,Lee2010,Peshkov2014,Bertin2017},  
to derive the hydrodynamic EOM of the model. 
\rd{We then solved these equations at mean-field level, 
which yielded disordered and ordered phases 
similarly to ordinary active matter, 
and then studied the linear stability of these solutions.}   

On the one hand, we found that the disordered phase is stable at the linear level against small perturbations in an arbitrary direction. 
This prediction agrees with the conventional theory of active fluids 
\cite{Bertin2006,Peruani2008,Baskaran2008,Bertin2009}. 
On the other hand, we found that the ordered phase at the onset of collective motion is also stable at linear level against small perturbations in the longitudinal direction. 
This is fundamentally different from ordinary active fluids, 
where \rd{a characteristic banding} instability has been shown to emerge both analytically and numerically
\cite{Gregoire2004,Chate2008,Solon2013,Solon2015,
Bertin2006,Bertin2009,Baskaran2008,Peruani2008,Lee2010,Peshkov2014}.        
This instability, which makes the phase transition first-order in ordinary active matter, 
is thus suppressed in ALM,  
thus allowing for a critical disordered-ordered phase transition \cite{CairoliL}. 
\rd{Finally, we validated this prediction by finite size scaling analysis of simulation data and estimated numerically the critical exponents of the transition, which revealed an intriguing relation between the static properties of our ALM model and 
the equilibrium $n$-component system with long-range attractive interactions 
(in the classical regime). } 

The derivation of the hydrodynamic EOM for ALM, 
\rd{the linear stability analysis on the order and disorder phases,}
\rd{and the numerical characterization of the critical properties of the phase transition}
are our most important results, even more so because 
\rd{they elucidate the novel physics of active matter integrating anomalous diffusive motility and inter-particle interactions.}
In particular, the connection with the universality class of long-range attractive interactions will need to be further elucidated with analytical arguments; for instance, by deriving general formulas for the critical exponents as a function of the characteristic model parameter $\alpha$.  
One viable method to achieve this task can be to apply a dynamic renormalisation group approach, 
similarly to that recently introduced to study the universality class of the ordered phase and of the disorder-to-collective motion phase transition of incompressible active fluids \cite{Chen2015,Chen2016,Chen2018}. 

Another important result relates to the structure of the hydrodynamic EOM 
for ALM (see Eqs.~(\ref{rhoeq}, \ref{meandireq})), 
which we obtained by coarse-graining the microscopic dynamical model~\eqref{model}. 
Our calculation shows that these equations contain:  
(a) terms that have a counterpart in the hydrodynamic EOM of ordinary active matter 
and are integrated across the domain weighted by probability measures that can be easily interpreted;  
(b) higher order terms in the Fourier angular expansion 
that are absent from the hydrodynamic EOM of ordinary active matter,   
which not only can depend on highly non-linear fields 
but can also exhibit a non local integral structure 
specified by non trivial probability measures. 
While the former terms can be understood straightforwardly 
also by applying a first principles approach, 
the latter ones pose considerable challenge 
because in general they may require complex and non intuitive jump statistics 
to be formulated correctly.  
Our discussion can thus potentially serve as a starting point to develop a systematic first-principles derivation of the hydrodynamic equations for ALM systems.     

Overall, our results highlight that anomalous diffusion at single-particle level can fundamentally change the collective properties of active systems. 
Therefore, it will be interesting to apply considerations 
similar to those developed in this paper 
to other paradigmatic examples of collective behaviour, such as 
active turbulence 
\cite{Dombrowski2004,Hernandez-Ortiz2005,Sokolov2007,Aranson2007,Saintillan2007,Wolgemuth2008,
Sanchez2012,Wensink2012,Doostmohammadi2018}
and motility-induced phase separation \cite{Tailleur2008,Fily2012,Redner2013,Cates2015},   
and to further elucidate their relevance to the biological realm \cite{Zaburdaev2015}.

\section{Acknowledgements}

A.C. gratefully acknowledges funding under the Science Research Fellowship granted by the Royal
  Commission for the Exhibition of 1851
  and the High Throughput Computing service provided by Imperial College Research Computing Service, DOI: 10.14469/hpc/2232.

\appendix 
\numberwithin{equation}{section}

\section{Diffusional features of the dynamical model~(\ref{ni-model}) with L{\'e}vy stable distributed step sizes \label{app-sdynamics}} 

We first calculate the position statistics of non-interacting active L{\'e}vy particles projected along 
the arbitrary direction 
$\vec{n}(\psi)\equiv(\cos{\psi},\sin{\psi})$, 
which is defined as
$P(x,t)\equiv \prec  \la \delta(x-\vec{r}^{\prime}(t)\cdot\vec{n}(\psi)) \ra \succ$ 
with $\vec{r}^{\prime}(t)\equiv\vec{r}(t)-\vec{r}(0)$ and   
$\vec{r}$ prescribed by the Langevin equations~\eqref{ni-model}.  
Taking its Fourier transform, we can write   
\begin{align}
\hat P(k,t)&=\prec \langle e^{\,ik \vec{r}^{\prime}(t)\cdot\vec{n}(\psi)} \rangle \succ \notag\\
&=\prec\langle e^{\,ik\int_0^t \eta(t^{\prime})\cos{(\theta(t^{\prime})-\psi)}\diff{t^{\prime}}} \rangle\succ .
\end{align}
Exploiting the independence of $\eta$ and $\xi$, we can calculate the average over the L{\'e}vy noise first by using Eq.~\eqref{chF} with $h(t^{\prime})\equiv \mathrm{H}(t-t^{\prime})(-ik)\cos(\theta(t^{\prime})-\psi)$ 
($\mathrm{H}$ is the Heaviside function;
\rd{i.e., $\mathrm{H}(x)=1$ for $x \geq 0$; $\mathrm{H}(x)=0$ otherwise}). 
Therefore, we obtain 
$\hat P(k,t)=\langle e^{\,-k^{\alpha} D(t) } \rangle$
with the auxiliary function
\begin{align}
D(t)&\equiv \int_0^t |\cos{(\theta(t^{\prime})-\psi)}|^{\alpha} e^{\,-i\frac{\alpha\pi}{2}\text{sign}(\cos{(\theta(t^{\prime})-\psi)})} \diff{t^{\prime}}. 
\label{si-X-stats}
\end{align}
The remaining Brownian average can be calculated with the Feynman-Kac equation \cite{Majumdar2005} (not shown). 
The projected statistics is thus formally given by  
\begin{align}
P(x,t)&=\frac{1}{\pi}\Re 
\int_{0}^{\infty} e^{\, -i k x}
\langle e^{\, - |k|^{\alpha} D(t) } \rangle
\diff{k} . 
\end{align}
To get its asymptotic tail behaviour, we expand the Brownian functional and rescale the Fourier variable as
\begin{align}
&P(x,t)-\delta(x) \notag\\
&\sim |x|^{-(1+\alpha)} 
\lim_{\epsilon\to0}
\left[ - \frac{1}{\pi}\Re \int_{0}^{\infty} e^{\,-i k - \epsilon k} 
|k|^{\alpha} \langle D(t) \rangle  \diff{k} \right]   \notag\\
&\sim -|x|^{-(1+\alpha)} \Gamma(1+\alpha)
\frac{1}{\pi}\Re\left[ e^{\,-i \frac{\pi}{2}(1+\alpha)}
\langle D(t) \rangle  \right] 
\notag\\ &
\sim |x|^{-(1+\alpha)} \Gamma(1+\alpha)
\frac{\sin{(\pi\alpha)}}{\pi^2} C(t),  
\end{align}
where the infinitesimal coefficient $e^{-\epsilon k}$ ensures the convergence of the expansion \cite{Fouxon2017} 
and $C(t)$ is the positive real coefficient 
\begin{equation}
C(t)\equiv \int_0^t \int_0^{\pi/2} \int_0^{2\pi} (\cos{\phi})^{\alpha} 
\frac{e^{-\frac{(\phi-\phi^{\prime})^2}{2\sigma t^{\prime}}}}{\sqrt{2\pi\sigma t^{\prime}}}  \diff{\phi^{\prime}} \diff{\phi} \diff{t^{\prime}} . 
\end{equation} 
\rd{This is calculated by assuming uniformly distributed initial angles for the active particles and by recalling that the transition probability for the angular variables is Gaussian.}    
This argument shows that the projected position statistics exhibits power-law asymptotic tail behaviour 
$P(x,t)\sim |x|^{-(1+\alpha)}$ for $|x|\gg1$.


Similarly, we can calculate the two-dimensional position statistics  
$P(\vec{x},t)\equiv \prec \la \delta(\vec{x}-\vec{r}^{\prime}(t))\ra \succ$. 
Its Fourier transform is 
\begin{align}
\hat P(\vec{k},t)&=\prec\langle e^{\,i \vec{k}\cdot\vec{r}^{\prime}(t)} \rangle\succ \notag\\
&=\prec\langle e^{\,i\int_0^t \eta(t^{\prime})[\vec{k}\cdot\vec{n}(\theta(t^{\prime}))]\diff{t^{\prime}}} \rangle\succ .
\end{align}
Setting 
$h(t^{\prime})\equiv \mathrm{H}(t-t^{\prime})[-i\, \vec{k}\cdot\vec{n}(\theta(t^{\prime}))]$
in Eq.~\eqref{chF} we obtain 
\begin{align}
\hat P(\vec{k},t)&=\langle e^{\,-\int_0^t [-i\,\vec{k}\cdot\vec{n}(\theta(t^{\prime}))]^{\alpha}\diff{t^{\prime}}} \rangle
\ .
\end{align}
Taking its inverse Fourier transform, 
and setting $\vec{k}\equiv k \vec{n}(\psi)$, 
we obtain the formal equation for the distribution  
\begin{align}
P(\vec{x},t)&=\frac{1}{2\pi}\int_{\mathbb{R}^2} \diff^{2} {\vec{k}} \, 
e^{\,-i\vec{k}\cdot\vec{x}} \langle e^{\,-k^{\alpha} D(t)} \rangle . 
\end{align}
Using this result, we can calculate the fractional moments 
$\prec\langle |\vec{x}|^{\delta} \rangle\succ  \equiv \int_{\mathbb{R}^2} |\vec{x}|^{\delta} P(\vec{x},t)\diff^{2}{\vec{x}}$ for $0<\delta<\alpha$. 
In details, using polar coordinates for both integrals, 
and performing the change of variables $k^{\prime}\equiv k [D(t)]^{1/\alpha}$, $x^{\prime}\equiv x / [D(t)]^{1/\alpha}$ we obtain 
\begin{multline}
\prec\langle |\vec{x}|^{\delta} \rangle \succ=
\frac{1}{2\pi}\int_0^{\infty} \diff{x^{\prime}} x^{\prime\,1+\delta} \int_0^{2\pi} \diff{{\phi}}
\int_0^{\infty} k^{\prime} \diff{k^{\prime}} 
\times \\ \times 
\int_0^{2\pi} \diff{{\phi^{\prime}}}
e^{\,-i k^{\prime} x^{\prime} \cos{({\phi}-{\phi^{\prime}})} -k^{\prime\,\alpha}}
\left\langle [D(t)]^{\delta/\alpha} \right\rangle
.   
\end{multline}
To solve this equation analytically, one needs to calculate the full statistics of the corresponding Brownian functional $W(y,t)\equiv\langle \delta(y-D(t)) \rangle$, which however we do not pursue here. Nevertheless, recalling that Brownian functionals scale linearly for long times \cite{Godreche2001}, i.e., in Laplace space $\mathcal{L}\{W(y,t)\}(s)=g(y s)$ with $g$ a suitable scaling function, we can write 
$\mathcal{L}\{\langle [D(t)]^{\delta/\alpha} \rangle\}(s)=s^{-1-\delta/\alpha}\int_{-\infty}^{\infty} y^{\delta/\alpha} g(y) \diff{y}$.  
The inverse Laplace transform of this equation
yields the scaling behaviour 
$\left\langle [D(t)]^{\delta/\alpha} \right\rangle\sim t^{\delta/\alpha}$.  
Upon rescaling the fractional moment, we obtain that the pseudo mean-square displacement 
$\prec\langle |\vec{x}|^{\delta} \rangle\succ^{2/\delta}$
scales as $t^{2/\alpha}$.

\rd{
\section{Details of the simulation protocol \label{si-sec-numerics}}

\begin{figure}[!b]
\includegraphics[width=86mm,keepaspectratio]{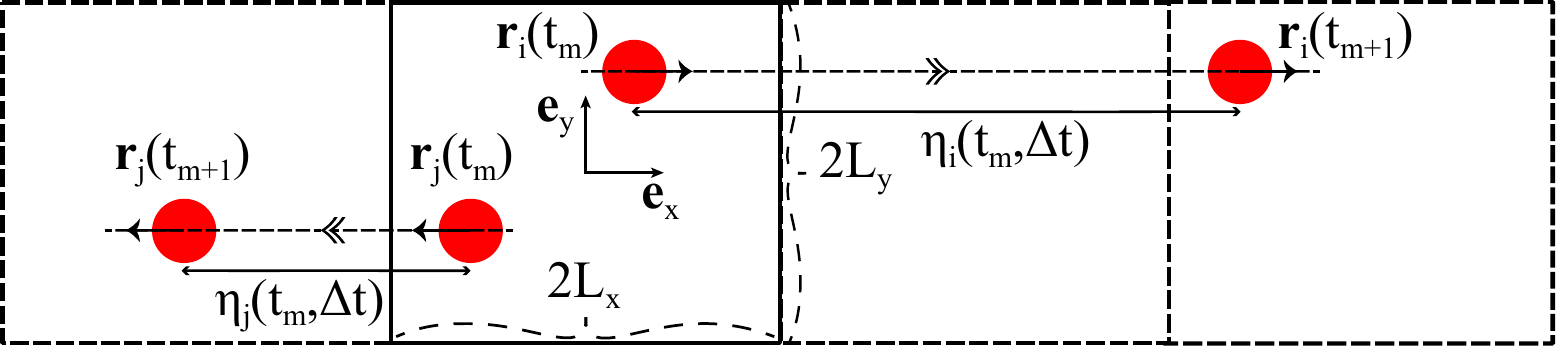}
\caption{\rd{
Schematic of jump event for two active L\'evy particles illustrating the rules to impose periodic boundary conditions.
}}\label{fig:pboundary}
\end{figure}

\subsection{Model numerical implementation}

To simulate the dynamics of active L\'evy particles 
from the initial time $t_0=0$ to the final time $t_T=M$, 
we consider a time discretization $\{t_m\}$ with time step $\Delta t$ 
(thus $m=0,\ldots,T$ with $T\equiv M/\Delta t$).  
The Langevin equations~\eqref{model} are discretized as 
\begin{align}
\vec{r}_i(t_{m+1})&=\vec{r}_i(t_m) + \eta_i(t_m,\Delta t) \vec{n}(\theta_i(t_m)), 
\label{dmodel1} \\ 
\theta_i(t_{m+1})&=\theta_i(t_m) + F_{i}(t_m)\Delta t + \xi_i(t_m,\Delta t) , 
\label{dmodel2}
\end{align}
where $\eta_i(t_m,\Delta t)$ is a random variable sampled from a one-sided positive L\'evy distribution with characteristic L\'evy symbol generically specified by Eq.~\eqref{LevySymbol}  
(see below the algorithm for generating in particular stable variates with stability parameter $0<\alpha<1$),  
$\xi_i(t_m,\Delta t)$ is a Gaussian random variable with null mean and variance $2\sigma\Delta t$, and  
$F_i(t_m)$ is the short-range alignment interaction exerted on particle $i$-th (see Eq.~\eqref{force-model}).
Note the choice of It\^o prescription for the multiplicative noise term in the rhs of Eq.~\eqref{dmodel1}. 

At each time step all particle trajectories are updated; 
in equivalent terms, for each simulation we then perform $T$ particle sweeps   
(in the simulations presented $\Delta t=0.1$). 
Therefore, the predictions on the macroscopic behaviour of the model do not depend on $\Delta t$, as long as we account for a total number of particle sweeps sufficient to ensure that the system is equilibrated.
 
Let us consider, in full generality, that numerical simulations of the Eqs.~(\ref{dmodel1}, \ref{dmodel2}) are performed in an asymmetric box of size $2 L_x \times 2L_y$ 
with periodic boundary conditions. 
Imposing these conditions for the dynamics described here requires more careful consideration with respect to the case of self-propelled particles 
(where one only needs to implement the rules presented, among other textbooks, in \cite{Rapaport2004}). 
In fact, because the L\'evy statistics allows for potentially large particle displacements between successive time steps, particles can jump from one time step to the next one in one of several consecutive adjacent boxes outside the simulation region (i.e., not only the first one) (see schematic in Fig.~\ref{fig:pboundary}).    
Therefore, the rules to impose periodic boundary conditions in our microscopic ALM model are adapted as follows. 
Consider, e.g., the $x$-component of the $i$-th particle position vector $r_{i x}$  
that satisfies the condition $-L_x \leq r_{i x} \leq L_x$; then, 
\begin{itemize}
\item{if $r_{i x}\geq L_x$, replace it by $r_{i x} - (1+n)2L_x$ 
with the integer $n=\lfloor (r_{i x}-L_x)/2L_x \rfloor$;} 
\item{otherwise, if $r_{i x}<-L_x$, replace it by $r_{i x} + (1+n)2L_x$ with the integer $n=\lfloor |r_{i x}+L_x|/2L_x \rfloor$. } 
\end{itemize}
The same rules with $L_x$ replaced by $L_y$ hold for the $y$-component of the position vector $r_{i y}$, which is also assumed to satisfy the condition $-L_y \leq r_{i y} \leq L_y$. 
Similarly, when evaluating the distance between particles $i$-th and $j$-th, 
these rules are applied to each of the components of the distance vector 
$\vec{r}_{ij}=\vec{r}_{i}-\vec{r}_{j}$ 
(clearly in this case $n=0$ \cite{Rapaport2004}).   
Finally, to efficiently evaluate the inter-particle force $F_{i}$ in the rhs of Eq.~\eqref{dmodel2} we implement the cell subdivision method \cite{Rapaport2004}. 


\subsection{Generation of one-sided positive L\'evy stable random variables}

Random variates $\eta(t_i,\Delta t)$ following a one-sided positive L{\'e}vy stable distribution with stability parameter $0<\alpha<1$ are obtained by implementing the algorithm below~\cite{Janicki1993,Kleinhans2007}:  
\begin{enumerate}[label=1.\Alph*]
\item{We generate a random variable $V_i$ uniformly distributed on $(-\pi/2 , \pi/2)$. This is realized by sampling a uniformly distributed variate $u_i^{(1)}$ on $(0,1)$ (here we use the ran2 algorithm presented in~\cite{Press2007}) and then setting $V_i=\pi (u_i^{(1)} -1/2 )$.}
\item{We generate a second independent random variable $W_i$ exponentially distributed with mean $1$. This is obtained similarly by sampling another uniformly distributed variate $u_i^{(2)}$ on $(0,1)$, independent on $u_i^{(1)}$, and setting $W_i=-\log{(u_i^{(2)})}$.}
\item{The L{\'e}vy distributed variate is given by     
\begin{align}
\eta(t_i,\Delta t)&=(\Delta t)^{1/\alpha}\frac{\sin{\left[\alpha\left(V_i+\frac{\pi}{2}\right)\right]}}{\left[\cos{(V_i)}\right]^{1/\alpha}}\times \notag\\
& \quad \times 
\left\{\frac{\cos{\left[V_i-\alpha\left(V_i+\frac{\pi}{2}\right)\right]}}{W_i}\right\}^{(1-\alpha)/\alpha}.
\end{align}
}
\end{enumerate} 
This algorithm also applies to the special case $\alpha=1$, where it yields 
$\eta(t_i,\Delta t)=\Delta t$ as expected (see Sec.~\ref{sec-single-particle}).

To avoid sampling random variables too large for the double-precision floating point format of typical calculators, we introduce a tempering $H \gg L$ 
by employing the rejection method~\cite{Baeumer2010}
(in the simulations presented we set $H=10^{10}$):  
\begin{enumerate}[label=2.\Alph*]
\item{We generate the L{\'e}vy distributed variate $\eta(t_i,\Delta t)$ as detailed above; then}
\item{we sample a second random variable $Y_i$ from an exponential distribution with mean $1/H$ (again as detailed above);} 
\item{finally, if $\eta(t_i,\Delta t)\leq Y_i$, we keep it; 
otherwise, we reject both variables and repeat steps 2.A and 2.B.} 
\end{enumerate}

We refer to the original references~\cite{Janicki1993,Kleinhans2007,Baeumer2010} for the numerical verification of these algorithms. An algorithm similar to 1.A -- 1.C yielding fully general L\'evy stable distributed random variables can be found in \cite{Weron1996}. 

}

\section{Large-scale limit of the integral term~(\ref{MoperatorF}) \label{sec-lscale}}

To simplify the notation, we consider a general test function $h$. 
Since the probability density $P$ is also a test function, the argument presented below also applies to it.  
We write the integral term  as   
\begin{align}
&\frac{1}{\pi d^2}\int_{\mathbb{R}^2} \diff{\vec{r}} \int_{\mathbb{R}^2} \diff{\vec{r}^\prime} \mathrm{H}(d-|\vec{r}-\vec{r}^{\prime}|)h(\vec{r}^{\prime}) \notag\\
&\,\,\,\, =\frac{1}{2\pi d^2}\int_{\mathbb{R}^2} \diff{\vec{r}_1} \int_{\mathbb{R}^2} \diff{\vec{r}_2}  \mathrm{H}(d-|\vec{r}_2|)h\!\left(\frac{\vec{r}_1-\vec{r}_2}{2}\right) \notag\\
&\,\,\,\, =\frac{1}{2\pi d^2}\int_{\mathbb{R}^2}\diff{\vec{r}_1} \int_0^{d}\diff{r_2} r_2  \int_{-\pi}^{\pi}\diff{\phi} h\!\left(\frac{\vec{r}_1-r_2\vec{e}_r}{2}\right)    
\end{align}
where first we use the change of variables $(\vec{r}_1,\vec{r}_2)\equiv (\vec{r}+\vec{r}^{\prime},\vec{r}-\vec{r}^{\prime})$ with Jacobian $1/2$ and second we express the integral over $\vec{r}_2$ in polar coordinates $(r_2,\phi)$ with 
$\vec{e}_r\equiv (\cos{\phi},\sin{\phi})$.   
Again changing variables as $r_2\equiv d r_2^{\prime} $, we obtain 
\begin{align}
&\frac{1}{\pi d^2}\int_{\mathbb{R}^2} \diff{\vec{r}} \int_{\mathbb{R}^2} \diff{\vec{r}^\prime} \mathrm{H}(d-|\vec{r}-\vec{r}^{\prime}|)h(\vec{r}^{\prime}) \notag\\
&\,\,\,\, =\frac{1}{2\pi}\int_{\mathbb{R}^2}\diff{\vec{r}_1} \int_0^{1}\diff{r_2^{\prime}} r_2^{\prime}  \int_{-\pi}^{\pi}\diff{\phi} h\!\left(\frac{\vec{r}_1-dr_2^{\prime}\vec{e}_r}{2}\right) .    
\end{align}
It is now straightforward to take the limit $d \to 0$. 
In this limit, $h$ can be taken outside the polar integral, and this latter one can be solved exactly. Therefore, we demonstrate the following relation 
\begin{align}
&\lim_{d \to 0} \frac{1}{\pi d^2}\int_{\mathbb{R}^2} \diff{\vec{r}} \int_{\mathbb{R}^2} \diff{\vec{r}^\prime} \mathrm{H}(d-|\vec{r}-\vec{r}^{\prime}|)h(\vec{r}^{\prime}) \notag\\
&\,\,\,\, =\frac{1}{2}\int_{\mathbb{R}^2}\diff{\vec{r}_1} h\!\left(\frac{\vec{r}_1}{2}\right)      
\equiv  
\int_{\mathbb{R}^2}\diff{\vec{r}} \int_{\mathbb{R}^2}\diff{\vec{r}^{\prime}} \delta(\vec{r}-\vec{r}^{\prime}) h(\vec{r}^{\prime}) .  
\end{align}
As this relation holds for any test function $h$, the outer integral can be dropped; 
therefore, this relation is equivalent to 
\begin{equation}
\lim_{d \to 0} \frac{1}{\pi d^2} \int_{\mathbb{R}^2} \diff{\vec{r}^\prime} \mathrm{H}(d-|\vec{r}-\vec{r}^{\prime}|)h(\vec{r}^{\prime})=
\int_{\mathbb{R}^2}\diff{\vec{r}^{\prime}} \delta(\vec{r}-\vec{r}^{\prime}) h(\vec{r}^{\prime}) .  
\label{LSrelation}     
\end{equation}
Applying this relation to Eq.~\eqref{MoperatorF} yields its large-scale limit~\eqref{Moperator}.  


\rd{
\section{Exact formulas for the coefficients $\Upsilon$ for L{\'e}vy stable distributed step sizes \label{sec:exact-Upsilon}}

We consider Eq.~\eqref{UpsilonC} with $\nu$ specified by Eq.~\eqref{stable-measure}, i.e.,
\begin{equation}
\Upsilon_{m}\equiv \frac{(-i)^{m}}{2\pi}
\int_{-\pi}^{\pi} e^{\,i \phi^{\prime} m}
(-i \cos{\phi^{\prime}})^{\alpha} \diff{\phi^{\prime}}. 
\end{equation}
The remaining angular integral can be solved analytically. 
For $m=0$ we find
\begin{equation}
\int_{-\pi}^{\pi} (-i \cos{\phi^{\prime}})^{\alpha}\diff{\phi^{\prime}}=2\sqrt{\pi}\cos{\left(\frac{\pi\alpha}{2}\right)}\frac{\Gamma\!\left(\frac{1+\alpha}{2}\right)}{\Gamma\!\left(1+\frac{\alpha}{2}\right)} ,
\label{Ups0}
\end{equation}
such that  
\begin{align}
\Upsilon_{0}(\alpha)&=
\frac{\cos{\left(\frac{\pi\alpha}{2}\right)}}{\sqrt{\pi}}\frac{\Gamma\!\left(\frac{1+\alpha}{2}\right)}{\Gamma\!\left(1+\frac{\alpha}{2}\right)}.  
\label{upsilon_0} 
\end{align}
For $m\neq0$ we can write  
\begin{multline}
\int_{-\pi}^{\pi} e^{\,i \phi^{\prime} m}(-i \cos{\phi^{\prime}})^{\alpha}\diff{\phi^{\prime}}= \\ 
2(-i)^{\alpha}\int_{0}^{\pi/2} \cos{(\phi^{\prime} m)}(\cos{\phi^{\prime}})^{\alpha}\diff{\phi^{\prime}} \\ 
+2 i^{\alpha} \int_{\pi/2}^{\pi} \cos{(\phi^{\prime} m)}|\cos{\phi^{\prime}}|^{\alpha} \diff{\phi^{\prime}}.  
\label{intalpha}
\end{multline}
The first integral in its rhs is equal to     
\begin{align}
&\int_{0}^{\pi/2} \cos{(\phi^{\prime} m)} (\cos{\phi^{\prime}})^{\alpha}\diff{\phi^{\prime}}= 
\notag \\
&\qquad \frac{i 2^{-1-\alpha}}{m-\alpha}
{}_2 F_1\!\left(\frac{m-\alpha}{2},-\alpha;1+\frac{m-\alpha}{2};-1 \right) \notag \\
& \qquad - \frac{i 2^{-1-\alpha}}{m+\alpha} 
{}_2 F_1\!\left(-\frac{m+\alpha}{2},-\alpha;1-\frac{m+\alpha}{2};-1 \right) \notag \\ 
& \qquad + \frac{i^{m}}{2}\left(-\frac{i}{2}\right)^{1+\alpha} 
\frac{\Gamma(1+\alpha)\Gamma\!\left(\frac{m-\alpha}{2}\right)}{\Gamma\!\left(1+\frac{m+\alpha}{2}\right)}
\notag \\
& \qquad + \frac{(-i)^{m}}{2}\left(-\frac{i}{2}\right)^{1+\alpha}
\frac{\Gamma(1+\alpha)\Gamma\!\left(-\frac{m+\alpha}{2}\right)}{\Gamma\!\left(1-\frac{m-\alpha}{2}\right)}; 
\end{align}
likewise, the second one is given by 
\begin{align}
& \int_{\pi/2}^{\pi} \cos{(\phi^{\prime} m)} |\cos{\phi^{\prime}}|^{\alpha}\diff{\phi^{\prime}}= 
\notag \\
&\qquad \frac{i 2^{-1-\alpha} (-1)^{1+m}}{m-\alpha}
{}_2 F_1\!\left(\frac{m-\alpha}{2},-\alpha;1+\frac{m-\alpha}{2};-1 \right) \notag \\
&\qquad + \frac{i 2^{-1-\alpha} (-1)^m}{m+\alpha}
{}_2 F_1\!\left(-\frac{m+\alpha}{2},-\alpha;1-\frac{m+\alpha}{2};-1 \right) \notag \\
&\qquad + \frac{i^{m}}{2}\left(\frac{i}{2}\right)^{1+\alpha} 
\frac{\Gamma(1+\alpha)\Gamma\!\left(\frac{m-\alpha}{2}\right)}{\Gamma\!\left(1+\frac{m+\alpha}{2}\right)} \notag \\
&\qquad + \frac{(-i)^{m}}{2}\left(\frac{i}{2}\right)^{1+\alpha}
\frac{\Gamma(1+\alpha)\Gamma\!\left(-\frac{m+\alpha}{2}\right)}{\Gamma\!\left(1-\frac{m-\alpha}{2}\right)} 
\end{align}
with ${}_2F_1$ the Gaussian hypergeometric function.
Combining these results into Eq.~\eqref{intalpha}, we finally obtain: 
\begin{multline}
\int_{-\pi}^{\pi} e^{\,i \phi^{\prime} m}(-i \cos{\phi^{\prime}})^{\alpha}\diff{\phi^{\prime}}= \frac{i^{\alpha+1}}{2^{\alpha}} \left\{ \frac{[(-1)^{\alpha}-(-1)^m]}{m-\alpha}\times \right. \\
{}_2 F_1\!\left(\frac{m-\alpha}{2},-\alpha;1+\frac{m-\alpha}{2};-1 \right) 
+ \frac{[(-1)^m - (-1)^{\alpha}]}{m+\alpha} 
\\ \times
\left. {}_2 F_1\!\left(-\frac{m+\alpha}{2},-\alpha;1-\frac{m+\alpha}{2};-1 \right)
\right\}.
\label{rep1}
\end{multline}
Using the relations 
\begin{subequations}
\begin{align}
i^{\alpha}[(-1)^m-(-1)^{\alpha}]&=
2 i^{1+m}\sin{\left(\frac{\pi}{2}(m+\alpha)\right)}, \\
i^{\alpha}[(-1)^m - (-1)^{\alpha}]&=2(-i)^{1+m}\sin{\left(\frac{\pi}{2}(m-\alpha)\right)}, 
\end{align}
\end{subequations}
we finally obtain  
\begin{align}
&\Upsilon_{m}(\alpha)\equiv
\notag \\ & \qquad 
\frac{\sin{\left(\frac{\pi}{2}(m+\alpha)\right)}}{\pi 2^{\alpha}(m-\alpha)} 
{}_2 F_1\!\left(\frac{m-\alpha}{2},-\alpha;1+\frac{m-\alpha}{2};-1 \right) 
\notag\\ & \qquad
+(-1)^{m} \frac{\sin{\left(\frac{\pi}{2}(m-\alpha)\right)}}{\pi 2^{\alpha}(m+\alpha)} 
\times
\notag\\ &\qquad\qquad \times 
{}_2 F_1\!\left(-\frac{m+\alpha}{2},-\alpha;1-\frac{m+\alpha}{2};-1 \right).  
\label{upsilon_n}
\end{align}
}


\section{Derivation of the second angular mode~(\ref{eq:f2sol}) \label{derivationf2}}

Employing the approximation \rd{$\partial_t \hat f_2 \approx 0$}, Eq.~\eqref{f2eq} can be cast into the following equation 
\begin{align}
(\Upsilon_0 k^{\alpha}+4\sigma)\hat{f}_2 + \Upsilon_{4} k^{\alpha}e^{i4\psi}\hat{f}^*_{2} 
&=\hat F_1 + i \hat F_2 , 
\end{align} 
with the auxiliary complex function 
\begin{align}
\hat F_1 + i \hat F_2&=\gamma(\hat{f}_{1}\star\hat{f}_{1})-\Upsilon_{1} i k^{\alpha}e^{i \psi}\hat{f}_1 \notag\\ 
& \quad +\Upsilon_{2} k^{\alpha}e^{i 2\psi}\hat{f}_{0}  
+\Upsilon_{3} i k^{\alpha}e^{i3\psi}\hat{f}^*_{1}, 
\label{funcF}    
\end{align}
which can be solved by writing out real and imaginary parts of all the modes. 
Using the ansatz 
$\hat{f}_n=\hat{a}_n + i \hat{b}_n$, 
for $n>0$ (the mode $n=0$ is in fact real) 
yields the coupled linear equations
\begin{align}
[\Upsilon_0 k^{\alpha}+4\sigma + \Upsilon_{4} k^{\alpha}\cos{(4\psi)}] \hat{a}_2 + \Upsilon_{4} k^{\alpha}\sin{(4\psi)}\hat{b}_2 &= \hat F_1 , \label{2modEq1} \\
[\Upsilon_0 k^{\alpha}+4\sigma - \Upsilon_{4} k^{\alpha}\cos{(4\psi)}] \hat{b}_2 + \Upsilon_{4} k^{\alpha}\sin{(4\psi)}\hat{a}_2 &= \hat F_2 .
\label{2modEq2} 
\end{align}
To obtain the solution of the set of equations (\ref{2modEq1}, \ref{2modEq2}) 
we write them in vector form 
\begin{equation}
\vec{M}
\left(
\begin{matrix}
\hat{a}_2 \\ \hat{b}_2
\end{matrix}
\right)=
\left(
\begin{matrix}
\hat F_1 \\ \hat F_2
\end{matrix}
\right), 
\quad  
\vec{M}=
(4\sigma+\Upsilon_0 k^{\alpha})\vec{1}
+ \Upsilon_{4} k^{\alpha} \vec{R}(4\psi)
\label{veceq}
\end{equation}
\rd{with $\vec{1}$ the identity matrix in two dimensions 
and $\vec{R}$ the involutory matrix (i.e., $\vec{R}^2=\vec{1}$) in two dimensions} 
\begin{equation}
\rd{
\vec{R}(\phi)\equiv 
\left(
\begin{matrix}
\cos\phi & \sin\phi \\
\sin\phi & -\cos\phi 
\end{matrix}
\right)
.
}
\end{equation}  
In fact, the solution of the linear system \eqref{veceq} is readily given by    
\begin{equation}
\left(
\begin{matrix}
\hat{a}_2 \\ \hat{b}_2
\end{matrix}
\right)=
\vec{M}^{-1}
\left(
\begin{matrix}
\hat F_1 \\ \hat F_2
\end{matrix}
\right), 
\, 
\vec{M}^{-1}=
\left[
\hat D_0 \vec{1}
-\hat D_4 \vec{R}(4\psi)
\right]
\label{vecsol}
\end{equation}
with the auxiliary \rd{$(k,\alpha)$-dependent} coefficients   
\begin{align}
\hat D_0 &\equiv \frac{4\sigma+\Upsilon_0 k^{\alpha}}{(4\sigma+\Upsilon_0 k^{\alpha})^2-(\Upsilon_{4} k^{\alpha})^2}, \label{eq:D0} \\  
\hat D_4 &\equiv \frac{\Upsilon_4 k^{\alpha}}{(4\sigma+\Upsilon_0 k^{\alpha})^2-(\Upsilon_{4} k^{\alpha})^2} . \label{eq:D4} 
\end{align}
Thus, we obtain the solution 
\begin{equation}
\left(
\begin{matrix}
\hat{a}_2 \\ \hat{b}_2
\end{matrix}
\right)=
\hat D_0
\left(
\begin{matrix}
\hat F_1 \\ \hat F_2
\end{matrix}
\right)
- \hat D_4
\left(
\begin{matrix}
\cos{(4\psi)} \hat F_1 + \sin{(4\psi)} \hat F_2 \\ \sin{(4\psi)} \hat F_1 - \cos{(4\psi)} \hat F_2
\end{matrix}
\right),
\end{equation}
or equivalently 
in compact notation 
\begin{equation}
\hat{f}_2=\hat D_0 (\hat F_1 + i \hat F_2)- \hat D_4 e^{i4\psi}(\hat F_1 + i \hat F_2)^* . 
\label{tmpf2sol}
\end{equation} 
Finally, using Eq.~\eqref{funcF} the solution~\eqref{tmpf2sol} can be written explicitly in terms of the lower order angular modes, i.e.,   
\begin{align}
\hat{f}_2&=\gamma \hat D_0 (\hat{f}_{1}\star\hat{f}_{1})-\gamma \hat D_4 e^{i 4 \psi}(\hat{f}_{1}\star\hat{f}_{1})^* + \hat D_{2} k^{\alpha}e^{i 2\psi}\hat{f}_{0} \notag \\
&\quad - \hat D_{1} i k^{\alpha}e^{i \psi}\hat{f}_1 + \hat D_{3} i k^{\alpha}e^{i3\psi}\hat{f}^*_{1} 
\end{align}
with the additional \rd{$(k,\alpha)$-dependent} coefficients 
\begin{align}
\hat D_1 
&\equiv \frac{4\Upsilon_1\sigma + (\Upsilon_0\Upsilon_1-\Upsilon_4\Upsilon_3) k^{\alpha}}{(4\sigma+\Upsilon_0 k^{\alpha})^2-(\Upsilon_{4} k^{\alpha})^2}, \label{eq:D1} \\  
\hat D_2 
&\equiv \frac{\Upsilon_2}{4\sigma+(\Upsilon_0 + \Upsilon_{4}) k^{\alpha}} , \label{eq:D2} \\  
\hat D_3 
&\equiv \frac{4\Upsilon_3\sigma + (\Upsilon_0\Upsilon_3-\Upsilon_4\Upsilon_1) k^{\alpha}}{(4\sigma+\Upsilon_0 k^{\alpha})^2-(\Upsilon_{4} k^{\alpha})^2} . \label{eq:D3} 
\end{align}
\rd{In the hydrodynamic limit ($k \to 0$), these coefficients satisfy the following asymptotic formulas:  
\begin{align}
\hat D_0&= \frac{1}{4\sigma}-\frac{\Upsilon_0}{(4\sigma)^2}k^{\alpha} + \frac{\Upsilon_0^2 + \Upsilon_4^2}{(4\sigma)^3}k^{2\alpha} + \mathcal{O}(k^{3\alpha}),  \label{eq:A0} \\
\hat D_1&= \frac{\Upsilon_1}{4\sigma} - \frac{1}{(4\sigma)^2}(\Upsilon_0\Upsilon_1+\Upsilon_3\Upsilon_4)k^{\alpha} + \mathcal{O}(k^{2\alpha}),  \label{eq:A1} \\
\hat D_2&= \frac{\Upsilon_2}{4\sigma} - \frac{\Upsilon_2}{4\sigma}(\Upsilon_0+\Upsilon_4)k^{\alpha} + \mathcal{O}(k^{2\alpha}), \label{eq:A2} \\
\hat D_3&= \frac{\Upsilon_3}{4\sigma}-\frac{1}{(4\sigma)^2}(\Upsilon_0\Upsilon_3+\Upsilon_1\Upsilon_4)k^{\alpha}+ \mathcal{O}(k^{2\alpha}), \label{eq:A3} \\
\hat D_4&= \frac{\Upsilon_4}{(4\sigma)^2}k^{\alpha}+ \mathcal{O}(k^{2\alpha}). \label{eq:A4}    
\end{align}
Conversely, in the opposite limit ($k \to \infty $) we find 
\begin{align}
\hat D_0&= \frac{\Upsilon_0}{\Upsilon_0^2 - \Upsilon_2^2}\frac{1}{k^{\alpha}} + \mathcal{O}(1), \label{eq:B0} \\
\hat D_1&= \frac{\Upsilon_0\Upsilon_1 - \Upsilon_3\Upsilon_4}{\Upsilon_0^2 - \Upsilon_4^2}\frac{1}{k^{\alpha}} + \mathcal{O}(1),  \label{eq:B1} \\
\hat D_2&= \frac{\Upsilon_2}{\Upsilon_0+\Upsilon_4}\frac{1}{k^{\alpha}} + \mathcal{O}(1),  \label{eq:B2} \\
\hat D_3&= \frac{\Upsilon_0\Upsilon_3 - \Upsilon_1\Upsilon_4}{\Upsilon_0^2 - \Upsilon_2^2}\frac{1}{k^{\alpha}} + \mathcal{O}(1), \label{eq:B3} \\
\hat D_4&= \frac{\Upsilon_4}{\Upsilon_0^2 - \Upsilon_2^2}\frac{1}{k^{\alpha}} + \mathcal{O}(1).    \label{eq:B4}
\end{align}
}

\section{\rd{Derivation of the auxiliary relations presented in Table~\ref{identities}} \label{auxiliary}}

We will denote $\vec{e}_i$ ($i=\{x,y,z\}$) the standard basis of a Cartesian coordinate system.    
We also introduce the complex differential operators $\hat{\nabla}$ and $\hat{\nabla}^*$ \cite{Bertin2006,Bertin2009,Bertin2017}
\begin{equation}
\hat{\nabla}=\derpar{}{x}+i\derpar{}{y}, \qquad 
\hat{\nabla}^*=\derpar{}{x}-i\derpar{}{y}, 
\label{nabla}
\end{equation}
whose Fourier transforms are   
\begin{equation}
\hat{\nabla}_{\vec{k}}=-i k e^{i \psi}, \qquad \hat{\nabla}_{\vec{k}}^*=-i k e^{-i \psi}. 
\label{nablaF}
\end{equation}
\rd{For later convenience, we also recall the Pauli matrices
\begin{equation}
\boldsymbol{\sigma}_1\equiv 
\left(
\begin{matrix}
0 & 1 \\ 1 & 0
\end{matrix}
\right), \,\,\, 
\boldsymbol{\sigma}_2\equiv 
\left(
\begin{matrix}
0 & i \\ -i & 0
\end{matrix}
\right), \,\,\,
\boldsymbol{\sigma}_3\equiv 
\left(
\begin{matrix}
1 & 0 \\ 0 & -1
\end{matrix}
\right).
\label{eq:pauli}
\end{equation}
}

\subsection{Equation~(\ref{aux1})}

This relation is easily obtained by employing Eqs.~\eqref{nablaF}.
In details, 
\begin{align}
\hat{\nabla}_{\vec{k}}^* \hat f_1&=-i k_x \hat p_x -i k_y \hat p_y + k_x \hat p_y-k_y \hat p_x \label{eq:nablastar}\\
\hat{\nabla}_{\vec{k}} \hat f_1^*&=-i k_x \hat p_x -i k_y \hat p_y - k_x \hat p_y + k_y \hat p_x, \label{eq:nabla} 
\end{align}   
such that 
\begin{align}
\hat{\nabla}_{\vec{k}}^* \hat f_1 + \hat{\nabla}_{\vec{k}} \hat f_1^*&=-2 i (k_x \hat p_x + k_y \hat p_y)=-2i\vec{k}\cdot\hat{\vec{p}}. 
\label{eq:div} 
\end{align}   
Therefore, we can write 
\begin{align}
i k^{\alpha}(e^{-i\psi} \hat f_1 + e^{i\psi} \hat f_{1}^*)&=
-k^{\alpha-1}(\hat{\nabla}_{\vec{k}}^* \hat f_1 + \hat{\nabla}_{\vec{k}} \hat f_1^*) \notag\\
&=2k^{\alpha-1}(i\vec{k}\cdot\hat{\vec{p}}),  
\end{align}   
whose inverse Fourier transform yields $-2 \mathbb{I}^{1-\alpha}(\nabla\cdot\vec{p})$.

\subsection{Equation~(\ref{aux2}) \label{sec:aux2}}

Employing again the explicit definitions~\eqref{nablaF}, 
we can show that 
$k^2 e^{-i 2\psi} \hat h= -\hat{\nabla}_{\vec{k}}^{*2} \hat h$  
and that 
$\hat{\nabla}_{\vec{k}}^2 \hat h^*=(\hat{\nabla}_{\vec{k}}^{*2} \hat h)^*$ 
with $\hat h$ \rd{the Fourier transform of} an arbitrary complex function.
Setting $\hat h=\hat f_1\star \hat f_1$, we can then write
\begin{align}
& 
k^{\alpha}[e^{-i 2\psi}(\hat f_{1}\star \hat f_{1})+e^{i 2\psi}(\hat f_{1}\star \hat f_{1})^{*} ] \notag\\
&\qquad 
=-k^{\alpha-2}[ \hat{\nabla}_{\vec{k}}^{*2} (\hat f_{1}\star \hat f_{1}) + \hat{\nabla}_{\vec{k}}^2(\hat f_{1}\star \hat f_{1})^*]
\notag\\&\qquad 
=2 k^{\alpha-2} (i\vec{k}\cdot\hat{\vecxi}_0),  
\label{eq:frac-flux-fourier}
\end{align}
\rd{where we employed Eq.~\eqref{eq:div} with $\hat f_1$ substituted by 
$\hat{\nabla}_{\vec{k}}^{*} (\hat f_{1}\star \hat f_{1})$   
and we defined the auxiliary vector field 
$\vecxi_0=(\Re{(\hat{\nabla}^* f_{1}^{\,2})}, \Im{(\hat{\nabla}^* f_{1}^{\,2})})$. 
The inverse Fourier transform of Eq.~\eqref{eq:frac-flux-fourier} is thus     
$-2 \mrI^{2-\alpha} (\nabla\cdot\vecxi_0)$. 
To obtain an analytical expression for $\vecxi_0$, 
we calculate the complex derivative}
\begin{equation}
\hat{\nabla}^* f_{1}^{\,2}=\left[ \derpar{}{x}-i\derpar{}{y}\right](p_x^2-p_y^2+i 2 p_x p_y). 
\label{zeta0F1}
\end{equation}
Its real part can be manipulated as below   
\begin{align}
\Re{[\hat{\nabla}^* f_{1}^{\,2}]}
&=2\left(p_x\derpar{}{x}+p_y\derpar{}{y}\right)p_x+2p_x\left(\derpar{p_x}{x} + \derpar{p_y}{y} \right) \notag\\
& \quad 
-\derpar{}{x}(p_x^2+p_y^2), 
\label{zeta0F2}
\end{align}
which is clearly the $x$-component of Eq.~\eqref{si-xi0}. 
Likewise, the imaginary part of $\hat{\nabla}^* f_{1}^{\,2}$ can be shown \rd{to be} equal to its $y$-component. 
\rd{We note that this final calculation also proves Eq.~\eqref{aux4}.}

\subsection{Equation~(\ref{aux6})}

The inverse Fourier transform of Eq.~\eqref{eq:nabla} is given by   
\begin{equation}
\hat{\nabla} f_1^*=\nabla\cdot\vec{p}+i \left( \derpar{p_x}{y}-\derpar{p_y}{x} \right). 
\label{nablaf1star}
\end{equation}
Applying again the complex derivative, we obtain 
\begin{equation}
\hat{\nabla}^2f_1^*=\hat{\nabla}(\nabla\cdot\vec{p})+\left( -\derpar{}{y}+i\derpar{}{x}\right) \left( \derpar{p_x}{y} - \derpar{p_y}{x} \right). 
\label{nabla2f1star}
\end{equation}
On the one hand, the first term is identified with the gradient of $(\nabla\cdot\vec{p})$; on the other hand, the second term is identified with the curl of the vorticity vector $\vecomega$, i.e., with the term $-\nabla\times\vecomega$. 
In fact, we can calculate 
$\vecomega=(\partial_y p_x - \partial_x p_y)\vec{e}_{\rd{z}}$. 
Likewise, we can compute 
$\nabla\times\vecomega=(\vec{e}_{\rd{x}} \partial_y -\vec{e}_{\rd{y}}\partial_x)(\partial_y p_x - \partial_x p_y)$, 
which corresponds in the complex plane representation to the term
$(\partial_y -i \partial_x)(\partial_y p_x - \partial_x p_y)$. 
Therefore, we obtain the vector representation 
\begin{equation}
\hat{\nabla}^2f_1^* \rightarrow \nabla(\nabla\cdot\vec{p})-\nabla\times\vecomega=\nabla^2\vec{p}  , 
\label{lapR}
\end{equation}
where we employed the identity 
\begin{equation} 
\nabla^2\vec{p}=\nabla(\nabla\cdot\vec{p})-\nabla\times\vecomega .  
\label{veclaplace}
\end{equation}

\rd{
\subsection{Equation~(\ref{aux5})}

The inverse Fourier transform of the manipulated term~(\ref{aux5}) is  
$-\mrI^{1-\alpha}(\hat \nabla f_1)$. 
Using the definition~\eqref{nabla}, we obtain  
\begin{align}
\hat \nabla f_1 &= \derpar{p_x}{x} - \derpar{p_y}{y} + i \left( \derpar{p_y}{x} + \derpar{p_x}{y} \right) . 
\end{align}
Its vector representation is given by   
\begin{equation}
\hat \nabla f_1 \rightarrow \left(\derpar{}{x}\textbf{1} + i \derpar{}{y}\boldsymbol{\sigma}_2 \right)\vec{p}. 
\label{nablaf1}
\end{equation}

\subsection{Equation~(\ref{aux7})}

The inverse Fourier transform of the manipulated term~\eqref{aux7} yields 
$-\mrI^{2-\alpha} \hat \nabla^2 f_0$. 
Applying twice the complex derivative~\eqref{nabla}, we can calculate      
\begin{equation}
\hat \nabla^2 f_0= \left(\dersecpar{}{x} - \dersecpar{}{y}\right)f_0 + 2 i \derpar{}{x}\derpar{}{y}f_0 . 
\end{equation}
Its vector representation is thus given by   
\begin{equation}
\hat \nabla^2 f_0 \rightarrow \left(\derpar{}{x}\textbf{1} + i \derpar{}{y}\boldsymbol{\sigma}_2 \right)\nabla\rho. 
\label{nabla2f0}
\end{equation}

\subsection{Equation~(\ref{aux10})}

Recalling the definitions~\eqref{nablaF} and the relations obtained in Appendix~\ref{sec:aux2}, 
we can show that
\begin{equation}
i k^{\alpha} e^{i 3\psi} \hat f_1^*=k^{\alpha-3}\hat{\nabla}_{\vec{k}}^{3}\hat f_1^*. 
\end{equation} 
The inverse Fourier transform of this equation is  
$\mrI^{3-\alpha}(\hat \nabla^3 f_1^*)$. 
This complex derivative can be computed analytically by applying Eq.~\eqref{nabla} to~\eqref{nabla2f1star}. In details, we obtain    
\begin{align}
&\hat{\nabla}^3f_1^{*}
=\left(\dersecpar{}{x}-\dersecpar{}{y}\right)(\nabla\cdot\vec{p})-2\derpar{}{x}\derpar{}{y}\left(\derpar{p_x}{y}-\derpar{p_y}{x}\right)
\notag\\ & \quad
+i\left[ \left(\dersecpar{}{x}-\dersecpar{}{y}\right)\left(\derpar{p_x}{y}-\derpar{p_y}{x}\right)+2\derpar{}{x}\derpar{}{y}(\nabla\cdot\vec{p})\right]. 
\label{nabla3f1star}
\end{align}
Its vector representation can thus be written as   
\begin{align}
\hat \nabla^3 f_1 &\rightarrow \left(\derpar{}{x}\textbf{1} + i \derpar{}{y}\boldsymbol{\sigma}_2 \right)[\nabla(\nabla\cdot\vec{p})-(\nabla \times \vecomega)] \notag\\
&\rightarrow \left(\derpar{}{x}\textbf{1} + i \derpar{}{y}\boldsymbol{\sigma}_2 \right)\nabla^2\vec{p}. 
\label{nabla3f1}
\end{align}
}

\subsection{Equation~(\ref{aux8})}

\rd{Analogously to the previous subsection, we can write}
\begin{equation}
i k^{\alpha} e^{i 3\psi}(\hat f_1\star \hat f_1)^*=k^{\alpha-3}\hat{\nabla}_{\vec{k}}^{3}(\hat f_1\star \hat f_1)^*, 
\end{equation} 
whose Fourier inverse transform yields $\mrI^{3-\alpha}\hat{\nabla}^{3}f_1^{*2}$. 
Thus, we only need to compute the complex derivative  
\begin{equation}
\hat{\nabla}^{3}f_1^{*2}=2(3\hat{\nabla} f_1^* \hat{\nabla}^2 f_1^* + f_1^*\hat{\nabla}^3 f_1^*). 
\end{equation}
For the first term, using Eqs.~(\ref{nablaf1star}, \ref{nabla2f1star}) we find 
\begin{align}
\hat{\nabla}f_1^{*}\hat{\nabla}^2f_1^{*}
&=(\nabla\cdot\vec{p})\hat{\nabla}(\nabla\cdot\vec{p}) 
\notag\\ &\quad 
+(\nabla\cdot\vec{p})\left(-\derpar{}{y}+i\derpar{}{x}\right)\left(\derpar{p_x}{y}-\derpar{p_y}{x}\right) 
\notag\\ & \quad 
+\left(\derpar{p_x}{y}-\derpar{p_y}{x}\right)\left(-\derpar{}{y}+i\derpar{}{x}\right)(\nabla\cdot\vec{p}) 
\notag\\ & \quad
-\left(\derpar{p_x}{y}-\derpar{p_y}{x}\right)\hat{\nabla}\left(\derpar{p_x}{y}-\derpar{p_y}{x}\right)  .
\end{align}
The first and last terms are clearly identified with the gradient of the scalar field 
$(\nabla\cdot\vec{p})^2-|\vecomega|^2$; the second term is identified with 
$-(\nabla\cdot\vec{p})(\nabla\times\vecomega)$, 
and the third one with    
$\vecomega\times\nabla(\nabla\cdot\vec{p})$. 
Indeed, recalling that 
$\nabla(\nabla\cdot\vec{p})=(\vec{e}_{\rd{x}} \partial_x + \vec{e}_{\rd{y}} \partial_y)(\nabla\cdot\vec{p})$, we can compute  
$\vecomega\times\nabla(\nabla\cdot\vec{p})=(\partial_y p_x - \partial_x p_y)(-\vec{e}_{\rd{x}} \partial_y + \vec{e}_{\rd{y}} \partial_x)(\nabla\cdot\vec{p})$, which in the complex plane representation corresponds to the term appearing in the equation.  
Therefore, we write 
\begin{align}
\hat{\nabla}f_1^{*}\hat{\nabla}^2f_1^{*}&\rightarrow
\frac{1}{2}\nabla[(\nabla\cdot\vec{p})^2-|\vecomega|^2] 
\notag\\ & \quad 
-(\nabla\cdot\vec{p})(\nabla\times\vecomega)+\vecomega\times\nabla(\nabla\cdot\vec{p}).
\label{term32}
\end{align}

\rd{For the second term, multiplying Eq.~\eqref{nabla3f1star} by $f_1^{*}$ and rearranging the terms, we obtain}
\begin{align}
&f_1^{*}\hat{\nabla}^3f_1^{*}= 
\notag \\ & \qquad
\left( p_y\derpar{}{y}+p_x\derpar{}{x}\right) 
\left(-\derpar{}{y}+i\derpar{}{x} \right)\left(\derpar{p_x}{y}-\derpar{p_y}{x}\right) \notag \\ & \qquad
+ \left( p_y\derpar{}{y}+p_x\derpar{}{x}\right) 
\hat\nabla(\nabla\cdot\vec{p})
\notag \\ & \qquad
+\left( p_y\derpar{}{x}-p_x\derpar{}{y}\right) 
\hat\nabla\left(\derpar{p_x}{y}-\derpar{p_y}{x}\right) 
\notag \\ & \qquad 
+\left( p_y\derpar{}{x}-p_x\derpar{}{y}\right) \left(\derpar{}{y}-i\derpar{}{x} \right)(\nabla\cdot\vec{p}).
\label{f1nabla3f1}
\end{align}
The first two terms are identified with  
$(\vec{p}\cdot\nabla)[-\nabla\times\vecomega+\nabla(\nabla\cdot\vec{p})]$. 
By using Eq.~\eqref{veclaplace}, this is equal to  
$(\vec{p}\cdot\nabla)\nabla^2\vec{p}$. 
To determine the remaining terms, we introduce the vector operator 
$\vec{p}\times\nabla=(p_x\partial_y-p_y\partial_x)\vec{e}_{\rd{z}}$, 
such that we can show that 
$(\vec{p}\times\nabla)\times\nabla(\nabla\cdot\vec{p})=(p_y\partial_x-p_x\partial_y)(\vec{e}_{\rd{x}}\partial_y-\vec{e}_{\rd{y}}\partial_x)(\nabla\cdot\vec{p})$ and that  
$(\vec{p}\times\nabla)\times(\nabla\times\vecomega)=-(p_y\partial_x-p_x\partial_y)(\vec{e}_{\rd{x}}\partial_x+\vec{e}_{\rd{y}}\partial_y)(\partial_y p_x-\partial_x p_y)$.  
Therefore, the second term in the rhs of the equation is identified with
$(\vec{p}\times\nabla)\times\nabla(\nabla\cdot\vec{p})-(\vec{p}\times\nabla)\times(\nabla\times\vecomega)$, 
which is shown equal to 
$(\vec{p}\times\nabla)\times\nabla^2\vec{p}$ 
by recalling Eq.~\eqref{veclaplace}. 
Thus, we obtain 
\begin{equation}
f_1^{*}\hat{\nabla}^3 f_1^{*}\rightarrow
(\vec{p}\cdot\nabla)\nabla^2\vec{p}+(\vec{p}\times\nabla)\times\nabla^2\vec{p}.
\label{term33}
\end{equation} 
Combining Eqs.~(\ref{term32}, \ref{term33}), we obtain the vector representation  
\begin{align}
\frac{1}{2}\hat{\nabla}^{3}f_1^{*2}&\rightarrow\frac{3}{2}\nabla[(\nabla\cdot\vec{p})^2-|\vecomega|^2]+(\vec{p}\cdot\nabla)\nabla^2\vec{p}
\notag\\ &\qquad
+(\vec{p}\times\nabla)\times\nabla^2\vec{p} 
-3(\nabla\cdot\vec{p})(\nabla\times\vecomega)
\notag\\ &\qquad
+3[\vecomega\times\nabla(\nabla\cdot\vec{p})].  
\end{align}
This is further simplified by using the relations 
\begin{align}
(\nabla\cdot\vec{p})(\nabla\times\vecomega)&=\frac{1}{2}\nabla(\nabla\cdot\vec{p})^2-(\nabla\cdot\vec{p})\nabla^2\vec{p}, \\
\vecomega\times\nabla(\nabla\cdot\vec{p})&=\vecomega\times(\nabla\times\vecomega)+\vecomega\times\nabla^2\vec{p}, \\
\vecomega\times(\nabla\times\vecomega)&=\frac{1}{2}\nabla|\vecomega|^2-(\vecomega\cdot\nabla)\vecomega,
\end{align}
where the first and second one are derived by using Eq.~\eqref{veclaplace} and the third one by using the identity of vector calculus
\begin{align}
\nabla(\vec{A}\cdot\vec{B})&=(\vec{A}\cdot\nabla)\vec{B}+(\vec{B}\cdot\nabla)\vec{A} 
\notag\\ & \quad 
+\vec{A}\times(\nabla\times\vec{B})+\vec{B}\times(\nabla\times\vec{A})  
\end{align}
with $\vec{A}=\vec{B}=\vecomega$.
Using these relations we obtain the representation 
\begin{align}
\frac{1}{2}\hat{\nabla}^{3}f_1^{*2}&\rightarrow
(\vec{p}\cdot\nabla)\nabla^2\vec{p}+(\vec{p}\times\nabla)\times\nabla^2\vec{p}
+3[(\nabla\cdot\vec{p})\nabla^2\vec{p}
\notag\\ &\qquad
+(\vecomega\times\nabla^2\vec{p})-(\vecomega\cdot\nabla)\vecomega].  
\end{align}

\rd{
\subsection{Equation~(\ref{aux9})}

Using the first relation from Appendix~\ref{sec:aux2}  
we find  
\begin{equation}
[k^{\alpha} e^{i 4\psi}(\hat f_1 \star \hat f_1)^*]=
k^{\alpha-4}[\hat{\nabla}_{\vec{k}}^{4}(\hat f_1\star \hat f_1)^*], 
\end{equation} 
whose Fourier inverse transform yields $\mrI^{4-\alpha}(\hat{\nabla}^{4}f_1^{*2})$. 
Recalling that $\hat{\nabla}^3 f_1^{*2} \rightarrow 2(\vecxi_{1}+3\vecxi_{2})$ 
and applying to it the complex derivative $\hat \nabla$,  
we can write: 
\begin{align}
&\hat{\nabla}^{4}f_1^{*2}=
2\left[ \derpar{}{x}(\Xi_{1x}+3\Xi_{2x})- \derpar{}{x} (\Xi_{1y}+3\Xi_{2y}) \right]
\notag\\ & \, 
+ 2 i \left[ \derpar{}{x}(\Xi_{1y}+3\Xi_{2y}) + \derpar{}{y}(\Xi_{1x}+3\Xi_{2x}) \right] . 
\end{align}   
Its representation is obtained straightforwardly as 
\begin{align}
\hat{\nabla}^{4}f_1^{*2}& \rightarrow
2\left(\derpar{}{x}\textbf{1}+i \derpar{}{y}\boldsymbol{\sigma}_2 \right) (\vecxi_1+3\vecxi_2) .
\end{align}   
}

\section{\rd{Derivation of the auxiliary relations for the linear stability analysis} \label{auxiliary-lsa}}

\rd{We denote $\vec{e}_{\parallel}\equiv(\cos \upvartheta_{\parallel},\sin \upvartheta_{\parallel})$ the unit vector specifying the direction of spontaneous symmetry breaking, 
$\vec{e}_{\perp}\equiv(-\sin \upvartheta_{\parallel},\cos \upvartheta_{\parallel})$ the corresponding orthogonal vector and 
$\vec{e}_{\upvartheta}\equiv \vec{R}(\upvartheta)\vec{e}_{\parallel}$ the unit vector specifying the direction of the wave number vector $\vec{q}$ with the two-dimensional rotation matrix $\textbf{R}$ (the angle $\upvartheta$ is thus defined with respect to the direction of collective motion).}  

\subsection{Equations~(\ref{lsa1.3}) and (\ref{lsa2.1}) \label{sec:lsa_aux1}}

\rd{The Fourier transform of the term $\mrI^{1-\alpha}\delta\vecxi_0$ is 
\begin{equation}
\mcF\{\mrI^{1-\alpha}\delta\vecxi_0\}=k^{\alpha-1}\hat{\delta\vecxi_0},  \label{fourier1.1}
\end{equation}
with $\delta\vecxi_0$ as in Eq.~\eqref{dzeta0}.}
Taking its Fourier transform and using the ansatz~\eqref{perturbs} we obtain for the first term 
\begin{align}
\mcF\{(\nabla\cdot\delta\vec{p})\vec{p}^*\}&=
\rd{(-i\vec{k}\cdot\delta\vec{p}_0)[2\pi e^{s t}\delta(\vec{k}+\vec{q})] p^* \vec{e}_{\parallel}} \notag\\
&=
\rd{i q p^*(\vec{e}_{\upvartheta}\cdot\delta\vec{p}_0)[2\pi e^{s t}\delta(\vec{k}+\vec{q})]\vec{e}_{\parallel}} 
\end{align}
and for the second one\rd{, analogously,}  
\begin{align}
&\mcF\{(\nabla\times\delta\vec{p})\times\vec{p}^*\}=
\rd{i q p^* 
(\vec{e}_{\upvartheta+\frac{\pi}{2}}\cdot\delta\vec{p}_0)
[2\pi e^{s t}\delta(\vec{k}+\vec{q})] \vec{e}_{\perp}}.
\end{align}
\rd{Substituting these expressions into Eq.~\eqref{fourier1.1}, we find the Fourier transform of Eq.~\eqref{lsa2.1}  
\begin{align}
\mcF\{\mrI^{1-\alpha}\delta\vecxi_0\}&=\rd{i q^{\alpha} p^*
[(\vec{e}_{\upvartheta}\cdot\hat{\delta\vec{p}})\vec{e}_{\parallel}} 
\rd{+ (\vec{e}_{\upvartheta+\frac{\pi}{2}}\cdot\hat{\delta\vec{p}})\vec{e}_{\perp}]} .  
\end{align}
}
\rd{Analogously,} the Fourier transform of $\mrI^{2-\alpha}(\nabla\cdot\delta\vecxi_0)$ is 
\begin{equation}
\mcF\{\mrI^{2-\alpha}(\nabla\cdot\delta\vecxi_0)\}=k^{\alpha-2}[(-i\vec{k})\cdot\hat{\delta\vecxi_0}].  \label{fourier1.2}
\end{equation}
\rd{By employing the formulas just derived, we obtain} 
\begin{align}
\mcF\{\mrI^{2-\alpha}(\nabla\cdot\delta\vecxi_0)\}&=
\rd{-q^{\alpha} p^*
[(\vec{e}_{\upvartheta}\cdot\hat{\delta\vec{p}})(\vec{e}_{\upvartheta}\cdot\vec{e}_{\parallel})} \notag \\ 
& \quad  
\rd{+ (\vec{e}_{\upvartheta+\frac{\pi}{2}}\cdot\hat{\delta\vec{p}})(\vec{e}_{\upvartheta}\cdot\vec{e}_{\perp})]} .  
\end{align}

\subsection{Equation~(\ref{lsa2.1}) \label{sec:lsa_aux2}}

The Fourier transform of the term $\mrI^{3-\alpha}\delta\vecxi_1$ is given by 
\begin{equation}
\mcF\{\mrI^{3-\alpha}\delta\vecxi_1\}=k^{\alpha-3}\mcF\{\delta\vecxi_1\},  \label{fourier2}
\end{equation}
with $\delta\vecxi_1$ as in Eq.~\eqref{dzeta1}. 
Taking its Fourier transform and using the ansatz~\eqref{perturbs}, we obtain for the first term 
\begin{align}
&\mcF\{(\vec{p}^*\cdot\nabla)\nabla^2\delta\vec{p}\} \notag\\ 
&\qquad
=p^*(-i\vec{k}\cdot\rd{\vec{e}_{\parallel}})(-i\vec{k})^2\delta\vec{p}_0[2\pi e^{s t}\delta(\vec{k}+\vec{q})] \notag\\  
&\qquad 
=-i q^3 p^* \rd{(\vec{e}_{\upvartheta}\cdot\vec{e}_{\parallel})}\delta\vec{p}_0[2\pi e^{s t}\delta(\vec{k}+\vec{q})] ;
\label{dzeta1F1}
\end{align}
and for the second one  
\begin{align}
& \mcF\{(\vec{p}^*\times\nabla)\times\nabla^2\delta\vec{p}\}
=\rd{p^* k^2(i \vec{k}\cdot \vec{e}_{\perp}) \times} 
\notag\\ &\qquad \times
(-\delta p_{0y}\vec{e}_x+\delta p_{0x}\vec{e}_y)[2\pi e^{s t}\delta(\vec{k}+\vec{q})] \notag\\
&\,
=-i q^3 p^* \rd{(\vec{e}_{\upvartheta}\cdot \vec{e}_{\perp})\textbf{R}\left(\frac{\pi}{2}\right)\delta\vec{p}_0}[2\pi e^{s t}\delta(\vec{k}+\vec{q})] . 
\label{dzeta1F2}
\end{align}
Substituting these expressions into Eq.~\eqref{fourier2}, we find the Fourier transform of Eq.~\eqref{lsa2.1}  
\begin{align}
\mcF\{\mrI^{3-\alpha}\delta\vecxi_1\}&=
\rd{-i q^{\alpha} p^* 
\left[(\vec{e}_{\upvartheta}\cdot\vec{e}_{\parallel}) 
+(\vec{e}_{\upvartheta}\cdot\vec{e}_{\perp})\textbf{R}\left(\frac{\pi}{2}\right)
\right]\hat{\delta\vec{p}}} .  
\end{align}

\rd{
\subsection{Equation~(\ref{lsa2.3}) \label{sec:lsa_aux3}}

The Fourier transform of Eq.~\eqref{dQ1} can be written as
\begin{align}
&\mcF\left\{
\delta \text{Q}_{j}^{(1)} 
\right\}
=
-B_0 k^{\alpha} \hat{ \delta \mathrm{M}_{jl} } p^*_{l}
-B_0 k^{\alpha} \mathrm{M}^*_{jl} \hat{ \delta p_{l} } 
\notag \\ & \qquad 
+ \frac{\lambda_1}{2} k^{\alpha-1}  \hat{\mathrm{T}}_{jl} \hat{ \delta p_{l} } 
- \frac{\lambda_2}{4} k^{\alpha-2} \hat{\mathrm{T}}_{jl} (- i k_{l}) \hat{ \delta\rho } 
\notag \\ & \qquad
+ \frac{\lambda_3}{4}k^{\alpha-3} \hat{\mathrm{T}}_{jl} (- k^2) \hat{ \delta p_{l} } 
- B_4 k^{\alpha-4} \hat{\mathrm{T}}_{jl} \hat{ \delta\Xi}_{1 l}. 
\label{dQ1Fourier}
\end{align} 
This equation can be simplified if we recall that all Fourier transform of the vector fields are $\propto \delta(\vec{q}+\vec{k})$.   
Thus, we can make the substitution $\vec{k} \to -q \, \vec{e}_{\upvartheta}$ and write the Fourier transform of the differential operator $\textbf{T}$ as
\begin{equation}
\hat{\textbf{T}}= 
i q
\left(
\begin{matrix}
\cos (\upvartheta + \upvartheta_{\parallel}) &  - \sin (\upvartheta + \upvartheta_{\parallel}) \\
\sin (\upvartheta + \upvartheta_{\parallel}) & \cos (\upvartheta + \upvartheta_{\parallel})
\end{matrix}
\right) .
\end{equation}
 Therefore, Eq.~(\ref{dQ1Fourier}) simplifies to  
\begin{align}
&\mcF\left\{
\delta \text{Q}_{j}^{(1)} 
\right\}
=
-B_0 q^{\alpha} \hat{ \delta \mathrm{M}}_{jl} p^*_{l}
-B_0 q^{\alpha} \mathrm{M}^*_{jl} \hat{ \delta p_{l} } 
\notag \\ & \qquad 
+ i \frac{\lambda_1}{2} q^{\alpha}  \mathrm{R}_{jl}(\upvartheta + \upvartheta_{\parallel}) \hat{ \delta p_{l} } 
+ \frac{\lambda_2}{4} q^{\alpha} \mathrm{R}_{jl}(\upvartheta + \upvartheta_{\parallel}) e_{\upvartheta l} \hat{ \delta\rho } 
\notag \\ & \qquad
- i \frac{\lambda_3}{4}q^{\alpha} \mathrm{R}_{jl}(\upvartheta + \upvartheta_{\parallel}) \hat{ \delta p_{l} } 
- i B_4 q^{\alpha-3} \mathrm{R}_{jl}(\upvartheta + \upvartheta_{\parallel}) \hat{ \delta\Xi}_{1 l}.   
\label{dQ1Fourier2}
\end{align} 
In addition, we can show that 
$\hat{ \delta \mathrm{M}}_{jl} p^*_{l} + \mathrm{M}^*_{jl} \hat{ \delta p_{l}}=2 p^* \hat{\delta p_{l}}$.  
Therefore, using also Eqs.~(\ref{dzeta1F1}, \ref{dzeta1F2}), we obtain 
\begin{align}
&\mcF\left\{
\delta \text{Q}_{j}^{(1)}
\right\}
=
-2B_0 q^{\alpha} p^* \hat{ \delta p_{j} } 
+ i \frac{\lambda_1}{2} q^{\alpha}  \text{R}_{jl}(\upvartheta + \upvartheta_{\parallel}) \hat{ \delta p_{l} } 
\notag \\ & \qquad 
+ \frac{\lambda_2}{4} q^{\alpha} \hat{ \delta\rho } \, \text{R}_{jl}(\upvartheta + \upvartheta_{\parallel}) e_{\upvartheta l}
- i \frac{\lambda_3}{4}q^{\alpha} \text{R}_{jl}(\upvartheta + \upvartheta_{\parallel}) \hat{ \delta p_{l} } 
\notag \\ & \qquad
- B_4 q^{\alpha} p^* (\vec{e}_{\upvartheta}\cdot\vec{e}_{\perp}) \text{R}_{jm}(\upvartheta + \upvartheta_{\parallel}) \text{R}_{ml}\left(\frac{\pi}{2}\right) \hat{ \delta p_{l}} 
\notag \\ & \qquad
- B_4 q^{\alpha} p^* (\vec{e}_{\upvartheta}\cdot\vec{e}_{\parallel})  \text{R}_{jl}(\upvartheta + \upvartheta_{\parallel}) \hat{ \delta p_{l}}.  
\end{align} 
}


\begin{thebibliography}{150}%
\makeatletter
\providecommand \@ifxundefined [1]{%
 \@ifx{#1\undefined}
}%
\providecommand \@ifnum [1]{%
 \ifnum #1\expandafter \@firstoftwo
 \else \expandafter \@secondoftwo
 \fi
}%
\providecommand \@ifx [1]{%
 \ifx #1\expandafter \@firstoftwo
 \else \expandafter \@secondoftwo
 \fi
}%
\providecommand \natexlab [1]{#1}%
\providecommand \enquote  [1]{``#1''}%
\providecommand \bibnamefont  [1]{#1}%
\providecommand \bibfnamefont [1]{#1}%
\providecommand \citenamefont [1]{#1}%
\providecommand \href@noop [0]{\@secondoftwo}%
\providecommand \href [0]{\begingroup \@sanitize@url \@href}%
\providecommand \@href[1]{\@@startlink{#1}\@@href}%
\providecommand \@@href[1]{\endgroup#1\@@endlink}%
\providecommand \@sanitize@url [0]{\catcode `\\12\catcode `\$12\catcode
  `\&12\catcode `\#12\catcode `\^12\catcode `\_12\catcode `\%12\relax}%
\providecommand \@@startlink[1]{}%
\providecommand \@@endlink[0]{}%
\providecommand \url  [0]{\begingroup\@sanitize@url \@url }%
\providecommand \@url [1]{\endgroup\@href {#1}{\urlprefix }}%
\providecommand \urlprefix  [0]{URL }%
\providecommand \Eprint [0]{\href }%
\providecommand \doibase [0]{http://dx.doi.org/}%
\providecommand \selectlanguage [0]{\@gobble}%
\providecommand \bibinfo  [0]{\@secondoftwo}%
\providecommand \bibfield  [0]{\@secondoftwo}%
\providecommand \translation [1]{[#1]}%
\providecommand \BibitemOpen [0]{}%
\providecommand \bibitemStop [0]{}%
\providecommand \bibitemNoStop [0]{.\EOS\space}%
\providecommand \EOS [0]{\spacefactor3000\relax}%
\providecommand \BibitemShut  [1]{\csname bibitem#1\endcsname}%
\let\auto@bib@innerbib\@empty
\bibitem [{\citenamefont {Toner}\ \emph {et~al.}(2005)\citenamefont {Toner},
  \citenamefont {Tu},\ and\ \citenamefont {Ramaswamy}}]{Toner2005}%
  \BibitemOpen
  \bibfield  {author} {\bibinfo {author} {\bibfnamefont {John}\ \bibnamefont
  {Toner}}, \bibinfo {author} {\bibfnamefont {Yuhai}\ \bibnamefont {Tu}}, \
  and\ \bibinfo {author} {\bibfnamefont {Sriram}\ \bibnamefont {Ramaswamy}},\
  }\bibfield  {title} {\enquote {\bibinfo {title} {Hydrodynamics and phases of
  flocks},}\ }\href@noop {} {\bibfield  {journal} {\bibinfo  {journal} {Ann.
  Phys.}\ }\textbf {\bibinfo {volume} {318}},\ \bibinfo {pages} {170--244}
  (\bibinfo {year} {2005})}\BibitemShut {NoStop}%
\bibitem [{\citenamefont {Schweitzer}(2007)}]{Schweitzer2007}%
  \BibitemOpen
  \bibfield  {author} {\bibinfo {author} {\bibfnamefont {Frank}\ \bibnamefont
  {Schweitzer}},\ }\href@noop {} {\emph {\bibinfo {title} {Brownian agents and
  active particles: collective dynamics in the natural and social sciences}}}\
  (\bibinfo  {publisher} {Springer},\ \bibinfo {year} {2007})\BibitemShut
  {NoStop}%
\bibitem [{\citenamefont {Ramaswamy}(2010)}]{Ramaswamy2010}%
  \BibitemOpen
  \bibfield  {author} {\bibinfo {author} {\bibfnamefont {Sriram}\ \bibnamefont
  {Ramaswamy}},\ }\bibfield  {title} {\enquote {\bibinfo {title} {The mechanics
  and statistics of active matter},}\ }\href@noop {} {\bibfield  {journal}
  {\bibinfo  {journal} {Annu. Rev. Condens. Matter Phys.}\ }\textbf {\bibinfo
  {volume} {1}},\ \bibinfo {pages} {323--345} (\bibinfo {year}
  {2010})}\BibitemShut {NoStop}%
\bibitem [{\citenamefont {Marchetti}\ \emph {et~al.}(2013)\citenamefont
  {Marchetti}, \citenamefont {Joanny}, \citenamefont {Ramaswamy}, \citenamefont
  {Liverpool}, \citenamefont {Prost}, \citenamefont {Rao},\ and\ \citenamefont
  {Simha}}]{Marchetti2013}%
  \BibitemOpen
  \bibfield  {author} {\bibinfo {author} {\bibfnamefont {M~Cristina}\
  \bibnamefont {Marchetti}}, \bibinfo {author} {\bibfnamefont {J~F}\
  \bibnamefont {Joanny}}, \bibinfo {author} {\bibfnamefont {S}~\bibnamefont
  {Ramaswamy}}, \bibinfo {author} {\bibfnamefont {T~B}\ \bibnamefont
  {Liverpool}}, \bibinfo {author} {\bibfnamefont {J}~\bibnamefont {Prost}},
  \bibinfo {author} {\bibfnamefont {Madan}\ \bibnamefont {Rao}}, \ and\
  \bibinfo {author} {\bibfnamefont {R~Aditi}\ \bibnamefont {Simha}},\
  }\bibfield  {title} {\enquote {\bibinfo {title} {Hydrodynamics of soft active
  matter},}\ }\href@noop {} {\bibfield  {journal} {\bibinfo  {journal} {Rev.
  Mod. Phys.}\ }\textbf {\bibinfo {volume} {85}},\ \bibinfo {pages} {1143}
  (\bibinfo {year} {2013})}\BibitemShut {NoStop}%
\bibitem [{\citenamefont {Hauser}\ and\ \citenamefont
  {Schimansky-Geier}(2012)}]{Hauser2015}%
  \BibitemOpen
  \bibfield  {author} {\bibinfo {author} {\bibfnamefont {M.~J.~B.}\
  \bibnamefont {Hauser}}\ and\ \bibinfo {author} {\bibfnamefont {Lutz}\
  \bibnamefont {Schimansky-Geier}},\ }\bibfield  {title} {\enquote {\bibinfo
  {title} {Statistical physics of self-propelled particles},}\ }\href@noop {}
  {\bibfield  {journal} {\bibinfo  {journal} {Eur. Phys. J. Spec. Top.}\
  }\textbf {\bibinfo {volume} {202}},\ \bibinfo {pages} {1--162} (\bibinfo
  {year} {2012})}\BibitemShut {NoStop}%
\bibitem [{\citenamefont {Needleman}\ and\ \citenamefont
  {Dogic}(2017)}]{Needleman2017}%
  \BibitemOpen
  \bibfield  {author} {\bibinfo {author} {\bibfnamefont {Daniel}\ \bibnamefont
  {Needleman}}\ and\ \bibinfo {author} {\bibfnamefont {Zvonimir}\ \bibnamefont
  {Dogic}},\ }\bibfield  {title} {\enquote {\bibinfo {title} {Active matter at
  the interface between materials science and cell biology},}\ }\href@noop {}
  {\bibfield  {journal} {\bibinfo  {journal} {Nat. Rev. Mat.}\ }\textbf
  {\bibinfo {volume} {2}},\ \bibinfo {pages} {17048} (\bibinfo {year}
  {2017})}\BibitemShut {NoStop}%
\bibitem [{\citenamefont {Vicsek}\ \emph {et~al.}(1995)\citenamefont {Vicsek},
  \citenamefont {Czir{\'o}k}, \citenamefont {Ben-Jacob}, \citenamefont
  {Cohen},\ and\ \citenamefont {Shochet}}]{Vicsek1995}%
  \BibitemOpen
  \bibfield  {author} {\bibinfo {author} {\bibfnamefont {Tam{\'a}s}\
  \bibnamefont {Vicsek}}, \bibinfo {author} {\bibfnamefont {Andr{\'a}s}\
  \bibnamefont {Czir{\'o}k}}, \bibinfo {author} {\bibfnamefont {Eshel}\
  \bibnamefont {Ben-Jacob}}, \bibinfo {author} {\bibfnamefont {Inon}\
  \bibnamefont {Cohen}}, \ and\ \bibinfo {author} {\bibfnamefont {Ofer}\
  \bibnamefont {Shochet}},\ }\bibfield  {title} {\enquote {\bibinfo {title}
  {Novel type of phase transition in a system of self-driven particles},}\
  }\href@noop {} {\bibfield  {journal} {\bibinfo  {journal} {Phys. Rev. Lett.}\
  }\textbf {\bibinfo {volume} {75}},\ \bibinfo {pages} {1226} (\bibinfo {year}
  {1995})}\BibitemShut {NoStop}%
\bibitem [{\citenamefont {Toner}\ and\ \citenamefont {Tu}(1995)}]{Toner1995}%
  \BibitemOpen
  \bibfield  {author} {\bibinfo {author} {\bibfnamefont {John}\ \bibnamefont
  {Toner}}\ and\ \bibinfo {author} {\bibfnamefont {Yuhai}\ \bibnamefont {Tu}},\
  }\bibfield  {title} {\enquote {\bibinfo {title} {Long-range order in a
  two-dimensional dynamical xy model: how birds fly together},}\ }\href@noop {}
  {\bibfield  {journal} {\bibinfo  {journal} {Phys. Rev. Lett.}\ }\textbf
  {\bibinfo {volume} {75}},\ \bibinfo {pages} {4326} (\bibinfo {year}
  {1995})}\BibitemShut {NoStop}%
\bibitem [{\citenamefont {Toner}\ and\ \citenamefont {Tu}(1998)}]{Toner1998}%
  \BibitemOpen
  \bibfield  {author} {\bibinfo {author} {\bibfnamefont {John}\ \bibnamefont
  {Toner}}\ and\ \bibinfo {author} {\bibfnamefont {Yuhai}\ \bibnamefont {Tu}},\
  }\bibfield  {title} {\enquote {\bibinfo {title} {Flocks, herds, and schools:
  A quantitative theory of flocking},}\ }\href@noop {} {\bibfield  {journal}
  {\bibinfo  {journal} {Phys. Rev. E}\ }\textbf {\bibinfo {volume} {58}},\
  \bibinfo {pages} {4828} (\bibinfo {year} {1998})}\BibitemShut {NoStop}%
\bibitem [{\citenamefont {Vicsek}\ and\ \citenamefont
  {Zafeiris}(2012)}]{Vicsek2012}%
  \BibitemOpen
  \bibfield  {author} {\bibinfo {author} {\bibfnamefont {Tam{\'a}s}\
  \bibnamefont {Vicsek}}\ and\ \bibinfo {author} {\bibfnamefont {Anna}\
  \bibnamefont {Zafeiris}},\ }\bibfield  {title} {\enquote {\bibinfo {title}
  {Collective motion},}\ }\href@noop {} {\bibfield  {journal} {\bibinfo
  {journal} {Phys. Rep.}\ }\textbf {\bibinfo {volume} {517}},\ \bibinfo {pages}
  {71--140} (\bibinfo {year} {2012})}\BibitemShut {NoStop}%
\bibitem [{\citenamefont {Dombrowski}\ \emph {et~al.}(2004)\citenamefont
  {Dombrowski}, \citenamefont {Cisneros}, \citenamefont {Chatkaew},
  \citenamefont {Goldstein},\ and\ \citenamefont {Kessler}}]{Dombrowski2004}%
  \BibitemOpen
  \bibfield  {author} {\bibinfo {author} {\bibfnamefont {Christopher}\
  \bibnamefont {Dombrowski}}, \bibinfo {author} {\bibfnamefont {Luis}\
  \bibnamefont {Cisneros}}, \bibinfo {author} {\bibfnamefont {Sunita}\
  \bibnamefont {Chatkaew}}, \bibinfo {author} {\bibfnamefont {Raymond~E}\
  \bibnamefont {Goldstein}}, \ and\ \bibinfo {author} {\bibfnamefont {John~O}\
  \bibnamefont {Kessler}},\ }\bibfield  {title} {\enquote {\bibinfo {title}
  {Self-concentration and large-scale coherence in bacterial dynamics},}\
  }\href@noop {} {\bibfield  {journal} {\bibinfo  {journal} {Phys. Rev. Lett.}\
  }\textbf {\bibinfo {volume} {93}},\ \bibinfo {pages} {098103} (\bibinfo
  {year} {2004})}\BibitemShut {NoStop}%
\bibitem [{\citenamefont {Hernandez-Ortiz}\ \emph {et~al.}(2005)\citenamefont
  {Hernandez-Ortiz}, \citenamefont {Stoltz},\ and\ \citenamefont
  {Graham}}]{Hernandez-Ortiz2005}%
  \BibitemOpen
  \bibfield  {author} {\bibinfo {author} {\bibfnamefont {Juan~P}\ \bibnamefont
  {Hernandez-Ortiz}}, \bibinfo {author} {\bibfnamefont {Christopher~G}\
  \bibnamefont {Stoltz}}, \ and\ \bibinfo {author} {\bibfnamefont {Michael~D}\
  \bibnamefont {Graham}},\ }\bibfield  {title} {\enquote {\bibinfo {title}
  {Transport and collective dynamics in suspensions of confined swimming
  particles},}\ }\href@noop {} {\bibfield  {journal} {\bibinfo  {journal}
  {Phys. Rev. Lett.}\ }\textbf {\bibinfo {volume} {95}},\ \bibinfo {pages}
  {204501} (\bibinfo {year} {2005})}\BibitemShut {NoStop}%
\bibitem [{\citenamefont {Sokolov}\ \emph {et~al.}(2007)\citenamefont
  {Sokolov}, \citenamefont {Aranson}, \citenamefont {Kessler},\ and\
  \citenamefont {Goldstein}}]{Sokolov2007}%
  \BibitemOpen
  \bibfield  {author} {\bibinfo {author} {\bibfnamefont {Andrey}\ \bibnamefont
  {Sokolov}}, \bibinfo {author} {\bibfnamefont {Igor~S}\ \bibnamefont
  {Aranson}}, \bibinfo {author} {\bibfnamefont {John~O}\ \bibnamefont
  {Kessler}}, \ and\ \bibinfo {author} {\bibfnamefont {Raymond~E}\ \bibnamefont
  {Goldstein}},\ }\bibfield  {title} {\enquote {\bibinfo {title} {Concentration
  dependence of the collective dynamics of swimming bacteria},}\ }\href@noop {}
  {\bibfield  {journal} {\bibinfo  {journal} {Phys. Rev. Lett.}\ }\textbf
  {\bibinfo {volume} {98}},\ \bibinfo {pages} {158102} (\bibinfo {year}
  {2007})}\BibitemShut {NoStop}%
\bibitem [{\citenamefont {Aranson}\ \emph {et~al.}(2007)\citenamefont
  {Aranson}, \citenamefont {Sokolov}, \citenamefont {Kessler},\ and\
  \citenamefont {Goldstein}}]{Aranson2007}%
  \BibitemOpen
  \bibfield  {author} {\bibinfo {author} {\bibfnamefont {Igor~S}\ \bibnamefont
  {Aranson}}, \bibinfo {author} {\bibfnamefont {Andrey}\ \bibnamefont
  {Sokolov}}, \bibinfo {author} {\bibfnamefont {John~O}\ \bibnamefont
  {Kessler}}, \ and\ \bibinfo {author} {\bibfnamefont {Raymond~E}\ \bibnamefont
  {Goldstein}},\ }\bibfield  {title} {\enquote {\bibinfo {title} {Model for
  dynamical coherence in thin films of self-propelled microorganisms},}\
  }\href@noop {} {\bibfield  {journal} {\bibinfo  {journal} {Phys. Rev. E}\
  }\textbf {\bibinfo {volume} {75}},\ \bibinfo {pages} {040901(R)} (\bibinfo
  {year} {2007})}\BibitemShut {NoStop}%
\bibitem [{\citenamefont {Saintillan}\ and\ \citenamefont
  {Shelley}(2007)}]{Saintillan2007}%
  \BibitemOpen
  \bibfield  {author} {\bibinfo {author} {\bibfnamefont {David}\ \bibnamefont
  {Saintillan}}\ and\ \bibinfo {author} {\bibfnamefont {Michael~J}\
  \bibnamefont {Shelley}},\ }\bibfield  {title} {\enquote {\bibinfo {title}
  {Orientational order and instabilities in suspensions of self-locomoting
  rods},}\ }\href@noop {} {\bibfield  {journal} {\bibinfo  {journal} {Phys.
  Rev. Lett.}\ }\textbf {\bibinfo {volume} {99}},\ \bibinfo {pages} {058102}
  (\bibinfo {year} {2007})}\BibitemShut {NoStop}%
\bibitem [{\citenamefont {Wolgemuth}(2008)}]{Wolgemuth2008}%
  \BibitemOpen
  \bibfield  {author} {\bibinfo {author} {\bibfnamefont {Charles~W}\
  \bibnamefont {Wolgemuth}},\ }\bibfield  {title} {\enquote {\bibinfo {title}
  {Collective swimming and the dynamics of bacterial turbulence},}\ }\href@noop
  {} {\bibfield  {journal} {\bibinfo  {journal} {Biophys. J.}\ }\textbf
  {\bibinfo {volume} {95}},\ \bibinfo {pages} {1564--1574} (\bibinfo {year}
  {2008})}\BibitemShut {NoStop}%
\bibitem [{\citenamefont {Sanchez}\ \emph {et~al.}(2012)\citenamefont
  {Sanchez}, \citenamefont {Chen}, \citenamefont {DeCamp}, \citenamefont
  {Heymann},\ and\ \citenamefont {Dogic}}]{Sanchez2012}%
  \BibitemOpen
  \bibfield  {author} {\bibinfo {author} {\bibfnamefont {Tim}\ \bibnamefont
  {Sanchez}}, \bibinfo {author} {\bibfnamefont {Daniel~TN}\ \bibnamefont
  {Chen}}, \bibinfo {author} {\bibfnamefont {Stephen~J}\ \bibnamefont
  {DeCamp}}, \bibinfo {author} {\bibfnamefont {Michael}\ \bibnamefont
  {Heymann}}, \ and\ \bibinfo {author} {\bibfnamefont {Zvonimir}\ \bibnamefont
  {Dogic}},\ }\bibfield  {title} {\enquote {\bibinfo {title} {Spontaneous
  motion in hierarchically assembled active matter},}\ }\href@noop {}
  {\bibfield  {journal} {\bibinfo  {journal} {Nature}\ }\textbf {\bibinfo
  {volume} {491}},\ \bibinfo {pages} {431} (\bibinfo {year}
  {2012})}\BibitemShut {NoStop}%
\bibitem [{\citenamefont {Wensink}\ \emph {et~al.}(2012)\citenamefont
  {Wensink}, \citenamefont {Dunkel}, \citenamefont {Heidenreich}, \citenamefont
  {Drescher}, \citenamefont {Goldstein}, \citenamefont {L{\"o}wen},\ and\
  \citenamefont {Yeomans}}]{Wensink2012}%
  \BibitemOpen
  \bibfield  {author} {\bibinfo {author} {\bibfnamefont {Henricus~H}\
  \bibnamefont {Wensink}}, \bibinfo {author} {\bibfnamefont {J{\"o}rn}\
  \bibnamefont {Dunkel}}, \bibinfo {author} {\bibfnamefont {Sebastian}\
  \bibnamefont {Heidenreich}}, \bibinfo {author} {\bibfnamefont {Knut}\
  \bibnamefont {Drescher}}, \bibinfo {author} {\bibfnamefont {Raymond~E}\
  \bibnamefont {Goldstein}}, \bibinfo {author} {\bibfnamefont {Hartmut}\
  \bibnamefont {L{\"o}wen}}, \ and\ \bibinfo {author} {\bibfnamefont {Julia~M}\
  \bibnamefont {Yeomans}},\ }\bibfield  {title} {\enquote {\bibinfo {title}
  {Meso-scale turbulence in living fluids},}\ }\href@noop {} {\bibfield
  {journal} {\bibinfo  {journal} {Proc. Natl. Acad. Sci.}\ }\textbf {\bibinfo
  {volume} {109}},\ \bibinfo {pages} {14308--14313} (\bibinfo {year}
  {2012})}\BibitemShut {NoStop}%
\bibitem [{\citenamefont {Doostmohammadi}\ \emph {et~al.}(2018)\citenamefont
  {Doostmohammadi}, \citenamefont {Ign{\'e}s-Mullol}, \citenamefont {Yeomans},\
  and\ \citenamefont {Sagu{\'e}s}}]{Doostmohammadi2018}%
  \BibitemOpen
  \bibfield  {author} {\bibinfo {author} {\bibfnamefont {Amin}\ \bibnamefont
  {Doostmohammadi}}, \bibinfo {author} {\bibfnamefont {Jordi}\ \bibnamefont
  {Ign{\'e}s-Mullol}}, \bibinfo {author} {\bibfnamefont {Julia~M}\ \bibnamefont
  {Yeomans}}, \ and\ \bibinfo {author} {\bibfnamefont {Francesc}\ \bibnamefont
  {Sagu{\'e}s}},\ }\bibfield  {title} {\enquote {\bibinfo {title} {Active
  nematics},}\ }\href@noop {} {\bibfield  {journal} {\bibinfo  {journal} {Nat.
  Commun.}\ }\textbf {\bibinfo {volume} {9}},\ \bibinfo {pages} {3246}
  (\bibinfo {year} {2018})}\BibitemShut {NoStop}%
\bibitem [{\citenamefont {Tailleur}\ and\ \citenamefont
  {Cates}(2008)}]{Tailleur2008}%
  \BibitemOpen
  \bibfield  {author} {\bibinfo {author} {\bibfnamefont {J}~\bibnamefont
  {Tailleur}}\ and\ \bibinfo {author} {\bibfnamefont {M~E}\ \bibnamefont
  {Cates}},\ }\bibfield  {title} {\enquote {\bibinfo {title} {Statistical
  mechanics of interacting run-and-tumble bacteria},}\ }\href@noop {}
  {\bibfield  {journal} {\bibinfo  {journal} {Phys. Rev. Lett.}\ }\textbf
  {\bibinfo {volume} {100}},\ \bibinfo {pages} {218103} (\bibinfo {year}
  {2008})}\BibitemShut {NoStop}%
\bibitem [{\citenamefont {Fily}\ and\ \citenamefont
  {Marchetti}(2012)}]{Fily2012}%
  \BibitemOpen
  \bibfield  {author} {\bibinfo {author} {\bibfnamefont {Yaouen}\ \bibnamefont
  {Fily}}\ and\ \bibinfo {author} {\bibfnamefont {M~Cristina}\ \bibnamefont
  {Marchetti}},\ }\bibfield  {title} {\enquote {\bibinfo {title} {Athermal
  phase separation of self-propelled particles with no alignment},}\
  }\href@noop {} {\bibfield  {journal} {\bibinfo  {journal} {Phys. Rev. Lett.}\
  }\textbf {\bibinfo {volume} {108}},\ \bibinfo {pages} {235702} (\bibinfo
  {year} {2012})}\BibitemShut {NoStop}%
\bibitem [{\citenamefont {Redner}\ \emph {et~al.}(2013)\citenamefont {Redner},
  \citenamefont {Hagan},\ and\ \citenamefont {Baskaran}}]{Redner2013}%
  \BibitemOpen
  \bibfield  {author} {\bibinfo {author} {\bibfnamefont {Gabriel~S}\
  \bibnamefont {Redner}}, \bibinfo {author} {\bibfnamefont {Michael~F}\
  \bibnamefont {Hagan}}, \ and\ \bibinfo {author} {\bibfnamefont {Aparna}\
  \bibnamefont {Baskaran}},\ }\bibfield  {title} {\enquote {\bibinfo {title}
  {Structure and dynamics of a phase-separating active colloidal fluid},}\
  }\href@noop {} {\bibfield  {journal} {\bibinfo  {journal} {Phys. Rev. Lett.}\
  }\textbf {\bibinfo {volume} {110}},\ \bibinfo {pages} {055701} (\bibinfo
  {year} {2013})}\BibitemShut {NoStop}%
\bibitem [{\citenamefont {Cates}\ and\ \citenamefont
  {Tailleur}(2015)}]{Cates2015}%
  \BibitemOpen
  \bibfield  {author} {\bibinfo {author} {\bibfnamefont {Michael~E}\
  \bibnamefont {Cates}}\ and\ \bibinfo {author} {\bibfnamefont {Julien}\
  \bibnamefont {Tailleur}},\ }\bibfield  {title} {\enquote {\bibinfo {title}
  {Motility-induced phase separation},}\ }\href@noop {} {\bibfield  {journal}
  {\bibinfo  {journal} {Annu. Rev. Condens. Matter Phys.}\ }\textbf {\bibinfo
  {volume} {6}},\ \bibinfo {pages} {219--244} (\bibinfo {year}
  {2015})}\BibitemShut {NoStop}%
\bibitem [{\citenamefont {Lauga}\ and\ \citenamefont
  {Powers}(2009)}]{Lauga2009}%
  \BibitemOpen
  \bibfield  {author} {\bibinfo {author} {\bibfnamefont {Eric}\ \bibnamefont
  {Lauga}}\ and\ \bibinfo {author} {\bibfnamefont {Thomas~R}\ \bibnamefont
  {Powers}},\ }\bibfield  {title} {\enquote {\bibinfo {title} {The
  hydrodynamics of swimming microorganisms},}\ }\href@noop {} {\bibfield
  {journal} {\bibinfo  {journal} {Rep. Progr. Phys.}\ }\textbf {\bibinfo
  {volume} {72}},\ \bibinfo {pages} {096601} (\bibinfo {year}
  {2009})}\BibitemShut {NoStop}%
\bibitem [{\citenamefont {Koch}\ and\ \citenamefont
  {Subramanian}(2011)}]{Koch2011}%
  \BibitemOpen
  \bibfield  {author} {\bibinfo {author} {\bibfnamefont {Donald~L.}\
  \bibnamefont {Koch}}\ and\ \bibinfo {author} {\bibfnamefont {Ganesh}\
  \bibnamefont {Subramanian}},\ }\bibfield  {title} {\enquote {\bibinfo {title}
  {Collective hydrodynamics of swimming microorganisms: Living fluids},}\
  }\href@noop {} {\bibfield  {journal} {\bibinfo  {journal} {Annu. Rev. Fluid
  Mech.}\ }\textbf {\bibinfo {volume} {43}},\ \bibinfo {pages} {637--659}
  (\bibinfo {year} {2011})}\BibitemShut {NoStop}%
\bibitem [{\citenamefont {Elgeti}\ \emph {et~al.}(2015)\citenamefont {Elgeti},
  \citenamefont {Winkler},\ and\ \citenamefont {Gompper}}]{Elgeti2015}%
  \BibitemOpen
  \bibfield  {author} {\bibinfo {author} {\bibfnamefont {Jens}\ \bibnamefont
  {Elgeti}}, \bibinfo {author} {\bibfnamefont {Roland~G}\ \bibnamefont
  {Winkler}}, \ and\ \bibinfo {author} {\bibfnamefont {Gerhard}\ \bibnamefont
  {Gompper}},\ }\bibfield  {title} {\enquote {\bibinfo {title} {Physics of
  microswimmers?single particle motion and collective behavior: a review},}\
  }\href@noop {} {\bibfield  {journal} {\bibinfo  {journal} {Rep. Progr.
  Phys.}\ }\textbf {\bibinfo {volume} {78}},\ \bibinfo {pages} {056601}
  (\bibinfo {year} {2015})}\BibitemShut {NoStop}%
\bibitem [{\citenamefont {Poujade}\ \emph {et~al.}(2007)\citenamefont
  {Poujade}, \citenamefont {Grasland-Mongrain}, \citenamefont {Hertzog},
  \citenamefont {Jouanneau}, \citenamefont {Chavrier}, \citenamefont {Ladoux},
  \citenamefont {Buguin},\ and\ \citenamefont {Silberzan}}]{Poujade2007}%
  \BibitemOpen
  \bibfield  {author} {\bibinfo {author} {\bibfnamefont {M.}~\bibnamefont
  {Poujade}}, \bibinfo {author} {\bibfnamefont {E.}~\bibnamefont
  {Grasland-Mongrain}}, \bibinfo {author} {\bibfnamefont {A.}~\bibnamefont
  {Hertzog}}, \bibinfo {author} {\bibfnamefont {J.}~\bibnamefont {Jouanneau}},
  \bibinfo {author} {\bibfnamefont {P.}~\bibnamefont {Chavrier}}, \bibinfo
  {author} {\bibfnamefont {B.}~\bibnamefont {Ladoux}}, \bibinfo {author}
  {\bibfnamefont {A.}~\bibnamefont {Buguin}}, \ and\ \bibinfo {author}
  {\bibfnamefont {P.}~\bibnamefont {Silberzan}},\ }\bibfield  {title} {\enquote
  {\bibinfo {title} {Collective migration of an epithelial monolayer in
  response to a model wound},}\ }\href@noop {} {\bibfield  {journal} {\bibinfo
  {journal} {Proc. Natl. Acad. Sci.}\ }\textbf {\bibinfo {volume} {104}},\
  \bibinfo {pages} {15988--15993} (\bibinfo {year} {2007})}\BibitemShut
  {NoStop}%
\bibitem [{\citenamefont {Saw}\ \emph {et~al.}(2017)\citenamefont {Saw},
  \citenamefont {Doostmohammadi}, \citenamefont {Nier}, \citenamefont
  {Kocgozlu}, \citenamefont {Thampi}, \citenamefont {Toyama}, \citenamefont
  {Marcq}, \citenamefont {Lim}, \citenamefont {Yeomans},\ and\ \citenamefont
  {Ladoux}}]{Saw2017}%
  \BibitemOpen
  \bibfield  {author} {\bibinfo {author} {\bibfnamefont {Thuan~Beng}\
  \bibnamefont {Saw}}, \bibinfo {author} {\bibfnamefont {Amin}\ \bibnamefont
  {Doostmohammadi}}, \bibinfo {author} {\bibfnamefont {Vincent}\ \bibnamefont
  {Nier}}, \bibinfo {author} {\bibfnamefont {Leyla}\ \bibnamefont {Kocgozlu}},
  \bibinfo {author} {\bibfnamefont {Sumesh}\ \bibnamefont {Thampi}}, \bibinfo
  {author} {\bibfnamefont {Yusuke}\ \bibnamefont {Toyama}}, \bibinfo {author}
  {\bibfnamefont {Philippe}\ \bibnamefont {Marcq}}, \bibinfo {author}
  {\bibfnamefont {Chwee~Teck}\ \bibnamefont {Lim}}, \bibinfo {author}
  {\bibfnamefont {Julia~M}\ \bibnamefont {Yeomans}}, \ and\ \bibinfo {author}
  {\bibfnamefont {Benoit}\ \bibnamefont {Ladoux}},\ }\bibfield  {title}
  {\enquote {\bibinfo {title} {Topological defects in epithelia govern cell
  death and extrusion},}\ }\href@noop {} {\bibfield  {journal} {\bibinfo
  {journal} {Nature}\ }\textbf {\bibinfo {volume} {544}},\ \bibinfo {pages}
  {212} (\bibinfo {year} {2017})}\BibitemShut {NoStop}%
\bibitem [{\citenamefont {Kawaguchi}\ \emph {et~al.}(2017)\citenamefont
  {Kawaguchi}, \citenamefont {Kageyama},\ and\ \citenamefont
  {Sano}}]{Kawaguchi2017}%
  \BibitemOpen
  \bibfield  {author} {\bibinfo {author} {\bibfnamefont {Kyogo}\ \bibnamefont
  {Kawaguchi}}, \bibinfo {author} {\bibfnamefont {Ryoichiro}\ \bibnamefont
  {Kageyama}}, \ and\ \bibinfo {author} {\bibfnamefont {Masaki}\ \bibnamefont
  {Sano}},\ }\bibfield  {title} {\enquote {\bibinfo {title} {Topological
  defects control collective dynamics in neural progenitor cell cultures},}\
  }\href@noop {} {\bibfield  {journal} {\bibinfo  {journal} {Nature}\ }\textbf
  {\bibinfo {volume} {545}},\ \bibinfo {pages} {327} (\bibinfo {year}
  {2017})}\BibitemShut {NoStop}%
\bibitem [{\citenamefont {Blanch-Mercader}\ \emph {et~al.}(2018)\citenamefont
  {Blanch-Mercader}, \citenamefont {Yashunsky}, \citenamefont {Garcia},
  \citenamefont {Duclos}, \citenamefont {Giomi},\ and\ \citenamefont
  {Silberzan}}]{Blanch2018}%
  \BibitemOpen
  \bibfield  {author} {\bibinfo {author} {\bibfnamefont {C}~\bibnamefont
  {Blanch-Mercader}}, \bibinfo {author} {\bibfnamefont {V}~\bibnamefont
  {Yashunsky}}, \bibinfo {author} {\bibfnamefont {S}~\bibnamefont {Garcia}},
  \bibinfo {author} {\bibfnamefont {G}~\bibnamefont {Duclos}}, \bibinfo
  {author} {\bibfnamefont {L}~\bibnamefont {Giomi}}, \ and\ \bibinfo {author}
  {\bibfnamefont {P}~\bibnamefont {Silberzan}},\ }\bibfield  {title} {\enquote
  {\bibinfo {title} {Turbulent dynamics of epithelial cell cultures},}\
  }\href@noop {} {\bibfield  {journal} {\bibinfo  {journal} {Phys. Rev. Lett.}\
  }\textbf {\bibinfo {volume} {120}},\ \bibinfo {pages} {208101} (\bibinfo
  {year} {2018})}\BibitemShut {NoStop}%
\bibitem [{\citenamefont {Xi}\ \emph {et~al.}(2018)\citenamefont {Xi},
  \citenamefont {Saw}, \citenamefont {Delacour}, \citenamefont {Lim},\ and\
  \citenamefont {Ladoux}}]{Xi2018}%
  \BibitemOpen
  \bibfield  {author} {\bibinfo {author} {\bibfnamefont {Wang}\ \bibnamefont
  {Xi}}, \bibinfo {author} {\bibfnamefont {Thuan~Beng}\ \bibnamefont {Saw}},
  \bibinfo {author} {\bibfnamefont {Delphine}\ \bibnamefont {Delacour}},
  \bibinfo {author} {\bibfnamefont {Chwee~Teck}\ \bibnamefont {Lim}}, \ and\
  \bibinfo {author} {\bibfnamefont {Benoit}\ \bibnamefont {Ladoux}},\
  }\bibfield  {title} {\enquote {\bibinfo {title} {Material approaches to
  active tissue mechanics},}\ }\href@noop {} {\bibfield  {journal} {\bibinfo
  {journal} {Nat. Rev. Mat.}\ }\textbf {\bibinfo {volume} {4}},\ \bibinfo
  {pages} {23--44} (\bibinfo {year} {2018})}\BibitemShut {NoStop}%
\bibitem [{\citenamefont {Trepat}\ and\ \citenamefont
  {Sahai}(2018)}]{Trepat2018}%
  \BibitemOpen
  \bibfield  {author} {\bibinfo {author} {\bibfnamefont {Xavier}\ \bibnamefont
  {Trepat}}\ and\ \bibinfo {author} {\bibfnamefont {Erik}\ \bibnamefont
  {Sahai}},\ }\bibfield  {title} {\enquote {\bibinfo {title} {Mesoscale
  physical principles of collective cell organization},}\ }\href@noop {}
  {\bibfield  {journal} {\bibinfo  {journal} {Nat. Phys.}\ }\textbf {\bibinfo
  {volume} {14}},\ \bibinfo {pages} {671--682} (\bibinfo {year}
  {2018})}\BibitemShut {NoStop}%
\bibitem [{\citenamefont {MacKintosh}\ and\ \citenamefont
  {Schmidt}(2010)}]{Mackintosh2010}%
  \BibitemOpen
  \bibfield  {author} {\bibinfo {author} {\bibfnamefont {Frederick~C}\
  \bibnamefont {MacKintosh}}\ and\ \bibinfo {author} {\bibfnamefont
  {Christoph~F}\ \bibnamefont {Schmidt}},\ }\bibfield  {title} {\enquote
  {\bibinfo {title} {Active cellular materials},}\ }\href@noop {} {\bibfield
  {journal} {\bibinfo  {journal} {Curr. Opin. Cell Biol.}\ }\textbf {\bibinfo
  {volume} {22}},\ \bibinfo {pages} {29--35} (\bibinfo {year}
  {2010})}\BibitemShut {NoStop}%
\bibitem [{\citenamefont {Prost}\ \emph {et~al.}(2015)\citenamefont {Prost},
  \citenamefont {J{\"u}licher},\ and\ \citenamefont {Joanny}}]{Prost2015}%
  \BibitemOpen
  \bibfield  {author} {\bibinfo {author} {\bibfnamefont {Jacques}\ \bibnamefont
  {Prost}}, \bibinfo {author} {\bibfnamefont {Frank}\ \bibnamefont
  {J{\"u}licher}}, \ and\ \bibinfo {author} {\bibfnamefont
  {Jean-Fran{\c{c}}ois}\ \bibnamefont {Joanny}},\ }\bibfield  {title} {\enquote
  {\bibinfo {title} {Active gel physics},}\ }\href@noop {} {\bibfield
  {journal} {\bibinfo  {journal} {Nat. Phys.}\ }\textbf {\bibinfo {volume}
  {11}},\ \bibinfo {pages} {111} (\bibinfo {year} {2015})}\BibitemShut
  {NoStop}%
\bibitem [{\citenamefont {Helbing}(2001)}]{Helbing2001}%
  \BibitemOpen
  \bibfield  {author} {\bibinfo {author} {\bibfnamefont {Dirk}\ \bibnamefont
  {Helbing}},\ }\bibfield  {title} {\enquote {\bibinfo {title} {Traffic and
  related self-driven many-particle systems},}\ }\href@noop {} {\bibfield
  {journal} {\bibinfo  {journal} {Rev. Mod. Phys.}\ }\textbf {\bibinfo {volume}
  {73}},\ \bibinfo {pages} {1067} (\bibinfo {year} {2001})}\BibitemShut
  {NoStop}%
\bibitem [{\citenamefont {Buhl}\ \emph {et~al.}(2006)\citenamefont {Buhl},
  \citenamefont {Sumpter}, \citenamefont {Couzin}, \citenamefont {Hale},
  \citenamefont {Despland}, \citenamefont {Miller},\ and\ \citenamefont
  {Simpson}}]{Buhl2006}%
  \BibitemOpen
  \bibfield  {author} {\bibinfo {author} {\bibfnamefont {Jerome}\ \bibnamefont
  {Buhl}}, \bibinfo {author} {\bibfnamefont {David~JT}\ \bibnamefont
  {Sumpter}}, \bibinfo {author} {\bibfnamefont {Iain~D}\ \bibnamefont
  {Couzin}}, \bibinfo {author} {\bibfnamefont {Joe~J}\ \bibnamefont {Hale}},
  \bibinfo {author} {\bibfnamefont {Emma}\ \bibnamefont {Despland}}, \bibinfo
  {author} {\bibfnamefont {Edgar~R}\ \bibnamefont {Miller}}, \ and\ \bibinfo
  {author} {\bibfnamefont {Steve~J}\ \bibnamefont {Simpson}},\ }\bibfield
  {title} {\enquote {\bibinfo {title} {From disorder to order in marching
  locusts},}\ }\href@noop {} {\bibfield  {journal} {\bibinfo  {journal}
  {Science}\ }\textbf {\bibinfo {volume} {312}},\ \bibinfo {pages} {1402--1406}
  (\bibinfo {year} {2006})}\BibitemShut {NoStop}%
\bibitem [{\citenamefont {Ballerini}\ \emph {et~al.}(2008)\citenamefont
  {Ballerini}, \citenamefont {Cabibbo}, \citenamefont {Candelier},
  \citenamefont {Cavagna}, \citenamefont {Cisbani}, \citenamefont {Giardina},
  \citenamefont {Lecomte}, \citenamefont {Orlandi}, \citenamefont {Parisi},
  \citenamefont {Procaccini}, \citenamefont {Viale},\ and\ \citenamefont
  {Zdravkovic}}]{Ballerini2008}%
  \BibitemOpen
  \bibfield  {author} {\bibinfo {author} {\bibfnamefont {Michele}\ \bibnamefont
  {Ballerini}}, \bibinfo {author} {\bibfnamefont {Nicola}\ \bibnamefont
  {Cabibbo}}, \bibinfo {author} {\bibfnamefont {Raphael}\ \bibnamefont
  {Candelier}}, \bibinfo {author} {\bibfnamefont {Andrea}\ \bibnamefont
  {Cavagna}}, \bibinfo {author} {\bibfnamefont {Evaristo}\ \bibnamefont
  {Cisbani}}, \bibinfo {author} {\bibfnamefont {Irene}\ \bibnamefont
  {Giardina}}, \bibinfo {author} {\bibfnamefont {Vivien}\ \bibnamefont
  {Lecomte}}, \bibinfo {author} {\bibfnamefont {Alberto}\ \bibnamefont
  {Orlandi}}, \bibinfo {author} {\bibfnamefont {Giorgio}\ \bibnamefont
  {Parisi}}, \bibinfo {author} {\bibfnamefont {Andrea}\ \bibnamefont
  {Procaccini}}, \bibinfo {author} {\bibfnamefont {M}~\bibnamefont {Viale}}, \
  and\ \bibinfo {author} {\bibfnamefont {V}~\bibnamefont {Zdravkovic}},\
  }\bibfield  {title} {\enquote {\bibinfo {title} {Interaction ruling animal
  collective behavior depends on topological rather than metric distance:
  Evidence from a field study},}\ }\href@noop {} {\bibfield  {journal}
  {\bibinfo  {journal} {Proc. Natl. Acad. Sci.}\ }\textbf {\bibinfo {volume}
  {105}},\ \bibinfo {pages} {1232--1237} (\bibinfo {year} {2008})}\BibitemShut
  {NoStop}%
\bibitem [{\citenamefont {Feinerman}\ \emph {et~al.}(2018)\citenamefont
  {Feinerman}, \citenamefont {Pinkoviezky}, \citenamefont {Gelblum},
  \citenamefont {Fonio},\ and\ \citenamefont {Gov}}]{Feinerman2018}%
  \BibitemOpen
  \bibfield  {author} {\bibinfo {author} {\bibfnamefont {Ofer}\ \bibnamefont
  {Feinerman}}, \bibinfo {author} {\bibfnamefont {Itai}\ \bibnamefont
  {Pinkoviezky}}, \bibinfo {author} {\bibfnamefont {Aviram}\ \bibnamefont
  {Gelblum}}, \bibinfo {author} {\bibfnamefont {Ehud}\ \bibnamefont {Fonio}}, \
  and\ \bibinfo {author} {\bibfnamefont {Nir~S}\ \bibnamefont {Gov}},\
  }\bibfield  {title} {\enquote {\bibinfo {title} {The physics of cooperative
  transport in groups of ants},}\ }\href@noop {} {\bibfield  {journal}
  {\bibinfo  {journal} {Nat. Phys.}\ }\textbf {\bibinfo {volume} {14}},\
  \bibinfo {pages} {683--693} (\bibinfo {year} {2018})}\BibitemShut {NoStop}%
\bibitem [{\citenamefont {Bechinger}\ \emph {et~al.}(2016)\citenamefont
  {Bechinger}, \citenamefont {Di~Leonardo}, \citenamefont {L{\"o}wen},
  \citenamefont {Reichhardt}, \citenamefont {Volpe},\ and\ \citenamefont
  {Volpe}}]{Bechinger2016}%
  \BibitemOpen
  \bibfield  {author} {\bibinfo {author} {\bibfnamefont {Clemens}\ \bibnamefont
  {Bechinger}}, \bibinfo {author} {\bibfnamefont {Roberto}\ \bibnamefont
  {Di~Leonardo}}, \bibinfo {author} {\bibfnamefont {Hartmut}\ \bibnamefont
  {L{\"o}wen}}, \bibinfo {author} {\bibfnamefont {Charles}\ \bibnamefont
  {Reichhardt}}, \bibinfo {author} {\bibfnamefont {Giorgio}\ \bibnamefont
  {Volpe}}, \ and\ \bibinfo {author} {\bibfnamefont {Giovanni}\ \bibnamefont
  {Volpe}},\ }\bibfield  {title} {\enquote {\bibinfo {title} {Active particles
  in complex and crowded environments},}\ }\href@noop {} {\bibfield  {journal}
  {\bibinfo  {journal} {Rev. Mod. Phys.}\ }\textbf {\bibinfo {volume} {88}},\
  \bibinfo {pages} {045006} (\bibinfo {year} {2016})}\BibitemShut {NoStop}%
\bibitem [{\citenamefont {Brambilla}\ \emph {et~al.}(2013)\citenamefont
  {Brambilla}, \citenamefont {Ferrante}, \citenamefont {Birattari},\ and\
  \citenamefont {Dorigo}}]{Brambilla2013}%
  \BibitemOpen
  \bibfield  {author} {\bibinfo {author} {\bibfnamefont {Manuele}\ \bibnamefont
  {Brambilla}}, \bibinfo {author} {\bibfnamefont {Eliseo}\ \bibnamefont
  {Ferrante}}, \bibinfo {author} {\bibfnamefont {Mauro}\ \bibnamefont
  {Birattari}}, \ and\ \bibinfo {author} {\bibfnamefont {Marco}\ \bibnamefont
  {Dorigo}},\ }\bibfield  {title} {\enquote {\bibinfo {title} {Swarm robotics:
  a review from the swarm engineering perspective},}\ }\href@noop {} {\bibfield
   {journal} {\bibinfo  {journal} {Swarm Intell.}\ }\textbf {\bibinfo {volume}
  {7}},\ \bibinfo {pages} {1--41} (\bibinfo {year} {2013})}\BibitemShut
  {NoStop}%
\bibitem [{\citenamefont {Ndlec}\ \emph {et~al.}(1997)\citenamefont {Ndlec},
  \citenamefont {Surrey}, \citenamefont {Maggs},\ and\ \citenamefont
  {Leibler}}]{Nedlec1997}%
  \BibitemOpen
  \bibfield  {author} {\bibinfo {author} {\bibfnamefont {FJ}~\bibnamefont
  {Ndlec}}, \bibinfo {author} {\bibfnamefont {Thomas}\ \bibnamefont {Surrey}},
  \bibinfo {author} {\bibfnamefont {Anthony~C}\ \bibnamefont {Maggs}}, \ and\
  \bibinfo {author} {\bibfnamefont {Stanislas}\ \bibnamefont {Leibler}},\
  }\bibfield  {title} {\enquote {\bibinfo {title} {Self-organization of
  microtubules and motors},}\ }\href@noop {} {\bibfield  {journal} {\bibinfo
  {journal} {Nature}\ }\textbf {\bibinfo {volume} {389}},\ \bibinfo {pages}
  {305} (\bibinfo {year} {1997})}\BibitemShut {NoStop}%
\bibitem [{\citenamefont {Schaller}\ \emph {et~al.}(2010)\citenamefont
  {Schaller}, \citenamefont {Weber}, \citenamefont {Semmrich}, \citenamefont
  {Frey},\ and\ \citenamefont {Bausch}}]{Schaller2010}%
  \BibitemOpen
  \bibfield  {author} {\bibinfo {author} {\bibfnamefont {Volker}\ \bibnamefont
  {Schaller}}, \bibinfo {author} {\bibfnamefont {Christoph}\ \bibnamefont
  {Weber}}, \bibinfo {author} {\bibfnamefont {Christine}\ \bibnamefont
  {Semmrich}}, \bibinfo {author} {\bibfnamefont {Erwin}\ \bibnamefont {Frey}},
  \ and\ \bibinfo {author} {\bibfnamefont {Andreas~R}\ \bibnamefont {Bausch}},\
  }\bibfield  {title} {\enquote {\bibinfo {title} {Polar patterns of driven
  filaments},}\ }\href@noop {} {\bibfield  {journal} {\bibinfo  {journal}
  {Nature}\ }\textbf {\bibinfo {volume} {467}},\ \bibinfo {pages} {73}
  (\bibinfo {year} {2010})}\BibitemShut {NoStop}%
\bibitem [{\citenamefont {Butt}\ \emph {et~al.}(2010)\citenamefont {Butt},
  \citenamefont {Mufti}, \citenamefont {Humayun}, \citenamefont {Rosenthal},
  \citenamefont {Khan}, \citenamefont {Khan},\ and\ \citenamefont
  {Molloy}}]{Butt2010}%
  \BibitemOpen
  \bibfield  {author} {\bibinfo {author} {\bibfnamefont {Tariq}\ \bibnamefont
  {Butt}}, \bibinfo {author} {\bibfnamefont {Tabish}\ \bibnamefont {Mufti}},
  \bibinfo {author} {\bibfnamefont {Ahmad}\ \bibnamefont {Humayun}}, \bibinfo
  {author} {\bibfnamefont {Peter~B}\ \bibnamefont {Rosenthal}}, \bibinfo
  {author} {\bibfnamefont {Sohaib}\ \bibnamefont {Khan}}, \bibinfo {author}
  {\bibfnamefont {Shahid}\ \bibnamefont {Khan}}, \ and\ \bibinfo {author}
  {\bibfnamefont {Justin~E}\ \bibnamefont {Molloy}},\ }\bibfield  {title}
  {\enquote {\bibinfo {title} {Myosin motors drive long range alignment of
  actin filaments},}\ }\href@noop {} {\bibfield  {journal} {\bibinfo  {journal}
  {J. Biol. Chem.}\ }\textbf {\bibinfo {volume} {285}},\ \bibinfo {pages}
  {4964--4974} (\bibinfo {year} {2010})}\BibitemShut {NoStop}%
\bibitem [{\citenamefont {Chaikin}\ \emph {et~al.}(1995)\citenamefont
  {Chaikin}, \citenamefont {Lubensky},\ and\ \citenamefont
  {Witten}}]{Chaikin1995}%
  \BibitemOpen
  \bibfield  {author} {\bibinfo {author} {\bibfnamefont {Paul~M}\ \bibnamefont
  {Chaikin}}, \bibinfo {author} {\bibfnamefont {Tom~C}\ \bibnamefont
  {Lubensky}}, \ and\ \bibinfo {author} {\bibfnamefont {Thomas~A}\ \bibnamefont
  {Witten}},\ }\href@noop {} {\emph {\bibinfo {title} {Principles of condensed
  matter physics}}},\ Vol.~\bibinfo {volume} {1}\ (\bibinfo  {publisher}
  {Cambridge University Press},\ \bibinfo {year} {1995})\BibitemShut {NoStop}%
\bibitem [{\citenamefont {Bertin}\ \emph {et~al.}(2006)\citenamefont {Bertin},
  \citenamefont {Droz},\ and\ \citenamefont {Gr{\'e}goire}}]{Bertin2006}%
  \BibitemOpen
  \bibfield  {author} {\bibinfo {author} {\bibfnamefont {Eric}\ \bibnamefont
  {Bertin}}, \bibinfo {author} {\bibfnamefont {Michel}\ \bibnamefont {Droz}}, \
  and\ \bibinfo {author} {\bibfnamefont {Guillaume}\ \bibnamefont
  {Gr{\'e}goire}},\ }\bibfield  {title} {\enquote {\bibinfo {title} {Boltzmann
  and hydrodynamic description for self-propelled particles},}\ }\href@noop {}
  {\bibfield  {journal} {\bibinfo  {journal} {Phys. Rev. E}\ }\textbf {\bibinfo
  {volume} {74}},\ \bibinfo {pages} {022101} (\bibinfo {year}
  {2006})}\BibitemShut {NoStop}%
\bibitem [{\citenamefont {Peruani}\ \emph {et~al.}(2008)\citenamefont
  {Peruani}, \citenamefont {Deutsch},\ and\ \citenamefont
  {B{\"a}r}}]{Peruani2008}%
  \BibitemOpen
  \bibfield  {author} {\bibinfo {author} {\bibfnamefont {F.}~\bibnamefont
  {Peruani}}, \bibinfo {author} {\bibfnamefont {A.}~\bibnamefont {Deutsch}}, \
  and\ \bibinfo {author} {\bibfnamefont {M.}~\bibnamefont {B{\"a}r}},\
  }\bibfield  {title} {\enquote {\bibinfo {title} {A mean-field theory for
  self-propelled particles interacting by velocity alignment mechanisms},}\
  }\href@noop {} {\bibfield  {journal} {\bibinfo  {journal} {Eur. Phys. J.
  Spec. Top.}\ }\textbf {\bibinfo {volume} {157}},\ \bibinfo {pages} {111--122}
  (\bibinfo {year} {2008})}\BibitemShut {NoStop}%
\bibitem [{\citenamefont {Baskaran}\ and\ \citenamefont
  {Marchetti}(2008)}]{Baskaran2008}%
  \BibitemOpen
  \bibfield  {author} {\bibinfo {author} {\bibfnamefont {Aparna}\ \bibnamefont
  {Baskaran}}\ and\ \bibinfo {author} {\bibfnamefont {M~Cristina}\ \bibnamefont
  {Marchetti}},\ }\bibfield  {title} {\enquote {\bibinfo {title} {Hydrodynamics
  of self-propelled hard rods},}\ }\href@noop {} {\bibfield  {journal}
  {\bibinfo  {journal} {Phys. Rev. E}\ }\textbf {\bibinfo {volume} {77}},\
  \bibinfo {pages} {011920} (\bibinfo {year} {2008})}\BibitemShut {NoStop}%
\bibitem [{\citenamefont {Bertin}\ \emph {et~al.}(2009)\citenamefont {Bertin},
  \citenamefont {Droz},\ and\ \citenamefont {Gr{\'e}goire}}]{Bertin2009}%
  \BibitemOpen
  \bibfield  {author} {\bibinfo {author} {\bibfnamefont {Eric}\ \bibnamefont
  {Bertin}}, \bibinfo {author} {\bibfnamefont {Michel}\ \bibnamefont {Droz}}, \
  and\ \bibinfo {author} {\bibfnamefont {Guillaume}\ \bibnamefont
  {Gr{\'e}goire}},\ }\bibfield  {title} {\enquote {\bibinfo {title}
  {Hydrodynamic equations for self-propelled particles: microscopic derivation
  and stability analysis},}\ }\href@noop {} {\bibfield  {journal} {\bibinfo
  {journal} {J. Phys. A}\ }\textbf {\bibinfo {volume} {42}},\ \bibinfo {pages}
  {445001} (\bibinfo {year} {2009})}\BibitemShut {NoStop}%
\bibitem [{\citenamefont {Lee}(2010)}]{Lee2010}%
  \BibitemOpen
  \bibfield  {author} {\bibinfo {author} {\bibfnamefont {Chiu~Fan}\
  \bibnamefont {Lee}},\ }\bibfield  {title} {\enquote {\bibinfo {title}
  {Fluctuation-induced collective motion: A single-particle density
  analysis},}\ }\href@noop {} {\bibfield  {journal} {\bibinfo  {journal} {Phys.
  Rev. E}\ }\textbf {\bibinfo {volume} {81}},\ \bibinfo {pages} {031125}
  (\bibinfo {year} {2010})}\BibitemShut {NoStop}%
\bibitem [{\citenamefont {Peshkov}\ \emph {et~al.}(2014)\citenamefont
  {Peshkov}, \citenamefont {Bertin}, \citenamefont {Ginelli},\ and\
  \citenamefont {Chat{\'e}}}]{Peshkov2014}%
  \BibitemOpen
  \bibfield  {author} {\bibinfo {author} {\bibfnamefont {A.}~\bibnamefont
  {Peshkov}}, \bibinfo {author} {\bibfnamefont {E.}~\bibnamefont {Bertin}},
  \bibinfo {author} {\bibfnamefont {F.}~\bibnamefont {Ginelli}}, \ and\
  \bibinfo {author} {\bibfnamefont {H.}~\bibnamefont {Chat{\'e}}},\ }\bibfield
  {title} {\enquote {\bibinfo {title} {{Boltzmann-Ginzburg-Landau approach for
  continuous descriptions of generic {Vicsek}-like models}},}\ }\href@noop {}
  {\bibfield  {journal} {\bibinfo  {journal} {Eur. Phys. J. Spec. Top.}\
  }\textbf {\bibinfo {volume} {223}},\ \bibinfo {pages} {1315--1344} (\bibinfo
  {year} {2014})}\BibitemShut {NoStop}%
\bibitem [{\citenamefont {Th{\"u}roff}\ \emph {et~al.}(2014)\citenamefont
  {Th{\"u}roff}, \citenamefont {Weber},\ and\ \citenamefont
  {Frey}}]{Thuroff2014}%
  \BibitemOpen
  \bibfield  {author} {\bibinfo {author} {\bibfnamefont {Florian}\ \bibnamefont
  {Th{\"u}roff}}, \bibinfo {author} {\bibfnamefont {Christoph~A}\ \bibnamefont
  {Weber}}, \ and\ \bibinfo {author} {\bibfnamefont {Erwin}\ \bibnamefont
  {Frey}},\ }\bibfield  {title} {\enquote {\bibinfo {title} {Numerical
  treatment of the {Boltzmann} equation for self-propelled particle systems},}\
  }\href@noop {} {\bibfield  {journal} {\bibinfo  {journal} {Phys. Rev. X}\
  }\textbf {\bibinfo {volume} {4}},\ \bibinfo {pages} {041030} (\bibinfo {year}
  {2014})}\BibitemShut {NoStop}%
\bibitem [{\citenamefont {Bertin}(2017)}]{Bertin2017}%
  \BibitemOpen
  \bibfield  {author} {\bibinfo {author} {\bibfnamefont {Eric}\ \bibnamefont
  {Bertin}},\ }\bibfield  {title} {\enquote {\bibinfo {title} {Theoretical
  approaches to the steady-state statistical physics of interacting dissipative
  units},}\ }\href@noop {} {\bibfield  {journal} {\bibinfo  {journal} {J. Phys.
  A}\ }\textbf {\bibinfo {volume} {50}},\ \bibinfo {pages} {083001} (\bibinfo
  {year} {2017})}\BibitemShut {NoStop}%
\bibitem [{\citenamefont {Romanczuk}\ \emph {et~al.}(2012)\citenamefont
  {Romanczuk}, \citenamefont {B{\"a}r}, \citenamefont {Ebeling}, \citenamefont
  {Lindner},\ and\ \citenamefont {Schimansky-Geier}}]{Romanczuk2012}%
  \BibitemOpen
  \bibfield  {author} {\bibinfo {author} {\bibfnamefont {Pawel}\ \bibnamefont
  {Romanczuk}}, \bibinfo {author} {\bibfnamefont {Markus}\ \bibnamefont
  {B{\"a}r}}, \bibinfo {author} {\bibfnamefont {Werner}\ \bibnamefont
  {Ebeling}}, \bibinfo {author} {\bibfnamefont {Benjamin}\ \bibnamefont
  {Lindner}}, \ and\ \bibinfo {author} {\bibfnamefont {Lutz}\ \bibnamefont
  {Schimansky-Geier}},\ }\bibfield  {title} {\enquote {\bibinfo {title} {Active
  {Brownian} particles},}\ }\href@noop {} {\bibfield  {journal} {\bibinfo
  {journal} {Eur. Phys. J. Spec. Top.}\ }\textbf {\bibinfo {volume} {202}},\
  \bibinfo {pages} {1--162} (\bibinfo {year} {2012})}\BibitemShut {NoStop}%
\bibitem [{\citenamefont {Toner}(2012)}]{Toner2012}%
  \BibitemOpen
  \bibfield  {author} {\bibinfo {author} {\bibfnamefont {John}\ \bibnamefont
  {Toner}},\ }\bibfield  {title} {\enquote {\bibinfo {title} {Reanalysis of the
  hydrodynamic theory of fluid, polar-ordered flocks},}\ }\href@noop {}
  {\bibfield  {journal} {\bibinfo  {journal} {Phys. Rev. E}\ }\textbf {\bibinfo
  {volume} {86}},\ \bibinfo {pages} {031918} (\bibinfo {year}
  {2012})}\BibitemShut {NoStop}%
\bibitem [{\citenamefont {Simha}\ and\ \citenamefont
  {Ramaswamy}(2002)}]{Simha2002}%
  \BibitemOpen
  \bibfield  {author} {\bibinfo {author} {\bibfnamefont {R~Aditi}\ \bibnamefont
  {Simha}}\ and\ \bibinfo {author} {\bibfnamefont {Sriram}\ \bibnamefont
  {Ramaswamy}},\ }\bibfield  {title} {\enquote {\bibinfo {title} {Hydrodynamic
  fluctuations and instabilities in ordered suspensions of self-propelled
  particles},}\ }\href@noop {} {\bibfield  {journal} {\bibinfo  {journal}
  {Phys. Rev. Lett.}\ }\textbf {\bibinfo {volume} {89}},\ \bibinfo {pages}
  {058101} (\bibinfo {year} {2002})}\BibitemShut {NoStop}%
\bibitem [{\citenamefont {Hatwalne}\ \emph {et~al.}(2004)\citenamefont
  {Hatwalne}, \citenamefont {Ramaswamy}, \citenamefont {Rao},\ and\
  \citenamefont {Simha}}]{Hatwalne2004}%
  \BibitemOpen
  \bibfield  {author} {\bibinfo {author} {\bibfnamefont {Yashodhan}\
  \bibnamefont {Hatwalne}}, \bibinfo {author} {\bibfnamefont {Sriram}\
  \bibnamefont {Ramaswamy}}, \bibinfo {author} {\bibfnamefont {Madan}\
  \bibnamefont {Rao}}, \ and\ \bibinfo {author} {\bibfnamefont {R~Aditi}\
  \bibnamefont {Simha}},\ }\bibfield  {title} {\enquote {\bibinfo {title}
  {Rheology of active-particle suspensions},}\ }\href@noop {} {\bibfield
  {journal} {\bibinfo  {journal} {Phys. Rev. Lett.}\ }\textbf {\bibinfo
  {volume} {92}},\ \bibinfo {pages} {118101} (\bibinfo {year}
  {2004})}\BibitemShut {NoStop}%
\bibitem [{\citenamefont {Kruse}\ \emph {et~al.}(2004)\citenamefont {Kruse},
  \citenamefont {Joanny}, \citenamefont {J{\"u}licher}, \citenamefont {Prost},\
  and\ \citenamefont {Sekimoto}}]{Kruse2004}%
  \BibitemOpen
  \bibfield  {author} {\bibinfo {author} {\bibfnamefont {Karsten}\ \bibnamefont
  {Kruse}}, \bibinfo {author} {\bibfnamefont {Jean-Fran{\c{c}}ois}\
  \bibnamefont {Joanny}}, \bibinfo {author} {\bibfnamefont {Frank}\
  \bibnamefont {J{\"u}licher}}, \bibinfo {author} {\bibfnamefont {Jacques}\
  \bibnamefont {Prost}}, \ and\ \bibinfo {author} {\bibfnamefont {Ken}\
  \bibnamefont {Sekimoto}},\ }\bibfield  {title} {\enquote {\bibinfo {title}
  {Asters, vortices, and rotating spirals in active gels of polar filaments},}\
  }\href@noop {} {\bibfield  {journal} {\bibinfo  {journal} {Phys. Rev. Lett.}\
  }\textbf {\bibinfo {volume} {92}},\ \bibinfo {pages} {078101} (\bibinfo
  {year} {2004})}\BibitemShut {NoStop}%
\bibitem [{\citenamefont {Kruse}\ \emph {et~al.}(2005)\citenamefont {Kruse},
  \citenamefont {Joanny}, \citenamefont {J{\"u}licher}, \citenamefont {Prost},\
  and\ \citenamefont {Sekimoto}}]{Kruse2005}%
  \BibitemOpen
  \bibfield  {author} {\bibinfo {author} {\bibfnamefont {Karsten}\ \bibnamefont
  {Kruse}}, \bibinfo {author} {\bibfnamefont {Jean-Francois}\ \bibnamefont
  {Joanny}}, \bibinfo {author} {\bibfnamefont {Frank}\ \bibnamefont
  {J{\"u}licher}}, \bibinfo {author} {\bibfnamefont {Jacques}\ \bibnamefont
  {Prost}}, \ and\ \bibinfo {author} {\bibfnamefont {Ken}\ \bibnamefont
  {Sekimoto}},\ }\bibfield  {title} {\enquote {\bibinfo {title} {Generic theory
  of active polar gels: a paradigm for cytoskeletal dynamics},}\ }\href@noop {}
  {\bibfield  {journal} {\bibinfo  {journal} {Eur. Phys. J. E}\ }\textbf
  {\bibinfo {volume} {16}},\ \bibinfo {pages} {5--16} (\bibinfo {year}
  {2005})}\BibitemShut {NoStop}%
\bibitem [{\citenamefont {J{\"u}licher}\ \emph {et~al.}(2018)\citenamefont
  {J{\"u}licher}, \citenamefont {Grill},\ and\ \citenamefont
  {Salbreux}}]{Julicher2018}%
  \BibitemOpen
  \bibfield  {author} {\bibinfo {author} {\bibfnamefont {Frank}\ \bibnamefont
  {J{\"u}licher}}, \bibinfo {author} {\bibfnamefont {Stephan~W}\ \bibnamefont
  {Grill}}, \ and\ \bibinfo {author} {\bibfnamefont {Guillaume}\ \bibnamefont
  {Salbreux}},\ }\bibfield  {title} {\enquote {\bibinfo {title} {Hydrodynamic
  theory of active matter},}\ }\href@noop {} {\bibfield  {journal} {\bibinfo
  {journal} {Rep Prog Phys}\ }\textbf {\bibinfo {volume} {81}},\ \bibinfo
  {pages} {076601} (\bibinfo {year} {2018})}\BibitemShut {NoStop}%
\bibitem [{\citenamefont {Gardiner}(2009)}]{Gardiner2009}%
  \BibitemOpen
  \bibfield  {author} {\bibinfo {author} {\bibfnamefont {Crispin}\ \bibnamefont
  {Gardiner}},\ }\href@noop {} {\emph {\bibinfo {title} {{Stochastic Methods: A
  Handbook for the Natural and Social Sciences}}}}\ (\bibinfo  {publisher}
  {Springer, Berlin, Germany},\ \bibinfo {year} {2009})\BibitemShut {NoStop}%
\bibitem [{\citenamefont {Gr{\'e}goire}\ and\ \citenamefont
  {Chat{\'e}}(2004)}]{Gregoire2004}%
  \BibitemOpen
  \bibfield  {author} {\bibinfo {author} {\bibfnamefont {Guillaume}\
  \bibnamefont {Gr{\'e}goire}}\ and\ \bibinfo {author} {\bibfnamefont {Hugues}\
  \bibnamefont {Chat{\'e}}},\ }\bibfield  {title} {\enquote {\bibinfo {title}
  {Onset of collective and cohesive motion},}\ }\href@noop {} {\bibfield
  {journal} {\bibinfo  {journal} {Phys. Rev. Lett.}\ }\textbf {\bibinfo
  {volume} {92}},\ \bibinfo {pages} {025702} (\bibinfo {year}
  {2004})}\BibitemShut {NoStop}%
\bibitem [{\citenamefont {Chat{\'e}}\ \emph {et~al.}(2008)\citenamefont
  {Chat{\'e}}, \citenamefont {Ginelli}, \citenamefont {Gr{\'e}goire},\ and\
  \citenamefont {Raynaud}}]{Chate2008}%
  \BibitemOpen
  \bibfield  {author} {\bibinfo {author} {\bibfnamefont {Hugues}\ \bibnamefont
  {Chat{\'e}}}, \bibinfo {author} {\bibfnamefont {Francesco}\ \bibnamefont
  {Ginelli}}, \bibinfo {author} {\bibfnamefont {Guillaume}\ \bibnamefont
  {Gr{\'e}goire}}, \ and\ \bibinfo {author} {\bibfnamefont {Franck}\
  \bibnamefont {Raynaud}},\ }\bibfield  {title} {\enquote {\bibinfo {title}
  {Collective motion of self-propelled particles interacting without
  cohesion},}\ }\href@noop {} {\bibfield  {journal} {\bibinfo  {journal} {Phys.
  Rev. E}\ }\textbf {\bibinfo {volume} {77}},\ \bibinfo {pages} {046113}
  (\bibinfo {year} {2008})}\BibitemShut {NoStop}%
\bibitem [{\citenamefont {Solon}\ and\ \citenamefont
  {Tailleur}(2013)}]{Solon2013}%
  \BibitemOpen
  \bibfield  {author} {\bibinfo {author} {\bibfnamefont {A~P}\ \bibnamefont
  {Solon}}\ and\ \bibinfo {author} {\bibfnamefont {Julien}\ \bibnamefont
  {Tailleur}},\ }\bibfield  {title} {\enquote {\bibinfo {title} {Revisiting the
  flocking transition using active spins},}\ }\href@noop {} {\bibfield
  {journal} {\bibinfo  {journal} {Phys. Rev. Lett.}\ }\textbf {\bibinfo
  {volume} {111}},\ \bibinfo {pages} {078101} (\bibinfo {year}
  {2013})}\BibitemShut {NoStop}%
\bibitem [{\citenamefont {Solon}\ and\ \citenamefont
  {Tailleur}(2015)}]{Solon2015}%
  \BibitemOpen
  \bibfield  {author} {\bibinfo {author} {\bibfnamefont {Alexandre~P}\
  \bibnamefont {Solon}}\ and\ \bibinfo {author} {\bibfnamefont {Julien}\
  \bibnamefont {Tailleur}},\ }\bibfield  {title} {\enquote {\bibinfo {title}
  {Flocking with discrete symmetry: The two-dimensional active ising model},}\
  }\href@noop {} {\bibfield  {journal} {\bibinfo  {journal} {Phys. Rev. E}\
  }\textbf {\bibinfo {volume} {92}},\ \bibinfo {pages} {042119} (\bibinfo
  {year} {2015})}\BibitemShut {NoStop}%
\bibitem [{\citenamefont {Szabo}\ \emph {et~al.}(2006)\citenamefont {Szabo},
  \citenamefont {Sz{\"o}ll{\"o}si}, \citenamefont {G{\"o}nci}, \citenamefont
  {Jur{\'a}nyi}, \citenamefont {Selmeczi},\ and\ \citenamefont
  {Vicsek}}]{Szabo2006}%
  \BibitemOpen
  \bibfield  {author} {\bibinfo {author} {\bibfnamefont {Balint}\ \bibnamefont
  {Szabo}}, \bibinfo {author} {\bibfnamefont {GJ}~\bibnamefont
  {Sz{\"o}ll{\"o}si}}, \bibinfo {author} {\bibfnamefont {B}~\bibnamefont
  {G{\"o}nci}}, \bibinfo {author} {\bibfnamefont {Zs}~\bibnamefont
  {Jur{\'a}nyi}}, \bibinfo {author} {\bibfnamefont {David}\ \bibnamefont
  {Selmeczi}}, \ and\ \bibinfo {author} {\bibfnamefont {Tam{\'a}s}\
  \bibnamefont {Vicsek}},\ }\bibfield  {title} {\enquote {\bibinfo {title}
  {Phase transition in the collective migration of tissue cells: experiment and
  model},}\ }\href@noop {} {\bibfield  {journal} {\bibinfo  {journal} {Phys.
  Rev. E}\ }\textbf {\bibinfo {volume} {74}},\ \bibinfo {pages} {061908}
  (\bibinfo {year} {2006})}\BibitemShut {NoStop}%
\bibitem [{\citenamefont {Czir{\'o}k}\ \emph {et~al.}(1997)\citenamefont
  {Czir{\'o}k}, \citenamefont {Stanley},\ and\ \citenamefont
  {Vicsek}}]{Czirok1997}%
  \BibitemOpen
  \bibfield  {author} {\bibinfo {author} {\bibfnamefont {Andr{\'a}s}\
  \bibnamefont {Czir{\'o}k}}, \bibinfo {author} {\bibfnamefont {H~Eugene}\
  \bibnamefont {Stanley}}, \ and\ \bibinfo {author} {\bibfnamefont {Tam{\'a}s}\
  \bibnamefont {Vicsek}},\ }\bibfield  {title} {\enquote {\bibinfo {title}
  {Spontaneously ordered motion of self-propelled particles},}\ }\href@noop {}
  {\bibfield  {journal} {\bibinfo  {journal} {J. Phys. A}\ }\textbf {\bibinfo
  {volume} {30}},\ \bibinfo {pages} {1375} (\bibinfo {year}
  {1997})}\BibitemShut {NoStop}%
\bibitem [{\citenamefont {Czir{\'o}k}\ \emph {et~al.}(1999)\citenamefont
  {Czir{\'o}k}, \citenamefont {Barab{\'a}si},\ and\ \citenamefont
  {Vicsek}}]{Czirok1999}%
  \BibitemOpen
  \bibfield  {author} {\bibinfo {author} {\bibfnamefont {Andr{\'a}s}\
  \bibnamefont {Czir{\'o}k}}, \bibinfo {author} {\bibfnamefont
  {Albert-L{\'a}szl{\'o}}\ \bibnamefont {Barab{\'a}si}}, \ and\ \bibinfo
  {author} {\bibfnamefont {Tam{\'a}s}\ \bibnamefont {Vicsek}},\ }\bibfield
  {title} {\enquote {\bibinfo {title} {Collective motion of self-propelled
  particles: Kinetic phase transition in one dimension},}\ }\href@noop {}
  {\bibfield  {journal} {\bibinfo  {journal} {Phys. Rev. Lett.}\ }\textbf
  {\bibinfo {volume} {82}},\ \bibinfo {pages} {209} (\bibinfo {year}
  {1999})}\BibitemShut {NoStop}%
\bibitem [{\citenamefont {Czir{\'o}k}\ and\ \citenamefont
  {Vicsek}(2000)}]{Czirok2000}%
  \BibitemOpen
  \bibfield  {author} {\bibinfo {author} {\bibfnamefont {Andr{\'a}s}\
  \bibnamefont {Czir{\'o}k}}\ and\ \bibinfo {author} {\bibfnamefont
  {Tam{\'a}s}\ \bibnamefont {Vicsek}},\ }\bibfield  {title} {\enquote {\bibinfo
  {title} {Collective behavior of interacting self-propelled particles},}\
  }\href@noop {} {\bibfield  {journal} {\bibinfo  {journal} {Physica A}\
  }\textbf {\bibinfo {volume} {281}},\ \bibinfo {pages} {17--29} (\bibinfo
  {year} {2000})}\BibitemShut {NoStop}%
\bibitem [{\citenamefont {Aldana}\ \emph {et~al.}(2007)\citenamefont {Aldana},
  \citenamefont {Dossetti}, \citenamefont {Huepe}, \citenamefont {Kenkre},\
  and\ \citenamefont {Larralde}}]{Aldana2007}%
  \BibitemOpen
  \bibfield  {author} {\bibinfo {author} {\bibfnamefont {Maximino}\
  \bibnamefont {Aldana}}, \bibinfo {author} {\bibfnamefont {Victor}\
  \bibnamefont {Dossetti}}, \bibinfo {author} {\bibfnamefont {Christian}\
  \bibnamefont {Huepe}}, \bibinfo {author} {\bibfnamefont {VM}~\bibnamefont
  {Kenkre}}, \ and\ \bibinfo {author} {\bibfnamefont {Hern{\'a}n}\ \bibnamefont
  {Larralde}},\ }\bibfield  {title} {\enquote {\bibinfo {title} {Phase
  transitions in systems of self-propelled agents and related network
  models},}\ }\href@noop {} {\bibfield  {journal} {\bibinfo  {journal} {Phys.
  Rev. Lett.}\ }\textbf {\bibinfo {volume} {98}},\ \bibinfo {pages} {095702}
  (\bibinfo {year} {2007})}\BibitemShut {NoStop}%
\bibitem [{\citenamefont {G{\"o}nci}\ \emph {et~al.}(2008)\citenamefont
  {G{\"o}nci}, \citenamefont {Nagy},\ and\ \citenamefont {Vicsek}}]{Gonci2008}%
  \BibitemOpen
  \bibfield  {author} {\bibinfo {author} {\bibfnamefont {Bal{\'a}zs}\
  \bibnamefont {G{\"o}nci}}, \bibinfo {author} {\bibfnamefont {M{\'a}t{\'e}}\
  \bibnamefont {Nagy}}, \ and\ \bibinfo {author} {\bibfnamefont {Tam{\'a}s}\
  \bibnamefont {Vicsek}},\ }\bibfield  {title} {\enquote {\bibinfo {title}
  {Phase transition in the scalar noise model of collective motion in three
  dimensions},}\ }\href@noop {} {\bibfield  {journal} {\bibinfo  {journal}
  {Eur. Phys. J. Spec. Top.}\ }\textbf {\bibinfo {volume} {157}},\ \bibinfo
  {pages} {53--59} (\bibinfo {year} {2008})}\BibitemShut {NoStop}%
\bibitem [{\citenamefont {Ginelli}(2016)}]{Ginelli2016}%
  \BibitemOpen
  \bibfield  {author} {\bibinfo {author} {\bibfnamefont {Francesco}\
  \bibnamefont {Ginelli}},\ }\bibfield  {title} {\enquote {\bibinfo {title}
  {{The Physics of the Vicsek model}},}\ }\href@noop {} {\bibfield  {journal}
  {\bibinfo  {journal} {Eur. Phys. J. Spec. Top.}\ }\textbf {\bibinfo {volume}
  {225}},\ \bibinfo {pages} {2099--2117} (\bibinfo {year} {2016})}\BibitemShut
  {NoStop}%
\bibitem [{\citenamefont {Gr{\'e}goire}\ \emph
  {et~al.}(2001{\natexlab{a}})\citenamefont {Gr{\'e}goire}, \citenamefont
  {Chat{\'e}},\ and\ \citenamefont {Tu}}]{Gregoire2001a}%
  \BibitemOpen
  \bibfield  {author} {\bibinfo {author} {\bibfnamefont {Guillaume}\
  \bibnamefont {Gr{\'e}goire}}, \bibinfo {author} {\bibfnamefont {Hugues}\
  \bibnamefont {Chat{\'e}}}, \ and\ \bibinfo {author} {\bibfnamefont {Yuhai}\
  \bibnamefont {Tu}},\ }\bibfield  {title} {\enquote {\bibinfo {title} {Comment
  on" particle diffusion in a quasi-two-dimensional bacterial bath"},}\
  }\href@noop {} {\bibfield  {journal} {\bibinfo  {journal} {Phys. Rev. Lett.}\
  }\textbf {\bibinfo {volume} {86}},\ \bibinfo {pages} {556} (\bibinfo {year}
  {2001}{\natexlab{a}})}\BibitemShut {NoStop}%
\bibitem [{\citenamefont {Gr{\'e}goire}\ \emph
  {et~al.}(2001{\natexlab{b}})\citenamefont {Gr{\'e}goire}, \citenamefont
  {Chat{\'e}},\ and\ \citenamefont {Tu}}]{Gregoire2001}%
  \BibitemOpen
  \bibfield  {author} {\bibinfo {author} {\bibfnamefont {Guillaume}\
  \bibnamefont {Gr{\'e}goire}}, \bibinfo {author} {\bibfnamefont {Hugues}\
  \bibnamefont {Chat{\'e}}}, \ and\ \bibinfo {author} {\bibfnamefont {Yuhai}\
  \bibnamefont {Tu}},\ }\bibfield  {title} {\enquote {\bibinfo {title} {Active
  and passive particles: Modeling beads in a bacterial bath},}\ }\href@noop {}
  {\bibfield  {journal} {\bibinfo  {journal} {Phys. Rev. E}\ }\textbf {\bibinfo
  {volume} {64}},\ \bibinfo {pages} {011902} (\bibinfo {year}
  {2001}{\natexlab{b}})}\BibitemShut {NoStop}%
\bibitem [{\citenamefont {Ginelli}\ and\ \citenamefont
  {Chat{\'e}}(2010)}]{Ginelli2010}%
  \BibitemOpen
  \bibfield  {author} {\bibinfo {author} {\bibfnamefont {Francesco}\
  \bibnamefont {Ginelli}}\ and\ \bibinfo {author} {\bibfnamefont {Hugues}\
  \bibnamefont {Chat{\'e}}},\ }\bibfield  {title} {\enquote {\bibinfo {title}
  {Relevance of metric-free interactions in flocking phenomena},}\ }\href@noop
  {} {\bibfield  {journal} {\bibinfo  {journal} {Phys. Rev. Lett.}\ }\textbf
  {\bibinfo {volume} {105}},\ \bibinfo {pages} {168103} (\bibinfo {year}
  {2010})}\BibitemShut {NoStop}%
\bibitem [{\citenamefont {Peshkov}\ , \citenamefont {Ngo}\ , \citenamefont {Bertin}\ ,  \citenamefont {Chat{\'e}\ and \citenamefont {Ginelli}}(2012)}]{Peshkov2012}%
  \BibitemOpen
  \bibfield  {author} {\bibinfo {author} {\bibfnamefont {A.}\ \bibnamefont {Peshkov}}\ ,
  \bibinfo {author} {\bibfnamefont {S.}\ \bibnamefont {Ngo}}\ , \bibinfo {author} {\bibfnamefont {E.}\ \bibnamefont {Bertin}}\ , \bibinfo {author}  {\bibfnamefont {H.}\
  \bibnamefont {Chat{\'e}}} \ and\ \bibinfo {author} {\bibfnamefont {F.}\
  \bibnamefont {Ginelli}},\ } \bibfield  {title} {\enquote {\bibinfo {title}
  {Continuous theory of active matter systems with metric-free interactions},}\ }\href@noop
  {} {\bibfield  {journal} {\bibinfo  {journal} {Phys. Rev. Lett.}\ }\textbf
  {\bibinfo {volume} {109}},\ \bibinfo {pages} {098101} (\bibinfo {year}
  {2012})}\BibitemShut {NoStop}%
\bibitem [{\citenamefont {Ramaswamy}\ \emph {et~al.}(2003)\citenamefont
  {Ramaswamy}, \citenamefont {Simha},\ and\ \citenamefont
  {Toner}}]{Ramaswamy2003}%
  \BibitemOpen
  \bibfield  {author} {\bibinfo {author} {\bibfnamefont {Sriram}\ \bibnamefont
  {Ramaswamy}}, \bibinfo {author} {\bibfnamefont {R~Aditi}\ \bibnamefont
  {Simha}}, \ and\ \bibinfo {author} {\bibfnamefont {John}\ \bibnamefont
  {Toner}},\ }\bibfield  {title} {\enquote {\bibinfo {title} {Active nematics
  on a substrate: {Giant} number fluctuations and long-time tails},}\
  }\href@noop {} {\bibfield  {journal} {\bibinfo  {journal} {EPL}\ }\textbf
  {\bibinfo {volume} {62}},\ \bibinfo {pages} {196} (\bibinfo {year}
  {2003})}\BibitemShut {NoStop}%
\bibitem [{\citenamefont {Chat{\'e}}\ \emph {et~al.}(2006)\citenamefont
  {Chat{\'e}}, \citenamefont {Ginelli},\ and\ \citenamefont
  {Montagne}}]{Chate2006}%
  \BibitemOpen
  \bibfield  {author} {\bibinfo {author} {\bibfnamefont {Hugues}\ \bibnamefont
  {Chat{\'e}}}, \bibinfo {author} {\bibfnamefont {Francesco}\ \bibnamefont
  {Ginelli}}, \ and\ \bibinfo {author} {\bibfnamefont {Ra{\'u}l}\ \bibnamefont
  {Montagne}},\ }\bibfield  {title} {\enquote {\bibinfo {title} {Simple model
  for active nematics: quasi-long-range order and giant fluctuations},}\
  }\href@noop {} {\bibfield  {journal} {\bibinfo  {journal} {Phys. Rev. Lett.}\
  }\textbf {\bibinfo {volume} {96}},\ \bibinfo {pages} {180602} (\bibinfo
  {year} {2006})}\BibitemShut {NoStop}%
\bibitem [{\citenamefont {Ngo}\ \emph {et~al.}(2012)\citenamefont {Ngo},
  \citenamefont {Ginelli},\ and\ \citenamefont {Chat{\'e}}}]{Ngo2012}%
  \BibitemOpen
  \bibfield  {author} {\bibinfo {author} {\bibfnamefont {Sandrine}\
  \bibnamefont {Ngo}}, \bibinfo {author} {\bibfnamefont {Francesco}\
  \bibnamefont {Ginelli}}, \ and\ \bibinfo {author} {\bibfnamefont {Hugues}\
  \bibnamefont {Chat{\'e}}},\ }\bibfield  {title} {\enquote {\bibinfo {title}
  {Competing ferromagnetic and nematic alignment in self-propelled polar
  particles},}\ }\href@noop {} {\bibfield  {journal} {\bibinfo  {journal}
  {Phys. Rev. E}\ }\textbf {\bibinfo {volume} {86}},\ \bibinfo {pages}
  {050101(R)} (\bibinfo {year} {2012})}\BibitemShut {NoStop}%
\bibitem [{\citenamefont {Huber}\ \emph {et~al.}(2018)\citenamefont {Huber},
  \citenamefont {Suzuki}, \citenamefont {Kr{\"u}ger}, \citenamefont {Frey},\
  and\ \citenamefont {Bausch}}]{Huber2018}%
  \BibitemOpen
  \bibfield  {author} {\bibinfo {author} {\bibfnamefont {L}~\bibnamefont
  {Huber}}, \bibinfo {author} {\bibfnamefont {R}~\bibnamefont {Suzuki}},
  \bibinfo {author} {\bibfnamefont {T}~\bibnamefont {Kr{\"u}ger}}, \bibinfo
  {author} {\bibfnamefont {E}~\bibnamefont {Frey}}, \ and\ \bibinfo {author}
  {\bibfnamefont {AR}~\bibnamefont {Bausch}},\ }\bibfield  {title} {\enquote
  {\bibinfo {title} {Emergence of coexisting ordered states in active matter
  systems},}\ }\href@noop {} {\bibfield  {journal} {\bibinfo  {journal}
  {Science}\ }\textbf {\bibinfo {volume} {361}},\ \bibinfo {pages} {255--258}
  (\bibinfo {year} {2018})}\BibitemShut {NoStop}%
\bibitem [{\citenamefont {Cates}\ \emph {et~al.}(2010)\citenamefont {Cates},
  \citenamefont {Marenduzzo}, \citenamefont {Pagonabarraga},\ and\
  \citenamefont {Tailleur}}]{Cates2010}%
  \BibitemOpen
  \bibfield  {author} {\bibinfo {author} {\bibfnamefont {ME}~\bibnamefont
  {Cates}}, \bibinfo {author} {\bibfnamefont {D}~\bibnamefont {Marenduzzo}},
  \bibinfo {author} {\bibfnamefont {I}~\bibnamefont {Pagonabarraga}}, \ and\
  \bibinfo {author} {\bibfnamefont {J}~\bibnamefont {Tailleur}},\ }\bibfield
  {title} {\enquote {\bibinfo {title} {Arrested phase separation in reproducing
  bacteria creates a generic route to pattern formation},}\ }\href@noop {}
  {\bibfield  {journal} {\bibinfo  {journal} {Proc. Natl. Acad. Sci.}\ }\textbf
  {\bibinfo {volume} {107}},\ \bibinfo {pages} {11715--11720} (\bibinfo {year}
  {2010})}\BibitemShut {NoStop}%
\bibitem [{\citenamefont {Farrell}\ \emph {et~al.}(2012)\citenamefont
  {Farrell}, \citenamefont {Marchetti}, \citenamefont {Marenduzzo},\ and\
  \citenamefont {Tailleur}}]{Farrell2012}%
  \BibitemOpen
  \bibfield  {author} {\bibinfo {author} {\bibfnamefont {F~D~C}\ \bibnamefont
  {Farrell}}, \bibinfo {author} {\bibfnamefont {M~C}\ \bibnamefont
  {Marchetti}}, \bibinfo {author} {\bibfnamefont {D}~\bibnamefont
  {Marenduzzo}}, \ and\ \bibinfo {author} {\bibfnamefont {J}~\bibnamefont
  {Tailleur}},\ }\bibfield  {title} {\enquote {\bibinfo {title} {Pattern
  formation in self-propelled particles with density-dependent motility},}\
  }\href@noop {} {\bibfield  {journal} {\bibinfo  {journal} {Phys. Rev. Lett.}\
  }\textbf {\bibinfo {volume} {108}},\ \bibinfo {pages} {248101} (\bibinfo
  {year} {2012})}\BibitemShut {NoStop}%
\bibitem [{\citenamefont {Sumino}\ \emph {et~al.}(2012)\citenamefont {Sumino},
  \citenamefont {Nagai}, \citenamefont {Shitaka}, \citenamefont {Tanaka},
  \citenamefont {Yoshikawa}, \citenamefont {Chat{\'e}},\ and\ \citenamefont
  {Oiwa}}]{Sumino2012}%
  \BibitemOpen
  \bibfield  {author} {\bibinfo {author} {\bibfnamefont {Yutaka}\ \bibnamefont
  {Sumino}}, \bibinfo {author} {\bibfnamefont {Ken~H}\ \bibnamefont {Nagai}},
  \bibinfo {author} {\bibfnamefont {Yuji}\ \bibnamefont {Shitaka}}, \bibinfo
  {author} {\bibfnamefont {Dan}\ \bibnamefont {Tanaka}}, \bibinfo {author}
  {\bibfnamefont {Kenichi}\ \bibnamefont {Yoshikawa}}, \bibinfo {author}
  {\bibfnamefont {Hugues}\ \bibnamefont {Chat{\'e}}}, \ and\ \bibinfo {author}
  {\bibfnamefont {Kazuhiro}\ \bibnamefont {Oiwa}},\ }\bibfield  {title}
  {\enquote {\bibinfo {title} {Large-scale vortex lattice emerging from
  collectively moving microtubules},}\ }\href@noop {} {\bibfield  {journal}
  {\bibinfo  {journal} {Nature}\ }\textbf {\bibinfo {volume} {483}},\ \bibinfo
  {pages} {448} (\bibinfo {year} {2012})}\BibitemShut {NoStop}%
\bibitem [{\citenamefont {Attanasi}\ \emph {et~al.}(2014)\citenamefont
  {Attanasi}, \citenamefont {Cavagna}, \citenamefont {Del~Castello},
  \citenamefont {Giardina}, \citenamefont {Grigera}, \citenamefont {Jeli{\'c}},
  \citenamefont {Melillo}, \citenamefont {Parisi}, \citenamefont {Pohl},
  \citenamefont {Shen} \emph {et~al.}}]{Attanasi2014}%
  \BibitemOpen
  \bibfield  {author} {\bibinfo {author} {\bibfnamefont {Alessandro}\
  \bibnamefont {Attanasi}}, \bibinfo {author} {\bibfnamefont {Andrea}\
  \bibnamefont {Cavagna}}, \bibinfo {author} {\bibfnamefont {Lorenzo}\
  \bibnamefont {Del~Castello}}, \bibinfo {author} {\bibfnamefont {Irene}\
  \bibnamefont {Giardina}}, \bibinfo {author} {\bibfnamefont {Tomas~S}\
  \bibnamefont {Grigera}}, \bibinfo {author} {\bibfnamefont {Asja}\
  \bibnamefont {Jeli{\'c}}}, \bibinfo {author} {\bibfnamefont {Stefania}\
  \bibnamefont {Melillo}}, \bibinfo {author} {\bibfnamefont {Leonardo}\
  \bibnamefont {Parisi}}, \bibinfo {author} {\bibfnamefont {Oliver}\
  \bibnamefont {Pohl}}, \bibinfo {author} {\bibfnamefont {Edward}\ \bibnamefont
  {Shen}},  \emph {et~al.},\ }\bibfield  {title} {\enquote {\bibinfo {title}
  {Information transfer and behavioural inertia in starling flocks},}\
  }\href@noop {} {\bibfield  {journal} {\bibinfo  {journal} {Nat. Phys.}\
  }\textbf {\bibinfo {volume} {10}},\ \bibinfo {pages} {691} (\bibinfo {year}
  {2014})}\BibitemShut {NoStop}%
\bibitem [{\citenamefont {Cavagna}\ \emph {et~al.}(2015)\citenamefont
  {Cavagna}, \citenamefont {Del~Castello}, \citenamefont {Giardina},
  \citenamefont {Grigera}, \citenamefont {Jelic}, \citenamefont {Melillo},
  \citenamefont {Mora}, \citenamefont {Parisi}, \citenamefont {Silvestri},
  \citenamefont {Viale} \emph {et~al.}}]{Cavagna2015}%
  \BibitemOpen
  \bibfield  {author} {\bibinfo {author} {\bibfnamefont {Andrea}\ \bibnamefont
  {Cavagna}}, \bibinfo {author} {\bibfnamefont {Lorenzo}\ \bibnamefont
  {Del~Castello}}, \bibinfo {author} {\bibfnamefont {Irene}\ \bibnamefont
  {Giardina}}, \bibinfo {author} {\bibfnamefont {Tomas}\ \bibnamefont
  {Grigera}}, \bibinfo {author} {\bibfnamefont {Asja}\ \bibnamefont {Jelic}},
  \bibinfo {author} {\bibfnamefont {Stefania}\ \bibnamefont {Melillo}},
  \bibinfo {author} {\bibfnamefont {Thierry}\ \bibnamefont {Mora}}, \bibinfo
  {author} {\bibfnamefont {Leonardo}\ \bibnamefont {Parisi}}, \bibinfo {author}
  {\bibfnamefont {Edmondo}\ \bibnamefont {Silvestri}}, \bibinfo {author}
  {\bibfnamefont {Massimiliano}\ \bibnamefont {Viale}},  \emph {et~al.},\
  }\bibfield  {title} {\enquote {\bibinfo {title} {Flocking and turning: a new
  model for self-organized collective motion},}\ }\href@noop {} {\bibfield
  {journal} {\bibinfo  {journal} {J. Stat. Phys.}\ }\textbf {\bibinfo {volume}
  {158}},\ \bibinfo {pages} {601--627} (\bibinfo {year} {2015})}\BibitemShut
  {NoStop}%
\bibitem [{\citenamefont {Yang}\ and\ \citenamefont
  {Marchetti}(2015)}]{Yang2015}%
  \BibitemOpen
  \bibfield  {author} {\bibinfo {author} {\bibfnamefont {Xingbo}\ \bibnamefont
  {Yang}}\ and\ \bibinfo {author} {\bibfnamefont {M~Cristina}\ \bibnamefont
  {Marchetti}},\ }\bibfield  {title} {\enquote {\bibinfo {title} {Hydrodynamics
  of turning flocks},}\ }\href@noop {} {\bibfield  {journal} {\bibinfo
  {journal} {Phys. Rev. Lett.}\ }\textbf {\bibinfo {volume} {115}},\ \bibinfo
  {pages} {258101} (\bibinfo {year} {2015})}\BibitemShut {NoStop}%
\bibitem [{\citenamefont {Nagai}\ \emph {et~al.}(2015)\citenamefont {Nagai},
  \citenamefont {Sumino}, \citenamefont {Montagne}, \citenamefont {Aranson},\
  and\ \citenamefont {Chat{\'e}}}]{Nagai2015}%
  \BibitemOpen
  \bibfield  {author} {\bibinfo {author} {\bibfnamefont {Ken~H}\ \bibnamefont
  {Nagai}}, \bibinfo {author} {\bibfnamefont {Yutaka}\ \bibnamefont {Sumino}},
  \bibinfo {author} {\bibfnamefont {Raul}\ \bibnamefont {Montagne}}, \bibinfo
  {author} {\bibfnamefont {Igor~S}\ \bibnamefont {Aranson}}, \ and\ \bibinfo
  {author} {\bibfnamefont {Hugues}\ \bibnamefont {Chat{\'e}}},\ }\bibfield
  {title} {\enquote {\bibinfo {title} {Collective motion of self-propelled
  particles with memory},}\ }\href@noop {} {\bibfield  {journal} {\bibinfo
  {journal} {Phys. Rev. Lett.}\ }\textbf {\bibinfo {volume} {114}},\ \bibinfo
  {pages} {168001} (\bibinfo {year} {2015})}\BibitemShut {NoStop}%
\bibitem [{\citenamefont {K{\"u}mmel}\ \emph {et~al.}(2013)\citenamefont
  {K{\"u}mmel}, \citenamefont {ten Hagen}, \citenamefont {Wittkowski},
  \citenamefont {Buttinoni}, \citenamefont {Eichhorn}, \citenamefont {Volpe},
  \citenamefont {L{\"o}wen},\ and\ \citenamefont {Bechinger}}]{Kummel2013}%
  \BibitemOpen
  \bibfield  {author} {\bibinfo {author} {\bibfnamefont {Felix}\ \bibnamefont
  {K{\"u}mmel}}, \bibinfo {author} {\bibfnamefont {Borge}\ \bibnamefont {ten
  Hagen}}, \bibinfo {author} {\bibfnamefont {Raphael}\ \bibnamefont
  {Wittkowski}}, \bibinfo {author} {\bibfnamefont {Ivo}\ \bibnamefont
  {Buttinoni}}, \bibinfo {author} {\bibfnamefont {Ralf}\ \bibnamefont
  {Eichhorn}}, \bibinfo {author} {\bibfnamefont {Giovanni}\ \bibnamefont
  {Volpe}}, \bibinfo {author} {\bibfnamefont {Hartmut}\ \bibnamefont
  {L{\"o}wen}}, \ and\ \bibinfo {author} {\bibfnamefont {Clemens}\ \bibnamefont
  {Bechinger}},\ }\bibfield  {title} {\enquote {\bibinfo {title} {Circular
  motion of asymmetric self-propelling particles},}\ }\href@noop {} {\bibfield
  {journal} {\bibinfo  {journal} {Phys. Rev. Lett.}\ }\textbf {\bibinfo
  {volume} {110}},\ \bibinfo {pages} {198302} (\bibinfo {year}
  {2013})}\BibitemShut {NoStop}%
\bibitem [{\citenamefont {Kaiser}\ and\ \citenamefont
  {L{\"o}wen}(2013)}]{Kaiser2013}%
  \BibitemOpen
  \bibfield  {author} {\bibinfo {author} {\bibfnamefont {A}~\bibnamefont
  {Kaiser}}\ and\ \bibinfo {author} {\bibfnamefont {H}~\bibnamefont
  {L{\"o}wen}},\ }\bibfield  {title} {\enquote {\bibinfo {title} {Vortex arrays
  as emergent collective phenomena for circle swimmers},}\ }\href@noop {}
  {\bibfield  {journal} {\bibinfo  {journal} {Phys. Rev. E}\ }\textbf {\bibinfo
  {volume} {87}},\ \bibinfo {pages} {032712} (\bibinfo {year}
  {2013})}\BibitemShut {NoStop}%
\bibitem [{\citenamefont {Wensink}\ \emph {et~al.}(2014)\citenamefont
  {Wensink}, \citenamefont {Kantsler}, \citenamefont {Goldstein},\ and\
  \citenamefont {Dunkel}}]{Wensink2014}%
  \BibitemOpen
  \bibfield  {author} {\bibinfo {author} {\bibfnamefont {H~H}\ \bibnamefont
  {Wensink}}, \bibinfo {author} {\bibfnamefont {V}~\bibnamefont {Kantsler}},
  \bibinfo {author} {\bibfnamefont {R~E}\ \bibnamefont {Goldstein}}, \ and\
  \bibinfo {author} {\bibfnamefont {J}~\bibnamefont {Dunkel}},\ }\bibfield
  {title} {\enquote {\bibinfo {title} {Controlling active self-assembly through
  broken particle-shape symmetry},}\ }\href@noop {} {\bibfield  {journal}
  {\bibinfo  {journal} {Phys. Rev. E}\ }\textbf {\bibinfo {volume} {89}},\
  \bibinfo {pages} {010302(R)} (\bibinfo {year} {2014})}\BibitemShut {NoStop}%
\bibitem [{\citenamefont {Nguyen}\ \emph {et~al.}(2014)\citenamefont {Nguyen},
  \citenamefont {Klotsa}, \citenamefont {Engel},\ and\ \citenamefont
  {Glotzer}}]{Nguyen2014}%
  \BibitemOpen
  \bibfield  {author} {\bibinfo {author} {\bibfnamefont {Nguyen H~P}\
  \bibnamefont {Nguyen}}, \bibinfo {author} {\bibfnamefont {Daphne}\
  \bibnamefont {Klotsa}}, \bibinfo {author} {\bibfnamefont {Michael}\
  \bibnamefont {Engel}}, \ and\ \bibinfo {author} {\bibfnamefont {Sharon~C}\
  \bibnamefont {Glotzer}},\ }\bibfield  {title} {\enquote {\bibinfo {title}
  {Emergent collective phenomena in a mixture of hard shapes through active
  rotation},}\ }\href@noop {} {\bibfield  {journal} {\bibinfo  {journal} {Phys.
  Rev. Lett.}\ }\textbf {\bibinfo {volume} {112}},\ \bibinfo {pages} {075701}
  (\bibinfo {year} {2014})}\BibitemShut {NoStop}%
\bibitem [{\citenamefont {Denk}\ \emph {et~al.}(2016)\citenamefont {Denk},
  \citenamefont {Huber}, \citenamefont {Reithmann},\ and\ \citenamefont
  {Frey}}]{Denk2016}%
  \BibitemOpen
  \bibfield  {author} {\bibinfo {author} {\bibfnamefont {Jonas}\ \bibnamefont
  {Denk}}, \bibinfo {author} {\bibfnamefont {Lorenz}\ \bibnamefont {Huber}},
  \bibinfo {author} {\bibfnamefont {Emanuel}\ \bibnamefont {Reithmann}}, \ and\
  \bibinfo {author} {\bibfnamefont {Erwin}\ \bibnamefont {Frey}},\ }\bibfield
  {title} {\enquote {\bibinfo {title} {Active curved polymers form vortex
  patterns on membranes},}\ }\href@noop {} {\bibfield  {journal} {\bibinfo
  {journal} {Phys. Rev. Lett.}\ }\textbf {\bibinfo {volume} {116}},\ \bibinfo
  {pages} {178301} (\bibinfo {year} {2016})}\BibitemShut {NoStop}%
\bibitem [{\citenamefont {Liebchen}\ \emph {et~al.}(2016)\citenamefont
  {Liebchen}, \citenamefont {Cates},\ and\ \citenamefont
  {Marenduzzo}}]{Liebchen2016}%
  \BibitemOpen
  \bibfield  {author} {\bibinfo {author} {\bibfnamefont {Benno}\ \bibnamefont
  {Liebchen}}, \bibinfo {author} {\bibfnamefont {Michael~E}\ \bibnamefont
  {Cates}}, \ and\ \bibinfo {author} {\bibfnamefont {Davide}\ \bibnamefont
  {Marenduzzo}},\ }\bibfield  {title} {\enquote {\bibinfo {title} {Pattern
  formation in chemically interacting active rotors with self-propulsion},}\
  }\href@noop {} {\bibfield  {journal} {\bibinfo  {journal} {Soft Matter}\
  }\textbf {\bibinfo {volume} {12}},\ \bibinfo {pages} {7259--7264} (\bibinfo
  {year} {2016})}\BibitemShut {NoStop}%
\bibitem [{\citenamefont {Liebchen}\ and\ \citenamefont
  {Levis}(2017)}]{Liebchen2017}%
  \BibitemOpen
  \bibfield  {author} {\bibinfo {author} {\bibfnamefont {Benno}\ \bibnamefont
  {Liebchen}}\ and\ \bibinfo {author} {\bibfnamefont {Demian}\ \bibnamefont
  {Levis}},\ }\bibfield  {title} {\enquote {\bibinfo {title} {Collective
  behavior of chiral active matter: pattern formation and enhanced flocking},}\
  }\href@noop {} {\bibfield  {journal} {\bibinfo  {journal} {Phys. Rev. Lett.}\
  }\textbf {\bibinfo {volume} {119}},\ \bibinfo {pages} {058002} (\bibinfo
  {year} {2017})}\BibitemShut {NoStop}%
\bibitem [{\citenamefont {Mahault}\ \emph {et~al.}(2018)\citenamefont
  {Mahault}, \citenamefont {Jiang}, \citenamefont {Bertin}, \citenamefont {Ma},
  \citenamefont {Patelli}, \citenamefont {Shi},\ and\ \citenamefont
  {Chat{\'e}}}]{Mahault2018}%
  \BibitemOpen
  \bibfield  {author} {\bibinfo {author} {\bibfnamefont {B}~\bibnamefont
  {Mahault}}, \bibinfo {author} {\bibfnamefont {X-c}\ \bibnamefont {Jiang}},
  \bibinfo {author} {\bibfnamefont {E}~\bibnamefont {Bertin}}, \bibinfo
  {author} {\bibfnamefont {Y-q}\ \bibnamefont {Ma}}, \bibinfo {author}
  {\bibfnamefont {A}~\bibnamefont {Patelli}}, \bibinfo {author} {\bibfnamefont
  {X-q}\ \bibnamefont {Shi}}, \ and\ \bibinfo {author} {\bibfnamefont
  {H}~\bibnamefont {Chat{\'e}}},\ }\bibfield  {title} {\enquote {\bibinfo
  {title} {Self-propelled particles with velocity reversals and ferromagnetic
  alignment: Active matter class with second-order transition to
  quasi-long-range polar order},}\ }\href@noop {} {\bibfield  {journal}
  {\bibinfo  {journal} {Phys. Rev. Lett.}\ }\textbf {\bibinfo {volume} {120}},\
  \bibinfo {pages} {258002} (\bibinfo {year} {2018})}\BibitemShut {NoStop}%
\bibitem [{\citenamefont {Bouchaud}\ and\ \citenamefont
  {Georges}(1990)}]{Bouchaud1990}%
  \BibitemOpen
  \bibfield  {author} {\bibinfo {author} {\bibfnamefont {Jean-Philippe}\
  \bibnamefont {Bouchaud}}\ and\ \bibinfo {author} {\bibfnamefont {Antoine}\
  \bibnamefont {Georges}},\ }\bibfield  {title} {\enquote {\bibinfo {title}
  {Anomalous diffusion in disordered media: statistical mechanisms, models and
  physical applications},}\ }\href@noop {} {\bibfield  {journal} {\bibinfo
  {journal} {Phys. Rep.}\ }\textbf {\bibinfo {volume} {195}},\ \bibinfo {pages}
  {127--293} (\bibinfo {year} {1990})}\BibitemShut {NoStop}%
\bibitem [{\citenamefont {Metzler}\ and\ \citenamefont
  {Klafter}(2000)}]{Metzler2000}%
  \BibitemOpen
  \bibfield  {author} {\bibinfo {author} {\bibfnamefont {R.}~\bibnamefont
  {Metzler}}\ and\ \bibinfo {author} {\bibfnamefont {J.}~\bibnamefont
  {Klafter}},\ }\bibfield  {title} {\enquote {\bibinfo {title} {The random
  walk's guide to anomalous diffusion: a fractional dynamics approach},}\
  }\href@noop {} {\bibfield  {journal} {\bibinfo  {journal} {Phys. Rep.}\
  }\textbf {\bibinfo {volume} {339}},\ \bibinfo {pages} {1--77} (\bibinfo
  {year} {2000})}\BibitemShut {NoStop}%
\bibitem [{\citenamefont {Metzler}\ and\ \citenamefont
  {Klafter}(2004)}]{Metzler2004}%
  \BibitemOpen
  \bibfield  {author} {\bibinfo {author} {\bibfnamefont {Ralf}\ \bibnamefont
  {Metzler}}\ and\ \bibinfo {author} {\bibfnamefont {Joseph}\ \bibnamefont
  {Klafter}},\ }\bibfield  {title} {\enquote {\bibinfo {title} {The restaurant
  at the end of the random walk: recent developments in the description of
  anomalous transport by fractional dynamics},}\ }\href@noop {} {\bibfield
  {journal} {\bibinfo  {journal} {J. Phys. A}\ }\textbf {\bibinfo {volume}
  {37}},\ \bibinfo {pages} {R161} (\bibinfo {year} {2004})}\BibitemShut
  {NoStop}%
\bibitem [{\citenamefont {Klages}\ \emph {et~al.}(2008)\citenamefont {Klages},
  \citenamefont {Radons},\ and\ \citenamefont {Sokolov}}]{Klages2008}%
  \BibitemOpen
  \bibfield  {author} {\bibinfo {author} {\bibfnamefont {R.}~\bibnamefont
  {Klages}}, \bibinfo {author} {\bibfnamefont {G.}~\bibnamefont {Radons}}, \
  and\ \bibinfo {author} {\bibfnamefont {I.~M.}\ \bibnamefont {Sokolov}},\
  }\href@noop {} {\emph {\bibinfo {title} {Anomalous transport: foundations and
  applications}}}\ (\bibinfo  {publisher} {John Wiley \& Sons},\ \bibinfo
  {year} {2008})\BibitemShut {NoStop}%
\bibitem [{\citenamefont {Zaburdaev}\ \emph {et~al.}(2015)\citenamefont
  {Zaburdaev}, \citenamefont {Denisov},\ and\ \citenamefont
  {Klafter}}]{Zaburdaev2015}%
  \BibitemOpen
  \bibfield  {author} {\bibinfo {author} {\bibfnamefont {V.}~\bibnamefont
  {Zaburdaev}}, \bibinfo {author} {\bibfnamefont {S.}~\bibnamefont {Denisov}},
  \ and\ \bibinfo {author} {\bibfnamefont {J.}~\bibnamefont {Klafter}},\
  }\bibfield  {title} {\enquote {\bibinfo {title} {L{\'e}vy walks},}\
  }\href@noop {} {\bibfield  {journal} {\bibinfo  {journal} {Rev. Mod. Phys.}\
  }\textbf {\bibinfo {volume} {87}},\ \bibinfo {pages} {483} (\bibinfo {year}
  {2015})}\BibitemShut {NoStop}%
\bibitem [{\citenamefont {Cairoli}\ and\ \citenamefont
  {Baule}(2015)}]{Cairoli2015}%
  \BibitemOpen
  \bibfield  {author} {\bibinfo {author} {\bibfnamefont {A.}~\bibnamefont
  {Cairoli}}\ and\ \bibinfo {author} {\bibfnamefont {A.}~\bibnamefont
  {Baule}},\ }\bibfield  {title} {\enquote {\bibinfo {title} {{Anomalous
  Processes with General Waiting Times: Functionals and Multipoint
  Structure}},}\ }\href@noop {} {\bibfield  {journal} {\bibinfo  {journal}
  {Phys. Rev. Lett.}\ }\textbf {\bibinfo {volume} {115}},\ \bibinfo {pages}
  {110601} (\bibinfo {year} {2015})}\BibitemShut {NoStop}%
\bibitem [{\citenamefont {Dieterich}\ \emph {et~al.}(2008)\citenamefont
  {Dieterich}, \citenamefont {Klages}, \citenamefont {Preuss},\ and\
  \citenamefont {Schwab}}]{Dieterich2008}%
  \BibitemOpen
  \bibfield  {author} {\bibinfo {author} {\bibfnamefont {Peter}\ \bibnamefont
  {Dieterich}}, \bibinfo {author} {\bibfnamefont {Rainer}\ \bibnamefont
  {Klages}}, \bibinfo {author} {\bibfnamefont {Roland}\ \bibnamefont {Preuss}},
  \ and\ \bibinfo {author} {\bibfnamefont {Albrecht}\ \bibnamefont {Schwab}},\
  }\bibfield  {title} {\enquote {\bibinfo {title} {Anomalous dynamics of cell
  migration},}\ }\href@noop {} {\bibfield  {journal} {\bibinfo  {journal}
  {Proc. Natl. Acad. Sci.}\ }\textbf {\bibinfo {volume} {105}},\ \bibinfo
  {pages} {459--463} (\bibinfo {year} {2008})}\BibitemShut {NoStop}%
\bibitem [{\citenamefont {Harris}\ \emph {et~al.}(2012)\citenamefont {Harris},
  \citenamefont {Banigan}, \citenamefont {Christian}, \citenamefont {Konradt},
  \citenamefont {Wojno}, \citenamefont {Norose}, \citenamefont {Wilson},
  \citenamefont {John}, \citenamefont {Weninger}, \citenamefont {Luster} \emph
  {et~al.}}]{Harris2012}%
  \BibitemOpen
  \bibfield  {author} {\bibinfo {author} {\bibfnamefont {Tajie~H}\ \bibnamefont
  {Harris}}, \bibinfo {author} {\bibfnamefont {Edward~J}\ \bibnamefont
  {Banigan}}, \bibinfo {author} {\bibfnamefont {David~A}\ \bibnamefont
  {Christian}}, \bibinfo {author} {\bibfnamefont {Christoph}\ \bibnamefont
  {Konradt}}, \bibinfo {author} {\bibfnamefont {Elia D~Tait}\ \bibnamefont
  {Wojno}}, \bibinfo {author} {\bibfnamefont {Kazumi}\ \bibnamefont {Norose}},
  \bibinfo {author} {\bibfnamefont {Emma~H}\ \bibnamefont {Wilson}}, \bibinfo
  {author} {\bibfnamefont {Beena}\ \bibnamefont {John}}, \bibinfo {author}
  {\bibfnamefont {Wolfgang}\ \bibnamefont {Weninger}}, \bibinfo {author}
  {\bibfnamefont {Andrew~D}\ \bibnamefont {Luster}},  \emph {et~al.},\
  }\bibfield  {title} {\enquote {\bibinfo {title} {Generalized l{\'e}vy walks
  and the role of chemokines in migration of effector cd8+ t cells},}\
  }\href@noop {} {\bibfield  {journal} {\bibinfo  {journal} {Nature}\ }\textbf
  {\bibinfo {volume} {486}},\ \bibinfo {pages} {545} (\bibinfo {year}
  {2012})}\BibitemShut {NoStop}%
\bibitem [{\citenamefont {Bressloff}\ and\ \citenamefont
  {Newby}(2013)}]{Bressloff2013}%
  \BibitemOpen
  \bibfield  {author} {\bibinfo {author} {\bibfnamefont {Paul~C}\ \bibnamefont
  {Bressloff}}\ and\ \bibinfo {author} {\bibfnamefont {Jay~M}\ \bibnamefont
  {Newby}},\ }\bibfield  {title} {\enquote {\bibinfo {title} {Stochastic models
  of intracellular transport},}\ }\href@noop {} {\bibfield  {journal} {\bibinfo
   {journal} {Rev. Mod. Phys.}\ }\textbf {\bibinfo {volume} {85}},\ \bibinfo
  {pages} {135} (\bibinfo {year} {2013})}\BibitemShut {NoStop}%
\bibitem [{\citenamefont {H{\"o}fling}\ and\ \citenamefont
  {Franosch}(2013)}]{Hofling2013}%
  \BibitemOpen
  \bibfield  {author} {\bibinfo {author} {\bibfnamefont {F.}~\bibnamefont
  {H{\"o}fling}}\ and\ \bibinfo {author} {\bibfnamefont {T.}~\bibnamefont
  {Franosch}},\ }\bibfield  {title} {\enquote {\bibinfo {title} {Anomalous
  transport in the crowded world of biological cells},}\ }\href@noop {}
  {\bibfield  {journal} {\bibinfo  {journal} {Rep. Progr. Phys.}\ }\textbf
  {\bibinfo {volume} {76}},\ \bibinfo {pages} {046602} (\bibinfo {year}
  {2013})}\BibitemShut {NoStop}%
\bibitem [{\citenamefont {Sokolov}(2012)}]{Sokolov2012}%
  \BibitemOpen
  \bibfield  {author} {\bibinfo {author} {\bibfnamefont {Igor~M}\ \bibnamefont
  {Sokolov}},\ }\bibfield  {title} {\enquote {\bibinfo {title} {Models of
  anomalous diffusion in crowded environments},}\ }\href@noop {} {\bibfield
  {journal} {\bibinfo  {journal} {Soft Matter}\ }\textbf {\bibinfo {volume}
  {8}},\ \bibinfo {pages} {9043--9052} (\bibinfo {year} {2012})}\BibitemShut
  {NoStop}%
\bibitem [{\citenamefont {Meroz}\ and\ \citenamefont
  {Sokolov}(2015)}]{Meroz2015}%
  \BibitemOpen
  \bibfield  {author} {\bibinfo {author} {\bibfnamefont {Yasmine}\ \bibnamefont
  {Meroz}}\ and\ \bibinfo {author} {\bibfnamefont {Igor~M}\ \bibnamefont
  {Sokolov}},\ }\bibfield  {title} {\enquote {\bibinfo {title} {A toolbox for
  determining subdiffusive mechanisms},}\ }\href@noop {} {\bibfield  {journal}
  {\bibinfo  {journal} {Phys. Rep.}\ }\textbf {\bibinfo {volume} {573}},\
  \bibinfo {pages} {1--29} (\bibinfo {year} {2015})}\BibitemShut {NoStop}%
\bibitem [{\citenamefont {Viswanathan}\ \emph {et~al.}(1999)\citenamefont
  {Viswanathan}, \citenamefont {Buldyrev}, \citenamefont {Havlin},
  \citenamefont {Da~Luz}, \citenamefont {Raposo},\ and\ \citenamefont
  {Stanley}}]{Viswanathan1999}%
  \BibitemOpen
  \bibfield  {author} {\bibinfo {author} {\bibfnamefont {Gandimohan~M}\
  \bibnamefont {Viswanathan}}, \bibinfo {author} {\bibfnamefont {Sergey~V}\
  \bibnamefont {Buldyrev}}, \bibinfo {author} {\bibfnamefont {Shlomo}\
  \bibnamefont {Havlin}}, \bibinfo {author} {\bibfnamefont {MGE}\ \bibnamefont
  {Da~Luz}}, \bibinfo {author} {\bibfnamefont {EP}~\bibnamefont {Raposo}}, \
  and\ \bibinfo {author} {\bibfnamefont {H~Eugene}\ \bibnamefont {Stanley}},\
  }\bibfield  {title} {\enquote {\bibinfo {title} {Optimizing the success of
  random searches},}\ }\href@noop {} {\bibfield  {journal} {\bibinfo  {journal}
  {Nature}\ }\textbf {\bibinfo {volume} {401}},\ \bibinfo {pages} {911}
  (\bibinfo {year} {1999})}\BibitemShut {NoStop}%
\bibitem [{\citenamefont {Lomholt}\ \emph {et~al.}(2008)\citenamefont
  {Lomholt}, \citenamefont {Tal}, \citenamefont {Metzler},\ and\ \citenamefont
  {Joseph}}]{Lomholt2008levy}%
  \BibitemOpen
  \bibfield  {author} {\bibinfo {author} {\bibfnamefont {Michael~A}\
  \bibnamefont {Lomholt}}, \bibinfo {author} {\bibfnamefont {Koren}\
  \bibnamefont {Tal}}, \bibinfo {author} {\bibfnamefont {Ralf}\ \bibnamefont
  {Metzler}}, \ and\ \bibinfo {author} {\bibfnamefont {Klafter}\ \bibnamefont
  {Joseph}},\ }\bibfield  {title} {\enquote {\bibinfo {title} {L{\'e}vy
  strategies in intermittent search processes are advantageous},}\ }\href@noop
  {} {\bibfield  {journal} {\bibinfo  {journal} {Proc. Natl. Acad. Sci.}\ }
  (\bibinfo {year} {2008})}\BibitemShut {NoStop}%
\bibitem [{\citenamefont {B{\'e}nichou}\ \emph {et~al.}(2011)\citenamefont
  {B{\'e}nichou}, \citenamefont {Loverdo}, \citenamefont {Moreau},\ and\
  \citenamefont {Voituriez}}]{Benichou2011}%
  \BibitemOpen
  \bibfield  {author} {\bibinfo {author} {\bibfnamefont {Olivier}\ \bibnamefont
  {B{\'e}nichou}}, \bibinfo {author} {\bibfnamefont {C}~\bibnamefont
  {Loverdo}}, \bibinfo {author} {\bibfnamefont {M}~\bibnamefont {Moreau}}, \
  and\ \bibinfo {author} {\bibfnamefont {R}~\bibnamefont {Voituriez}},\
  }\bibfield  {title} {\enquote {\bibinfo {title} {Intermittent search
  strategies},}\ }\href@noop {} {\bibfield  {journal} {\bibinfo  {journal}
  {Rev. Mod. Phys.}\ }\textbf {\bibinfo {volume} {83}},\ \bibinfo {pages} {81}
  (\bibinfo {year} {2011})}\BibitemShut {NoStop}%
\bibitem [{\citenamefont {Viswanathan}\ \emph {et~al.}(2011)\citenamefont
  {Viswanathan}, \citenamefont {Da~Luz}, \citenamefont {Raposo},\ and\
  \citenamefont {Stanley}}]{Viswanathan2011}%
  \BibitemOpen
  \bibfield  {author} {\bibinfo {author} {\bibfnamefont {Gandhimohan~M}\
  \bibnamefont {Viswanathan}}, \bibinfo {author} {\bibfnamefont {Marcos~GE}\
  \bibnamefont {Da~Luz}}, \bibinfo {author} {\bibfnamefont {Ernesto~P}\
  \bibnamefont {Raposo}}, \ and\ \bibinfo {author} {\bibfnamefont {H~Eugene}\
  \bibnamefont {Stanley}},\ }\href@noop {} {\emph {\bibinfo {title} {The
  physics of foraging: an introduction to random searches and biological
  encounters}}}\ (\bibinfo  {publisher} {Cambridge University Press},\ \bibinfo
  {year} {2011})\BibitemShut {NoStop}%
\bibitem [{\citenamefont {Cavagna}\ \emph {et~al.}(2013)\citenamefont
  {Cavagna}, \citenamefont {Queir{\'o}s}, \citenamefont {Giardina},
  \citenamefont {Stefanini},\ and\ \citenamefont {Viale}}]{Cavagna2013}%
  \BibitemOpen
  \bibfield  {author} {\bibinfo {author} {\bibfnamefont {Andrea}\ \bibnamefont
  {Cavagna}}, \bibinfo {author} {\bibfnamefont {SM~Duarte}\ \bibnamefont
  {Queir{\'o}s}}, \bibinfo {author} {\bibfnamefont {Irene}\ \bibnamefont
  {Giardina}}, \bibinfo {author} {\bibfnamefont {Fabio}\ \bibnamefont
  {Stefanini}}, \ and\ \bibinfo {author} {\bibfnamefont {Massimiliano}\
  \bibnamefont {Viale}},\ }\bibfield  {title} {\enquote {\bibinfo {title}
  {Diffusion of individual birds in starling flocks},}\ }\href@noop {}
  {\bibfield  {journal} {\bibinfo  {journal} {Proc. R. Soc. B}\ }\textbf
  {\bibinfo {volume} {280}},\ \bibinfo {pages} {20122484} (\bibinfo {year}
  {2013})}\BibitemShut {NoStop}%
\bibitem [{\citenamefont {Murakami}\ \emph {et~al.}(2015)\citenamefont
  {Murakami}, \citenamefont {Niizato}, \citenamefont {Tomaru}, \citenamefont
  {Nishiyama},\ and\ \citenamefont {Gunji}}]{Murakami2015}%
  \BibitemOpen
  \bibfield  {author} {\bibinfo {author} {\bibfnamefont {Hisashi}\ \bibnamefont
  {Murakami}}, \bibinfo {author} {\bibfnamefont {Takayuki}\ \bibnamefont
  {Niizato}}, \bibinfo {author} {\bibfnamefont {Takenori}\ \bibnamefont
  {Tomaru}}, \bibinfo {author} {\bibfnamefont {Yuta}\ \bibnamefont
  {Nishiyama}}, \ and\ \bibinfo {author} {\bibfnamefont {Yukio-Pegio}\
  \bibnamefont {Gunji}},\ }\bibfield  {title} {\enquote {\bibinfo {title}
  {Inherent noise appears as a {L{\'e}vy} walk in fish schools},}\ }\href@noop
  {} {\bibfield  {journal} {\bibinfo  {journal} {Sci. Rep.}\ }\textbf {\bibinfo
  {volume} {5}} (\bibinfo {year} {2015})}\BibitemShut {NoStop}%
\bibitem [{\citenamefont {Ariel}\ \emph {et~al.}(2015)\citenamefont {Ariel},
  \citenamefont {Rabani}, \citenamefont {Benisty}, \citenamefont {Partridge},
  \citenamefont {Harshey},\ and\ \citenamefont {Be'Er}}]{Ariel2015}%
  \BibitemOpen
  \bibfield  {author} {\bibinfo {author} {\bibfnamefont {Gil}\ \bibnamefont
  {Ariel}}, \bibinfo {author} {\bibfnamefont {Amit}\ \bibnamefont {Rabani}},
  \bibinfo {author} {\bibfnamefont {Sivan}\ \bibnamefont {Benisty}}, \bibinfo
  {author} {\bibfnamefont {Jonathan~D}\ \bibnamefont {Partridge}}, \bibinfo
  {author} {\bibfnamefont {Rasika~M}\ \bibnamefont {Harshey}}, \ and\ \bibinfo
  {author} {\bibfnamefont {Avraham}\ \bibnamefont {Be'Er}},\ }\bibfield
  {title} {\enquote {\bibinfo {title} {Swarming bacteria migrate by {L{\'e}vy}
  walk},}\ }\href@noop {} {\bibfield  {journal} {\bibinfo  {journal} {Nat.
  Commun.}\ }\textbf {\bibinfo {volume} {6}} (\bibinfo {year}
  {2015})}\BibitemShut {NoStop}%
\bibitem [{\citenamefont {Fedotov}\ and\ \citenamefont
  {Korabel}(2017)}]{Fedotov2017}%
  \BibitemOpen
  \bibfield  {author} {\bibinfo {author} {\bibfnamefont {Sergei}\ \bibnamefont
  {Fedotov}}\ and\ \bibinfo {author} {\bibfnamefont {Nickolay}\ \bibnamefont
  {Korabel}},\ }\bibfield  {title} {\enquote {\bibinfo {title} {Emergence of
  l{\'e}vy walks in systems of interacting individuals},}\ }\href@noop {}
  {\bibfield  {journal} {\bibinfo  {journal} {Phys. Rev. E}\ }\textbf {\bibinfo
  {volume} {95}},\ \bibinfo {pages} {030107(R)} (\bibinfo {year}
  {2017})}\BibitemShut {NoStop}%
\bibitem [{\citenamefont {Gnedenko}\ and\ \citenamefont
  {Kolmogorov}(1954)}]{Gnedenko1954}%
  \BibitemOpen
  \bibfield  {author} {\bibinfo {author} {\bibfnamefont {B.~V.}\ \bibnamefont
  {Gnedenko}}\ and\ \bibinfo {author} {\bibfnamefont {A.~N.}\ \bibnamefont
  {Kolmogorov}},\ }\href@noop {} {\emph {\bibinfo {title} {{Limit distributions
  for sums of independent random variables}}}}\ (\bibinfo  {publisher}
  {Addison-Wesley, Cambridge, United States},\ \bibinfo {year}
  {1954})\BibitemShut {NoStop}%
\bibitem [{\citenamefont {Cairoli}\ and\ \citenamefont {Lee}()}]{CairoliL}%
  \BibitemOpen
  \bibfield  {author} {\bibinfo {author} {\bibfnamefont {Andrea}\ \bibnamefont
  {Cairoli}}\ and\ \bibinfo {author} {\bibfnamefont {Chiu~Fan}\ \bibnamefont
  {Lee}},\ }\href@noop {} {\enquote {\bibinfo {title} {Hydrodynamics of active
  {L}{\'e}vy matter},}\ }\bibinfo {note} {The accompanying Letter}\BibitemShut
  {NoStop}%
\bibitem [{\citenamefont {Mandelbrot}(1982)}]{Mandelbrot1982}%
  \BibitemOpen
  \bibfield  {author} {\bibinfo {author} {\bibfnamefont {Benoit~B}\
  \bibnamefont {Mandelbrot}},\ }\href@noop {} {\emph {\bibinfo {title} {The
  fractal geometry of nature}}},\ Vol.~\bibinfo {volume} {2}\ (\bibinfo
  {publisher} {WH freeman New York},\ \bibinfo {year} {1982})\BibitemShut
  {NoStop}%
\bibitem [{\citenamefont {Hughes}\ \emph {et~al.}(1981)\citenamefont {Hughes},
  \citenamefont {Shlesinger},\ and\ \citenamefont {Montroll}}]{Hughes1981}%
  \BibitemOpen
  \bibfield  {author} {\bibinfo {author} {\bibfnamefont {Barry~D}\ \bibnamefont
  {Hughes}}, \bibinfo {author} {\bibfnamefont {Michael~F}\ \bibnamefont
  {Shlesinger}}, \ and\ \bibinfo {author} {\bibfnamefont {Elliott~W}\
  \bibnamefont {Montroll}},\ }\bibfield  {title} {\enquote {\bibinfo {title}
  {Random walks with self-similar clusters},}\ }\href@noop {} {\bibfield
  {journal} {\bibinfo  {journal} {Proc. Natl. Acad. Sci.}\ }\textbf {\bibinfo
  {volume} {78}},\ \bibinfo {pages} {3287--3291} (\bibinfo {year}
  {1981})}\BibitemShut {NoStop}%
\bibitem [{\citenamefont {Shlesinger}\ \emph {et~al.}(1995)\citenamefont
  {Shlesinger}, \citenamefont {Zaslavsky},\ and\ \citenamefont
  {Frisch}}]{Shlesinger1995}%
  \BibitemOpen
  \bibfield  {author} {\bibinfo {author} {\bibfnamefont {Micheal~F}\
  \bibnamefont {Shlesinger}}, \bibinfo {author} {\bibfnamefont {George~M}\
  \bibnamefont {Zaslavsky}}, \ and\ \bibinfo {author} {\bibfnamefont {Uriel}\
  \bibnamefont {Frisch}},\ }\href@noop {} {\emph {\bibinfo {title} {L{\'e}vy
  flights and related topics in {Physics}}}},\ \bibinfo {series} {Lecture notes
  in {Physics}}, Vol.\ \bibinfo {volume} {450}\ (\bibinfo  {publisher}
  {Springer-Verlag},\ \bibinfo {year} {1995})\BibitemShut {NoStop}%
\bibitem [{\citenamefont {Gro{\ss}mann}\ \emph {et~al.}(2016)\citenamefont
  {Gro{\ss}mann}, \citenamefont {Peruani},\ and\ \citenamefont
  {B{\"a}r}}]{Grossmann2016}%
  \BibitemOpen
  \bibfield  {author} {\bibinfo {author} {\bibfnamefont {R.}~\bibnamefont
  {Gro{\ss}mann}}, \bibinfo {author} {\bibfnamefont {F.}~\bibnamefont
  {Peruani}}, \ and\ \bibinfo {author} {\bibfnamefont {M.}~\bibnamefont
  {B{\"a}r}},\ }\bibfield  {title} {\enquote {\bibinfo {title} {Superdiffusion,
  large-scale synchronization, and topological defects},}\ }\href@noop {}
  {\bibfield  {journal} {\bibinfo  {journal} {Phys. Rev. E}\ }\textbf {\bibinfo
  {volume} {93}},\ \bibinfo {pages} {040102(R)} (\bibinfo {year}
  {2016})}\BibitemShut {NoStop}%
\bibitem [{\citenamefont {Estrada-Rodriguez}\ and\ \citenamefont
  {Gimperlein}(2018)}]{Estrada2018}%
  \BibitemOpen
  \bibfield  {author} {\bibinfo {author} {\bibfnamefont {Gissell}\ \bibnamefont
  {Estrada-Rodriguez}}\ and\ \bibinfo {author} {\bibfnamefont {Heiko}\
  \bibnamefont {Gimperlein}},\ }\bibfield  {title} {\enquote {\bibinfo {title}
  {Swarming of interacting robots with {L}{\'e}vy strategies: a macroscopic
  description},}\ }\href@noop {} {\bibfield  {journal} {\bibinfo  {journal}
  {arXiv preprint arXiv:1807.10124}\ } (\bibinfo {year} {2018})}\BibitemShut
  {NoStop}%
\bibitem [{\citenamefont {Huang}(1987)}]{Huang1987}%
  \BibitemOpen
  \bibfield  {author} {\bibinfo {author} {\bibfnamefont {Kerson}\ \bibnamefont
  {Huang}},\ }\href@noop {} {\emph {\bibinfo {title} {Statistical
  Mechanics}}},\ \bibinfo {edition} {2nd}\ ed.\ (\bibinfo  {publisher} {New
  York: John Wiley \& Sons},\ \bibinfo {year} {1987})\BibitemShut {NoStop}%
\bibitem [{\citenamefont {Applebaum}(2009)}]{Applebaum2009}%
  \BibitemOpen
  \bibfield  {author} {\bibinfo {author} {\bibfnamefont {David}\ \bibnamefont
  {Applebaum}},\ }\href@noop {} {\emph {\bibinfo {title} {L{\'e}vy processes
  and stochastic calculus}}}\ (\bibinfo  {publisher} {Cambridge university
  press},\ \bibinfo {year} {2009})\BibitemShut {NoStop}%
\bibitem [{\citenamefont {Tankov}(2003)}]{Tankov2003}%
  \BibitemOpen
  \bibfield  {author} {\bibinfo {author} {\bibfnamefont {Peter}\ \bibnamefont
  {Tankov}},\ }\href@noop {} {\emph {\bibinfo {title} {Financial modelling with
  jump processes}}},\ Vol.~\bibinfo {volume} {2}\ (\bibinfo  {publisher} {CRC
  press},\ \bibinfo {year} {2003})\BibitemShut {NoStop}%
\bibitem [{\citenamefont {Mikusinski}(1959)}]{Mikusinski1959}%
  \BibitemOpen
  \bibfield  {author} {\bibinfo {author} {\bibfnamefont {J}~\bibnamefont
  {Mikusinski}},\ }\bibfield  {title} {\enquote {\bibinfo {title} {On the
  function whose laplace transform is $\exp{?sa}$},}\ }\href@noop {}
  {\bibfield  {journal} {\bibinfo  {journal} {Stud. Math.}\ }\textbf {\bibinfo
  {volume} {18}},\ \bibinfo {pages} {191} (\bibinfo {year} {1959})}\BibitemShut
  {NoStop}%
\bibitem [{\citenamefont {Bendler}(1984)}]{Bendler1984}%
  \BibitemOpen
  \bibfield  {author} {\bibinfo {author} {\bibfnamefont {John~T}\ \bibnamefont
  {Bendler}},\ }\bibfield  {title} {\enquote {\bibinfo {title} {L{\'e}vy
  (stable) probability densities and mechanical relaxation in solid
  polymers},}\ }\href@noop {} {\bibfield  {journal} {\bibinfo  {journal} {J.
  Stat. Phys.}\ }\textbf {\bibinfo {volume} {36}},\ \bibinfo {pages} {625--637}
  (\bibinfo {year} {1984})}\BibitemShut {NoStop}%
\bibitem [{\citenamefont {Hilfer}(2002)}]{Hilfer2002}%
  \BibitemOpen
  \bibfield  {author} {\bibinfo {author} {\bibfnamefont {R}~\bibnamefont
  {Hilfer}},\ }\bibfield  {title} {\enquote {\bibinfo {title} {H-function
  representations for stretched exponential relaxation and non-{Debye}
  susceptibilities in glassy systems},}\ }\href@noop {} {\bibfield  {journal}
  {\bibinfo  {journal} {Phys. Rev. E}\ }\textbf {\bibinfo {volume} {65}},\
  \bibinfo {pages} {061510} (\bibinfo {year} {2002})}\BibitemShut {NoStop}%
\bibitem [{\citenamefont {Penson}\ and\ \citenamefont
  {G{\'o}rska}(2010)}]{Penson2010}%
  \BibitemOpen
  \bibfield  {author} {\bibinfo {author} {\bibfnamefont {KA}~\bibnamefont
  {Penson}}\ and\ \bibinfo {author} {\bibfnamefont {K}~\bibnamefont
  {G{\'o}rska}},\ }\bibfield  {title} {\enquote {\bibinfo {title} {Exact and
  explicit probability densities for one-sided {L\'e}vy stable
  distributions},}\ }\href@noop {} {\bibfield  {journal} {\bibinfo  {journal}
  {Phys. Rev. Lett.}\ }\textbf {\bibinfo {volume} {105}},\ \bibinfo {pages}
  {210604} (\bibinfo {year} {2010})}\BibitemShut {NoStop}%
\bibitem [{\citenamefont {Fogedby}(1994)}]{Fogedby1994}%
  \BibitemOpen
  \bibfield  {author} {\bibinfo {author} {\bibfnamefont {Hans~C}\ \bibnamefont
  {Fogedby}},\ }\bibfield  {title} {\enquote {\bibinfo {title} {Langevin
  equations for continuous time {L{\'e}vy} flights},}\ }\href@noop {}
  {\bibfield  {journal} {\bibinfo  {journal} {Phys. Rev. E}\ }\textbf {\bibinfo
  {volume} {50}},\ \bibinfo {pages} {1657} (\bibinfo {year}
  {1994})}\BibitemShut {NoStop}%
\bibitem [{\citenamefont {Brilliantov}\ and\ \citenamefont
  {P{\"o}schel}(2010)}]{Brilliantov2010}%
  \BibitemOpen
  \bibfield  {author} {\bibinfo {author} {\bibfnamefont {Nikolai~V}\
  \bibnamefont {Brilliantov}}\ and\ \bibinfo {author} {\bibfnamefont
  {Thorsten}\ \bibnamefont {P{\"o}schel}},\ }\href@noop {} {\emph {\bibinfo
  {title} {Kinetic theory of granular gases}}}\ (\bibinfo  {publisher} {Oxford
  University Press},\ \bibinfo {year} {2010})\BibitemShut {NoStop}%
\bibitem [{\citenamefont {Vlasov}(1968)}]{Vlasov1968}%
  \BibitemOpen
  \bibfield  {author} {\bibinfo {author} {\bibfnamefont
  {Anatoly~Aleksandrovich}\ \bibnamefont {Vlasov}},\ }\bibfield  {title}
  {\enquote {\bibinfo {title} {The vibrational properties of an electron
  gas},}\ }\href@noop {} {\bibfield  {journal} {\bibinfo  {journal}
  {Phys.-Uspekhi}\ }\textbf {\bibinfo {volume} {10}},\ \bibinfo {pages}
  {721--733} (\bibinfo {year} {1968})}\BibitemShut {NoStop}%
\bibitem [{\citenamefont {Samko}\ \emph {et~al.}(1993)\citenamefont {Samko},
  \citenamefont {Kilbas}, \citenamefont {Marichev} \emph {et~al.}}]{Samko1993}%
  \BibitemOpen
  \bibfield  {author} {\bibinfo {author} {\bibfnamefont {Stefan~G}\
  \bibnamefont {Samko}}, \bibinfo {author} {\bibfnamefont {Anatoly~A}\
  \bibnamefont {Kilbas}}, \bibinfo {author} {\bibfnamefont {Oleg~I}\
  \bibnamefont {Marichev}},  \emph {et~al.},\ }\href@noop {} {\emph {\bibinfo
  {title} {Fractional integrals and derivatives: Theory and Applications}}},\
  Vol.\ \bibinfo {volume} {1993}\ (\bibinfo  {publisher} {Gordon and Breach,
  Yverdon},\ \bibinfo {year} {1993})\BibitemShut {NoStop}%
\bibitem [{\citenamefont {Meerschaert}\ \emph {et~al.}(1999)\citenamefont
  {Meerschaert}, \citenamefont {Benson},\ and\ \citenamefont
  {B{\"a}umer}}]{Meerschaert1999}%
  \BibitemOpen
  \bibfield  {author} {\bibinfo {author} {\bibfnamefont {Mark~M}\ \bibnamefont
  {Meerschaert}}, \bibinfo {author} {\bibfnamefont {David~A}\ \bibnamefont
  {Benson}}, \ and\ \bibinfo {author} {\bibfnamefont {Boris}\ \bibnamefont
  {B{\"a}umer}},\ }\bibfield  {title} {\enquote {\bibinfo {title}
  {Multidimensional advection and fractional dispersion},}\ }\href@noop {}
  {\bibfield  {journal} {\bibinfo  {journal} {Phys. Rev. E}\ }\textbf {\bibinfo
  {volume} {59}},\ \bibinfo {pages} {5026} (\bibinfo {year}
  {1999})}\BibitemShut {NoStop}%
\bibitem [{\citenamefont {Taylor-King}\ \emph {et~al.}(2016)\citenamefont
  {Taylor-King}, \citenamefont {Klages}, \citenamefont {Fedotov},\ and\
  \citenamefont {Van~Gorder}}]{taylor2016}%
  \BibitemOpen
  \bibfield  {author} {\bibinfo {author} {\bibfnamefont {Jake~P}\ \bibnamefont
  {Taylor-King}}, \bibinfo {author} {\bibfnamefont {Rainer}\ \bibnamefont
  {Klages}}, \bibinfo {author} {\bibfnamefont {Sergei}\ \bibnamefont
  {Fedotov}}, \ and\ \bibinfo {author} {\bibfnamefont {Robert~A}\ \bibnamefont
  {Van~Gorder}},\ }\bibfield  {title} {\enquote {\bibinfo {title} {Fractional
  diffusion equation for an n-dimensional correlated {L}{\'e}vy walk},}\
  }\href@noop {} {\bibfield  {journal} {\bibinfo  {journal} {Phys. Rev. E}\
  }\textbf {\bibinfo {volume} {94}},\ \bibinfo {pages} {012104} (\bibinfo
  {year} {2016})}\BibitemShut {NoStop}%
\bibitem [{\citenamefont {Tarasov}(2011)}]{Tarasov2011}%
  \BibitemOpen
  \bibfield  {author} {\bibinfo {author} {\bibfnamefont {Vasily~E}\
  \bibnamefont {Tarasov}},\ }\href@noop {} {\emph {\bibinfo {title} {Fractional
  dynamics: applications of fractional calculus to dynamics of particles,
  fields and media}}}\ (\bibinfo  {publisher} {Springer Science \& Business
  Media},\ \bibinfo {year} {2011})\BibitemShut {NoStop}%
\bibitem [{\citenamefont {Fouxon}\ \emph {et~al.}(2017)\citenamefont {Fouxon},
  \citenamefont {Denisov}, \citenamefont {Zaburdaev},\ and\ \citenamefont
  {Barkai}}]{Fouxon2017}%
  \BibitemOpen
  \bibfield  {author} {\bibinfo {author} {\bibfnamefont {Itzhak}\ \bibnamefont
  {Fouxon}}, \bibinfo {author} {\bibfnamefont {Sergey}\ \bibnamefont
  {Denisov}}, \bibinfo {author} {\bibfnamefont {Vasily}\ \bibnamefont
  {Zaburdaev}}, \ and\ \bibinfo {author} {\bibfnamefont {Eli}\ \bibnamefont
  {Barkai}},\ }\bibfield  {title} {\enquote {\bibinfo {title} {Limit theorems
  for {L\'e}vy walks in d dimensions: rare and bulk fluctuations},}\
  }\href@noop {} {\bibfield  {journal} {\bibinfo  {journal} {J. Phys. A}\
  }\textbf {\bibinfo {volume} {50}},\ \bibinfo {pages} {154002} (\bibinfo
  {year} {2017})}\BibitemShut {NoStop}%
\bibitem [{\citenamefont {Landau}\ and\ \citenamefont
  {Binder}(2014)}]{Landau2014}%
  \BibitemOpen
  \bibfield  {author} {\bibinfo {author} {\bibfnamefont {David~P}\ \bibnamefont
  {Landau}}\ and\ \bibinfo {author} {\bibfnamefont {Kurt}\ \bibnamefont
  {Binder}},\ }\href@noop {} {\emph {\bibinfo {title} {{A Guide to Monte Carlo
  Simulations in Statistical Physics}}}}\ (\bibinfo  {publisher} {Cambridge
  University Press, England},\ \bibinfo {year} {2014})\BibitemShut {NoStop}%
\bibitem [{\citenamefont {Binder}(1981)}]{Binder1981}%
  \BibitemOpen
  \bibfield  {author} {\bibinfo {author} {\bibfnamefont {Kurt}\ \bibnamefont
  {Binder}},\ }\bibfield  {title} {\enquote {\bibinfo {title} {Finite size
  scaling analysis of ising model block distribution functions},}\ }\href@noop
  {} {\bibfield  {journal} {\bibinfo  {journal} {Zeitschrift f{\"u}r Physik B
  Condensed Matter}\ }\textbf {\bibinfo {volume} {43}},\ \bibinfo {pages}
  {119--140} (\bibinfo {year} {1981})}\BibitemShut {NoStop}%
\bibitem [{\citenamefont {Fisher}\ \emph {et~al.}(1972)\citenamefont {Fisher},
  \citenamefont {Ma},\ and\ \citenamefont {Nickel}}]{Fisher1972}%
  \BibitemOpen
  \bibfield  {author} {\bibinfo {author} {\bibfnamefont {Michael~E}\
  \bibnamefont {Fisher}}, \bibinfo {author} {\bibfnamefont {Shang-keng}\
  \bibnamefont {Ma}}, \ and\ \bibinfo {author} {\bibfnamefont {BG}~\bibnamefont
  {Nickel}},\ }\bibfield  {title} {\enquote {\bibinfo {title} {Critical
  exponents for long-range interactions},}\ }\href@noop {} {\bibfield
  {journal} {\bibinfo  {journal} {Phys. Rev. Lett.}\ }\textbf {\bibinfo
  {volume} {29}},\ \bibinfo {pages} {917} (\bibinfo {year} {1972})}\BibitemShut
  {NoStop}%
\bibitem [{\citenamefont {Suzuki}(1973)}]{Suzuki1973}%
  \BibitemOpen
  \bibfield  {author} {\bibinfo {author} {\bibfnamefont {Masuo}\ \bibnamefont
  {Suzuki}},\ }\bibfield  {title} {\enquote {\bibinfo {title} {Critical
  exponents for long-range interactions. i: Dimensionality, symmetry and
  potential-range},}\ }\href@noop {} {\bibfield  {journal} {\bibinfo  {journal}
  {Prog. Theor. Phys.}\ }\textbf {\bibinfo {volume} {49}},\ \bibinfo {pages}
  {424--441} (\bibinfo {year} {1973})}\BibitemShut {NoStop}%
\bibitem [{\citenamefont {Chen}\ \emph {et~al.}(2015)\citenamefont {Chen},
  \citenamefont {Toner},\ and\ \citenamefont {Lee}}]{Chen2015}%
  \BibitemOpen
  \bibfield  {author} {\bibinfo {author} {\bibfnamefont {Leiming}\ \bibnamefont
  {Chen}}, \bibinfo {author} {\bibfnamefont {John}\ \bibnamefont {Toner}}, \
  and\ \bibinfo {author} {\bibfnamefont {Chiu~Fan}\ \bibnamefont {Lee}},\
  }\bibfield  {title} {\enquote {\bibinfo {title} {Critical phenomenon of the
  order--disorder transition in incompressible active fluids},}\ }\href@noop {}
  {\bibfield  {journal} {\bibinfo  {journal} {New J. Phys.}\ }\textbf {\bibinfo
  {volume} {17}},\ \bibinfo {pages} {042002} (\bibinfo {year}
  {2015})}\BibitemShut {NoStop}%
\bibitem [{\citenamefont {Chen}\ \emph {et~al.}(2016)\citenamefont {Chen},
  \citenamefont {Lee},\ and\ \citenamefont {Toner}}]{Chen2016}%
  \BibitemOpen
  \bibfield  {author} {\bibinfo {author} {\bibfnamefont {Leiming}\ \bibnamefont
  {Chen}}, \bibinfo {author} {\bibfnamefont {Chiu~Fan}\ \bibnamefont {Lee}}, \
  and\ \bibinfo {author} {\bibfnamefont {John}\ \bibnamefont {Toner}},\
  }\bibfield  {title} {\enquote {\bibinfo {title} {Mapping two-dimensional
  polar active fluids to two-dimensional soap and one-dimensional
  sandblasting},}\ }\href@noop {} {\bibfield  {journal} {\bibinfo  {journal}
  {Nat. Commun.}\ }\textbf {\bibinfo {volume} {7}} (\bibinfo {year}
  {2016})}\BibitemShut {NoStop}%
\bibitem [{\citenamefont {Chen}\ \emph {et~al.}(2018)\citenamefont {Chen},
  \citenamefont {Lee},\ and\ \citenamefont {Toner}}]{Chen2018}%
  \BibitemOpen
  \bibfield  {author} {\bibinfo {author} {\bibfnamefont {Leiming}\ \bibnamefont
  {Chen}}, \bibinfo {author} {\bibfnamefont {Chiu~Fan}\ \bibnamefont {Lee}}, \
  and\ \bibinfo {author} {\bibfnamefont {John}\ \bibnamefont {Toner}},\
  }\bibfield  {title} {\enquote {\bibinfo {title} {Incompressible polar active
  fluids in the moving phase in dimensions d> 2},}\ }\href@noop {} {\bibfield
  {journal} {\bibinfo  {journal} {New J. Phys.}\ }\textbf {\bibinfo {volume}
  {20}},\ \bibinfo {pages} {113035} (\bibinfo {year} {2018})}\BibitemShut
  {NoStop}%
\bibitem [{\citenamefont {Partridge}\ and \citenamefont {Lee}(2019)}]{Partridge2019}%
  \BibitemOpen
  \bibfield  {author} {\bibinfo {author} {\bibfnamefont {B.}\ \bibnamefont
  {Partridge}}, \ and\ \bibinfo {author} {\bibfnamefont {C.~F.}\ \bibnamefont {Lee}}}
  \bibfield  {title} {\enquote {\bibinfo {title} {Critical motility-induced phase separation belongs to the Ising universality class},}\ }\href@noop {} {\bibfield
  {journal} {\bibinfo  {journal} {Phys. Rev. Lett.}\ }\textbf {\bibinfo {volume}
  {123}},\ \bibinfo {pages} {068002} (\bibinfo {year} {2019})}\BibitemShut
  {NoStop}%
\bibitem [{\citenamefont {Siebert}\ \emph {et~al.}(2018)}]{Siebert2018}%
  \BibitemOpen
  \bibfield  {author} {{\bibfnamefont {J.~T.}\ \bibnamefont
  {Siebert}}, \ \bibinfo {author} {\bibfnamefont {F.}\ \bibnamefont
  {Dittrich}}, \ \bibinfo {author} {\bibfnamefont {F.}\ \bibnamefont
  {Schmid}}, \ \bibinfo {author} {\bibfnamefont {K.}\ \bibnamefont
  {Binder}}, \ \bibinfo {author} {\bibfnamefont {T.}\ \bibnamefont
  {Speck}}, \ and\ \bibinfo {author} {\bibfnamefont {P.}\ \bibnamefont {Virnau}}}
  \bibfield  {title} {\enquote {\bibinfo {title} {Critical behavior of active Brownian particles},}\ }\href@noop {} {\bibfield
  {journal} {\bibinfo  {journal} {Phys. Rev. E}\ }\textbf {\bibinfo {volume}
  {98}},\ \bibinfo {pages} {030601(R)} (\bibinfo {year} {2018})}\BibitemShut
  {NoStop}%
\bibitem [{\citenamefont {Caballero}\ \emph {et~al.}(2018)}]{Caballero2018}%
  \BibitemOpen
  \bibfield  {author} {\bibinfo {author} {\bibfnamefont {F.}\ \bibnamefont
  {Caballero}}, \ \bibinfo {author} {\bibfnamefont {C.}\ \bibnamefont
  {Nardini}}, \ and\ \bibinfo {author} {\bibfnamefont {M.~E.}\ \bibnamefont {Cates}}}
  \bibfield  {title} {\enquote {\bibinfo {title} {From bulk to microphase separation in scalar active matter: A perturbative renormalization group analysis},}\ }\href@noop {} {\bibfield
  {journal} {\bibinfo  {journal} {J. Stat. Mech.}\ }\textbf {\bibinfo {volume}
  {12}},\ \bibinfo {pages} {123208} (\bibinfo {year} {2018})}\BibitemShut
  {NoStop}%
\bibitem [{\citenamefont {Nesbitt}\ \emph {et~al.}(2019)}]{Nesbitt2019}%
  \BibitemOpen
  \bibfield  {author} {\bibinfo {author} {\bibfnamefont {D.}\ \bibnamefont
  {Nesbitt}},\ \bibinfo {author} {\bibfnamefont {G.}\ \bibnamefont
  {Pruessner}}\ and\ \bibinfo {author} {\bibfnamefont {C.~F.}\ \bibnamefont
  {Lee}},\ }\href@noop {} {\enquote {\bibinfo {title} {Using a lattice Boltzmann method to uncover novel phase transitions in dry polar active fluids},}\ }\bibinfo {note} {arXiv:1902.00530}\BibitemShut
  {NoStop}%
\bibitem [{\citenamefont {Majumdar}(2005)}]{Majumdar2005}%
  \BibitemOpen
  \bibfield  {author} {\bibinfo {author} {\bibfnamefont {Satya~N}\ \bibnamefont
  {Majumdar}},\ }\bibfield  {title} {\enquote {\bibinfo {title} {{Brownian
  functionals in Physics and Computer Science}},}\ }\href@noop {} {\bibfield
  {journal} {\bibinfo  {journal} {Curr. Sci.}\ }\textbf {\bibinfo {volume}
  {88}} (\bibinfo {year} {2005})}\BibitemShut {NoStop}%
\bibitem [{\citenamefont {Godreche}\ and\ \citenamefont
  {Luck}(2001)}]{Godreche2001}%
  \BibitemOpen
  \bibfield  {author} {\bibinfo {author} {\bibfnamefont {C}~\bibnamefont
  {Godreche}}\ and\ \bibinfo {author} {\bibfnamefont {JM}~\bibnamefont
  {Luck}},\ }\bibfield  {title} {\enquote {\bibinfo {title} {Statistics of the
  occupation time of renewal processes},}\ }\href@noop {} {\bibfield  {journal}
  {\bibinfo  {journal} {J. Stat. Phys.}\ }\textbf {\bibinfo {volume} {104}},\
  \bibinfo {pages} {489--524} (\bibinfo {year} {2001})}\BibitemShut {NoStop}%
\bibitem [{\citenamefont {Rapaport}(2004)}]{Rapaport2004}%
  \BibitemOpen
  \bibfield  {author} {\bibinfo {author} {\bibfnamefont {Dennis~C}\
  \bibnamefont {Rapaport}},\ }\href@noop {} {\emph {\bibinfo {title} {The art
  of molecular dynamics simulation}}}\ (\bibinfo  {publisher} {Cambridge
  university press},\ \bibinfo {year} {2004})\BibitemShut {NoStop}%
\bibitem [{\citenamefont {Janicki}\ and\ \citenamefont
  {Weron}(1993)}]{Janicki1993}%
  \BibitemOpen
  \bibfield  {author} {\bibinfo {author} {\bibfnamefont {Aleksand}\
  \bibnamefont {Janicki}}\ and\ \bibinfo {author} {\bibfnamefont {Aleksander}\
  \bibnamefont {Weron}},\ }\href@noop {} {\emph {\bibinfo {title} {Simulation
  and chaotic behavior of alpha-stable stochastic processes}}},\ Vol.\ \bibinfo
  {volume} {178}\ (\bibinfo  {publisher} {CRC Press},\ \bibinfo {year}
  {1993})\BibitemShut {NoStop}%
\bibitem [{\citenamefont {Kleinhans}\ and\ \citenamefont
  {Friedrich}(2007)}]{Kleinhans2007}%
  \BibitemOpen
  \bibfield  {author} {\bibinfo {author} {\bibfnamefont {D}~\bibnamefont
  {Kleinhans}}\ and\ \bibinfo {author} {\bibfnamefont {R}~\bibnamefont
  {Friedrich}},\ }\bibfield  {title} {\enquote {\bibinfo {title}
  {Continuous-time random walks: Simulation of continuous trajectories},}\
  }\href@noop {} {\bibfield  {journal} {\bibinfo  {journal} {Phys. Rev. E}\
  }\textbf {\bibinfo {volume} {76}},\ \bibinfo {pages} {061102} (\bibinfo
  {year} {2007})}\BibitemShut {NoStop}%
\bibitem [{\citenamefont {Press}\ \emph {et~al.}(2007)\citenamefont {Press},
  \citenamefont {Teukolsky}, \citenamefont {Vetterling},\ and\ \citenamefont
  {Flannery}}]{Press2007}%
  \BibitemOpen
  \bibfield  {author} {\bibinfo {author} {\bibfnamefont {William~H}\
  \bibnamefont {Press}}, \bibinfo {author} {\bibfnamefont {Saul~A}\
  \bibnamefont {Teukolsky}}, \bibinfo {author} {\bibfnamefont {William~T}\
  \bibnamefont {Vetterling}}, \ and\ \bibinfo {author} {\bibfnamefont
  {Brian~P}\ \bibnamefont {Flannery}},\ }\href@noop {} {\emph {\bibinfo {title}
  {Numerical recipes 3rd edition: The art of scientific computing}}}\ (\bibinfo
   {publisher} {Cambridge university press},\ \bibinfo {year}
  {2007})\BibitemShut {NoStop}%
\bibitem [{\citenamefont {Baeumer}\ and\ \citenamefont
  {Meerschaert}(2010)}]{Baeumer2010}%
  \BibitemOpen
  \bibfield  {author} {\bibinfo {author} {\bibfnamefont {Boris}\ \bibnamefont
  {Baeumer}}\ and\ \bibinfo {author} {\bibfnamefont {Mark~M}\ \bibnamefont
  {Meerschaert}},\ }\bibfield  {title} {\enquote {\bibinfo {title} {Tempered
  stable l{\'e}vy motion and transient super-diffusion},}\ }\href@noop {}
  {\bibfield  {journal} {\bibinfo  {journal} {J. Comput. Appl. Math.}\ }\textbf
  {\bibinfo {volume} {233}},\ \bibinfo {pages} {2438--2448} (\bibinfo {year}
  {2010})}\BibitemShut {NoStop}%
\bibitem [{\citenamefont {Weron}(1996)}]{Weron1996}%
  \BibitemOpen
  \bibfield  {author} {\bibinfo {author} {\bibfnamefont {Rafa{\l}}\
  \bibnamefont {Weron}},\ }\bibfield  {title} {\enquote {\bibinfo {title} {On
  the {Chambers-Mallows-Stuck} method for simulating skewed stable random
  variables},}\ }\href@noop {} {\bibfield  {journal} {\bibinfo  {journal}
  {Stat. Probabil. Lett.}\ }\textbf {\bibinfo {volume} {28}},\ \bibinfo {pages}
  {165--171} (\bibinfo {year} {1996})}\BibitemShut {NoStop}%
\end{thebibliography}
\end{document}